\documentclass[a4paper,fleqn,usenatbib]{mnras}

\usepackage{mathptmx}

\usepackage[T1]{fontenc}
\usepackage{ae,aecompl}

\usepackage{graphicx}	
\usepackage{amsmath}	
\usepackage{amssymb}	
\usepackage{multirow}

\usepackage[english]{babel}
\usepackage{blindtext}

\usepackage[T1]{fontenc}
\usepackage{bm}

\DeclareMathAlphabet{\mathsfit}{\encodingdefault}{\sfdefault}{m}{sl}
\SetMathAlphabet{\mathsfit}{bold}{\encodingdefault}{\sfdefault}{bx}{sl}

\title[Modelling the CCE X-rays of $\eta$~Carinae]{Modelling the Central Constant Emission X-ray component of $\eta$~Carinae}

\author[C. M. P. Russell et al.]{Christopher M. P. Russell,$^{1}$\thanks{E-mail: crussell@udel.edu (CMPR)}
Michael F. Corcoran,$^{2,3}$
Kenji Hamaguchi,$^{2,4}$\newauthor
Thomas I. Madura,$^{3,5}$
Stanley P. Owocki$^{6}$
and D. John Hillier$^{7}$
\\
$^{1}$X-ray Astrophysics Laboratory, Code 662, NASA/Goddard Space Flight Center, Greenbelt, MD 20771, USA\\
$^{2}$CRESST and X-ray Astrophysics Laboratory, NASA/Goddard Space Flight Center, Greenbelt, MD 20771, USA\\
$^{3}$Universities Space Research Association, 7178 Columbia Gateway Drive, Columbia, MD 21044, USA\\
$^{4}$Department of Physics, University of Maryland, Baltimore County, 1000 Hilltop Circle, Baltimore, MD 21250, USA\\
$^{5}$CRESST and Astrophysics Science Division, Code 667, NASA/Goddard Space Flight Center, Greenbelt, MD 20771, USA\\
$^{6}$Bartol Research Institute, Department of Physics and Astronomy, University of Delaware, Newark, DE 19716, USA\\
$^{7}$Department of Physics and Astronomy, 
University of Pittsburgh, Pittsburgh, PA 15260, USA
}

\date{Accepted 2016 February 9. Received 2015 December 7.}

\pubyear{2016}

\begin{document}
\label{firstpage}
\pagerange{\pageref{firstpage}--\pageref{lastpage}}
\maketitle

\begin{abstract}
  The X-ray emission of $\eta$ Carinae shows multiple features at various spatial and temporal scales.  The central constant emission (CCE) component is centred on the binary and arises from spatial scales much smaller than the bipolar Homunculus nebula, but likely larger than the central wind--wind collision region between the stars as it does not vary over the $\sim$2--3 month X-ray minimum when it can be observed.  Using large-scale 3D smoothed particle hydrodynamics (SPH) simulations, we model both the colliding-wind region between the stars, and the region where the secondary wind collides with primary wind ejected from the previous periastron passage. The simulations extend out to one hundred semimajor axes and make two limiting assumptions (strong coupling and no coupling) about the influence of the primary radiation field on the secondary wind.  We perform 3D radiative transfer calculations on the SPH output to synthesize the X-ray emission, with the aim of reproducing the CCE spectrum.  For the preferred primary mass-loss rate $\dot{M}_A\approx8.5\times10^{-4}$\,M$_\odot$\,yr$^{-1}$, the model spectra well reproduce the observation as the strong- and no-coupling spectra bound the CCE observation for longitude of periastron $\omega\approx252^\circ$, and bound/converge on the observation for $\omega\approx90^\circ$.  This suggests that $\eta$ Carinae has moderate coupling between the primary radiation and secondary wind, that both the region between the stars and the comoving collision on the backside of the secondary generate the CCE, and that the CCE cannot place constraints on the binary's line of sight.  We also discuss comparisons with common X-ray fitting parameters.
\end{abstract}




\section{Introduction}


\noindent$\eta$ Carinae has been well observed in X-rays, providing important constraints on this complex binary system.  Extensive monitoring by the \textit{Rossi X-ray Timing Explorer (RXTE)} found a strongly periodic X-ray flux, indicating a binary orbit, where in dramatic fashion the X-ray flux plummets for $\sim2-3$~months around periastron passage \citep{Corcoran05,CorcoranP10}, consistent with the timing of a multitude of spectral changes in other wavebands \citep[e.g.][]{DamineliP08}.  The X-rays also provide some of the best constraints on the secondary star in $\eta$ Carinae's orbit.
It is not detected at optical and ultraviolet wavelengths because the primary star is so bright, and because ground-based spectra are contaminated by emission from the Homunculus nebula. Additionally, the broad lines from the optically thick wind of the primary, and the existence of multiple emission regions make it difficult to measure and interpret radial velocity variations.
Since the primary wind speed \citep[420~km~s$^{-1}$;][]{GrohP12} is much too low to produce the hard X-rays observed, they must be related to the secondary star.  From matching a single \textit{Chandra} spectrum, \cite{PittardCorcoran02} found that the secondary wind speed and mass-loss rate are $\sim$\,$3000$~km~s$^{-1}$ and $\sim$\,$10^{-5}$\,M$_\odot$\,yr$^{-1}$. 
To match the X-ray light curve, \citet{CorcoranP01} inferred the system to be highly eccentric ($e$\,$\sim$\,0.9), which has been confirmed by more complex hydrodynamic and radiative transfer modelling (\citealt{OkazakiP08}; \citealt{ParkinP09}; \citealt{ParkinP11}).
X-rays also play an important role in determining how the radiation fields of two stars drive both winds \citep{ParkinP11}, and how the X-rays produced ionize one or both winds, thus changing the wind accelerations and the collision dynamics \citep{SokerBehar06,ParkinSim13}.  Furthermore, the X-ray emission can be used to constrain the line of sight to the binary orbit as the X-ray absorption looking through the primary wind is much stronger than when looking through the secondary wind \citep{OkazakiP08}.

The spatial and temporal variations of the X-ray emission range from essentially temporally constant emission from beyond the Homunculus \citep{SewardP79}, to fluctuations every few days that are thought to arise from clumps in the primary wind impacting the wind-wind collision region between the stars \citep{MoffatCorcoran09}.  The focus of this work is the central constant emission (CCE) component identified by \citet{HamaguchiP07}, and confirmed in \citet{HamaguchiP14}, with the \textit{Chandra X-ray Observatory}.  The CCE is only observable over the $\sim$2-3-month X-ray minimum when the colliding-wind X-ray emission produced between the stars is diminished, so `constant' refers to this time-scale.  The emission is spatially unresolvable to \textit{Chandra} indicating that the emission originates within 0.5~arcsec $\approx$~1150~au of the central binary stars.

\citet{HamaguchiP07} proposed three explanations for the origin of the CCE: (1) inherent emission from embedded wind shocks of one or both winds, (2) a fast, polar flow of the primary wind interacting with the Little Homunculus, and/or (3) the secondary wind (either shocked and then cooled, or unshocked) flowing away from the system and then colliding with circumstellar material farther out.  The first two mechanisms are now ruled out, since observations during the 2009.0 and 2014.6 periastron passages revealed a hot, $kT>\sim$5\,keV, component to the CCE \citep{HamaguchiP14,HamaguchiP15}.  The hydrodynamic simulations of \citet{MaduraP13} provide a framework for the third method; the X-ray emission could come from the wind--wind collision between a shell of primary material ejected during the previous periastron passage and the secondary wind ejected during the current cycle.  \citet{RussellP11a,RussellP11b} noted the seed of this interaction when, just after periastron, the secondary star becomes completely embedded in primary wind, and therefore creates a hot, post-periastron bubble as secondary wind shocks with primary wind in all directions.  This includes the comoving shock of secondary wind catching up to primary wind on the back side of the secondary star.  \citet{MaduraP13} performed hydrodynamic simulations out to a much larger volume (a hundred semimajor axes =\,100$a$) and captured this comoving shock over an orbital cycle, thus showing this interaction is still occurring one cycle later and potentially generates the CCE emission.

This work aims to model the CCE X-ray emission by performing 3D X-ray radiative transfer calculations on large-volume ($r<100a$) hydrodynamic simulations of $\eta$~Carinae.  Section~\ref{sec:SPH} presents the hydrodynamic simulations, and Section~\ref{sec:X} details the radiative transfer calculations.  We discuss our results in Section~\ref{sec:D} and present our conclusions in Section~\ref{sec:C}.

\section{Smoothed Particle Hydrodynamics}\label{sec:SPH}

\subsection{Method}

We model the wind--wind structure of $\eta$ Carinae by using a 3D smoothed particle hydrodynamics (SPH) code originally developed by \citet{Benz90} and \citet{BateBonnellPrice95}, and first applied to a colliding wind system -- $\eta$ Carinae -- in \citet{OkazakiP08}.  The stars are described by sink particles \citep{BateBonnellPrice95} that orbit each other while continuously ejecting regular SPH particles to model the interacting stellar winds.  The current capabilities of the code, described more fully in \citet{Russell13} and \citet{MaduraP13}, include radiative cooling using the exact integration scheme \citep{Townsend09}\footnote{To mimic the heating of the gas from the stars \citep[e.g.][]{Drew89}, the simulations impose a floor temperature of 10~kK.  The winds are injected at 35~kK, which is inconsequential for the both gas dynamics and the comparison with observations since X-rays are produced from much larger temperatures.}, and accelerating the winds according to a $\beta$=1 velocity law, $v(r)=v_\infty(1-R/r)^\beta$, where $v(r)$ is the velocity at radius $r$, $v_\infty$ is the wind terminal velocity, and $R$ is the stellar radius. The acceleration is done in an `antigravity' fashion\footnote{Attempts to implement CAK \citep*{CastorAbbottKlein75} line-driving have been hampered by the noise in the velocity gradient computation in the SPH code.  Future work will explore a higher order smoothing kernel -- quintic spline instead of the cubic spline used here -- and a larger number of neighbours to reduce this noise.} -- the winds `fall' off their stars -- using a radially varying opacity $\kappa(r)$ that is tuned to pair with the flux $F(r)$ to produce the acceleration $g_\textrm{rad}(r)=\kappa(r)F(r)/c$ required for the $\beta$ velocity law, where $c$ is the speed of light.

In a binary system that accounts for the radiation fields of both stars, the total acceleration of a gas parcel from star $i$ with companion star $j$ is
\begin{equation}
  \textit{\textbf{g}}_\textrm{rad,i}(\textit{\textbf{r}}_i)\sim \kappa_i(r_i)F(r_i) \hat{\textit{\textbf{r}}}_i+\kappa_{ij}(r_{ij})F(r_j) \hat{\textit{\textbf{r}}}_j.
\end{equation}
$\kappa_{ij}$ determines how the radiation from star $j$ affects the wind of star $i$.  For systems such as $\eta$ Carinae where the temperatures of the stars are very different, calculating $\kappa_{ij}$ is challenging since the peak in the temperature of the radiation field of star $j$ is mismatched to the ionization states of the wind of star $i$ \citep[see e.g.][]{ParkinP09}.  As such, three straightforward possibilities for the coupling exist: 
$\kappa_{ij}(r_{ij})=\kappa_i(r_i)$ attaches the opacity to the wind properties, $\kappa_{ij}(r_{ij})=\kappa_j(r_j)$ sets the opacity to the radiation field properties, and $\kappa_{ij}=0$ means the radiation of star $j$ does not influence the wind of star $i$.  These three possibilities span the range of coupling strength and are referred to as the strong coupling if $\kappa_{ij}(r_{ij})=\max(\kappa_{i}(r_{i}),\kappa_{j}(r_{j}))$, the weak coupling if $\kappa_{ij}(r_{ij})=\min(\kappa_{i}(r_{i}),\kappa_{j}(r_{j}))$, and no coupling if $\kappa_{ij}=0$.  If the strong coupling is chosen for the radiation of star $j$ to interact with the wind of star $i$, then the weak coupling is chosen for the radiation of star $i$ to interact with the wind of star $j$.

In the context of $\eta$ Carinae, and in particular its X-ray emission, we are predominantly concerned with the influence of the much stronger primary radiation field on the secondary wind, and so will consider strong and weak in this context.  Since the radiation of the primary is much greater than the secondary, and the terminal speed of the secondary is much greater than the speed of the primary, $\kappa_B(r)>\kappa_A(r)$ where $A$ and $B$ stand for the primary and secondary, respectively, so $\kappa_{ij}(r_{ij})=\kappa_B(r_B)$, i.e.\ the opacity couples to the wind, is the strong coupling, while $\kappa_{ij}(r_{ij})=\kappa_A(r_A)$, i.e.\ the opacity couples to the radiation field, is the weak coupling.

The radiation of star $j$ influencing the wind of star $i$ goes under the term `radiative inhibition'\footnote{Note that this is different than the inhibition of the wind described by \citet{SokerBehar06}, where soft X-rays ionize from the shock ionize the companion wind around periastron and thus reduce the companion wind (an idea based on this phenomenon occurring in X-ray binaries; \citealt{StevensKallman90}).  To avoid further nomenclature discrepancies in the literature, we propose to call this phenomenon `ionization inhibition'.} since it was originally applied to the region between the stars where the radiation field of star $j$ acts in the opposite direction of the flow of the wind of star $i$, thus inhibiting the acceleration of the wind of star $i$ along the direction towards star $j$ \citep{StevensPollock94}.  For this work, however, the secondary wind on the side opposite the primary (once it has a line of sight to the primary star) will feel radiation from both the secondary and primary stars in the same direction, thus accelerating much (slightly) faster in the strong (weak) coupling case as compared to no coupling. Therefore, as opposed to using the terminology strong or weak radiative inhibition, we will specify strong or weak (or no) coupling.

Radiative braking \citep{OwockiGayley95,GayleyOwockiCranmer97}, wherein the radiation from the weak-wind star can suddenly decelerate a stronger wind and prevent the strong wind from otherwise impacting the weak-wind star, is not incorporated since this requires implementing the full radiative line force \citep{CastorAbbottKlein75} due to radiative braking's non-radial nature.  Based on the stellar, wind, and orbital parameters of $\eta$ Carinae in this work, a straightforward implementation of the radiative braking requirements shows a stable ram pressure balance between the winds occurs throughout the orbit, so radiative braking should not be very important.  However, accounting for the reduced wind velocities due to radiative inhibition and for orbital motion shows that radiative braking might happen for a short phase around periastron \citep{ParkinP11,MaduraP13}.  Since the shock of the weaker-wind, secondary star produces the X-ray emission in $\eta$ Carinae, whether radiative braking happens should have little effect on the X-ray emission at periastron since a wind that shocks very close to its surface at a fraction of its terminal speed (with radiative braking) and a wind that never initiates (no radiative braking) both produce effectively zero X-ray emission.

\citet{MaduraP13} computed a grid of SPH simulations spanning three mass-loss rates $\dot{M}_A$ = 8.5, 4.8, and $2.4\times10^{-4}$\,M$_\odot$\,yr$^{-1}$ to primarily investigate how the primary mass-loss rate affects the wind--wind interaction region.  These were all done with the strong coupling between the primary radiation field and the secondary wind.  To determine the maximum extent of the coupling's influence, we repeat the set of simulations with the most probable $\dot{M}_A=8.5\times10^{-4}$\,M$_\odot$\,yr$^{-1}$ with no coupling.  Table~\ref{ta:SPH} provides the stellar, wind, and orbital parameters of the simulations used in this work.
The simulations of \citet{MaduraP13} also spanned a range of resolutions by varying the outer boundary $r_\textrm{max}$ = 1.5$a$, 10$a$ and 100$a$ while keeping the number of SPH particles in a simulation approximately constant.  We embed the timesteps of the various resolutions inside one another to produce the highest possible accuracy of the hydrodynamic structure around $\eta$ Carinae for computing the model CCE spectra.  This increases the number of particles in the central $r<1.5a$ region to $\sim$100/1.5\,=\,67 times more than the $r_\textrm{max}=100a$ simulation, which equates to a resolution improvement of a factor of $67^{1/3}\sim4$.

\begin{table}
  \centering
  \caption{Stellar, wind, and orbital parameters of the SPH simulations (both from \citealt{MaduraP13} and new) used in this work.}
  \label{ta:SPH}
  \begin{tabular}{lccr} 
	\hline
	Parameter & Primary & Secondary & Reference\\
    & A & B & \\
	\hline
	$M$ (M$_\odot$) & 90 & 30 & H01; O08\\
    $R$ (R$_\odot$) & 90 & 30 & H01; H06\\
	$\dot{M}$ ($10^{-4}$\,M$_\odot$\,yr$^{-1}$)& 8.5, 4.8, and 2.5  & 0.14 & G12; P09\\
	$v_\infty$ (km s$^{-1}$) & 420 & 3000 & G12; PC02\\
    $\beta$ & 1 & 1 & H01; G12\\
    \cline{2-3}
    $e$ & \multicolumn{2}{c}{0.9} & C01 \\
    $P$ (d) [yr]& \multicolumn{2}{c}{2024 [5.54]} & C05 \\ 
    $a$ (au) [$10^{14}$ cm]& \multicolumn{2}{c}{15.4 [2.311]} & -- \\
    $D$ (kpc) & \multicolumn{2}{c}{2.3} & DH97 \\
    $r_\textrm{max}$ ($a$) & \multicolumn{2}{c}{1.5, 10, and 100} & -- \\
    \hline
  \end{tabular}
  \begin{flushleft}
    C01: \citealt{CorcoranP01}, 
    C05: \citealt{Corcoran05}, DH97: \citealt{DavidsonHumphreys97}, G12: \citealt{GrohP12}, H01: \citealt{HillierP01}, H06: \citealt{HillierP06}, O08: \citealt{OkazakiP08}, P09: \citealt{ParkinP09}, PC02: \citealt{PittardCorcoran02}
  \end{flushleft}
\end{table}

The new SPH simulations contain one notable improvement over the simulations from \citet{MaduraP13}.  To account for the mixing of particles, the opacity value of an individual particle was $\bar{\kappa}=(\rho_i\kappa_i+\rho_j\kappa_j)/(\rho_i+\rho_j)$ where $\rho_i$ and $\rho_j$ are the contribution to the density $\rho$ that is due to neighbouring particles ejected from stars $i$ and $j$, respectively.  Particles with homogeneous neighbours behave as expected (for a region of only star $i$ particles, $\rho_i=\rho$, $\rho_j=0$, so $\bar{\kappa}=\kappa_i$), but particles near the contact discontinuity between the two wind species have a transition region, where $\bar{\kappa}$ smoothly varies from $\kappa_i$ to $\kappa_j$ starting on the wind-$i$ side and moving to the wind-$j$ side.  Even though the species of particles remain separated on either side of the contact discontinuity, the average opacity implementation was not treating the particles as such.  The real crux of this average opacity issue is that the size of the transition region from $\kappa_i$ to $\kappa_j$ is resolution dependent (more particles lead to a smaller transition region), so different-resolution simulations show behaviour that varies beyond what is expected from just differing resolutions. The new method eliminates the opacity averaging -- all star $i$ particles have $\kappa_i$, all star $j$ particles have $\kappa_j$ -- so there is no transition region across a contact discontinuity.  If there are regions where the particles are mixed, then the volume-averaged region of the mixing still experiences an average opacity.  The end result of this improvement is that simulations of varying resolutions have more consistent radiative driving, and therefore they show more consistent hydrodynamic behaviour, e.g.\ the velocity profile of material accelerating from away from its star, and, by extension, the location and temperature of shocks between the winds.

Additionally, but less consequentially, the new simulations include the shadowing of the radiative influence on a gas parcel by the farther star if it is obscured by the opaque core of the nearer star.  A particle that is in the umbra, and hence does not have a line of sight to any portion of the farther star, experiences zero radiative driving from the farther star, while a particle in the penumbra, and hence only has a line of sight to a fraction of the far star, experiences a driving force proportional to the visible fractional area of the farther star.

Finally, the new SPH simulations also shut off the radiative acceleration for any particle with a temperature $T>10^6$~K as it is too ionized to be driven.  This, too, has a minimal effect.

\subsection{Results}

Figs.~\ref{fi:SPH100}--\ref{fi:SPH002} show the density (left column), temperature (centre-left column), 1 keV X-ray source function (centre-right column, described in the next section), and speed (right column) of the embedded SPH simulations in the orbital plane at periastron, zooming in with each successive figure.  The rows from top to bottom show model NC8.5, SC8.5, SC4.8, and SC2.4, the meaning of which is in Table~\ref{ta:Model}.  The major axis is the $x$-axis, the minor axis is the $y$-axis, and the orbital angular momentum axis is the $z$-axis (pointing out of the page).  The stars, which orbit counter-clockwise are oriented such that the primary (secondary) is on the right (left) at periastron, though the majority of its ejected wind is on the left (right) portion of the panels, shown in pink, cyan, and blue (red and orange) in the density panels.

\newcommand*{\facA}{0.5}
\newcommand*{\facAB}{2.01}
\begin{figure*}
  \includegraphics[width=\facA\columnwidth]{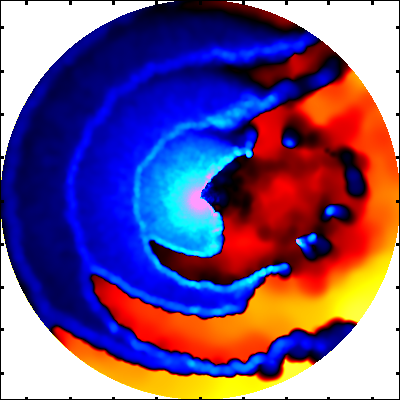}\put(-117,110){\fontfamily{phv}\selectfont dens}\put(-117,3){\fontfamily{phv}\selectfont NC8.5}%
  \includegraphics[width=\facA\columnwidth]{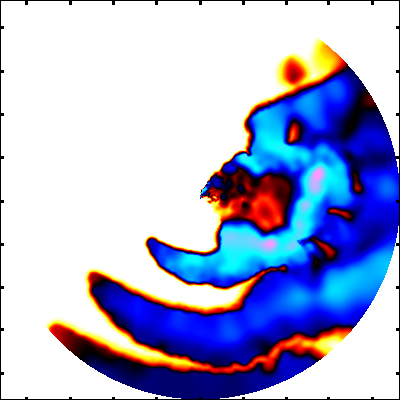}\put(-117,110){\fontfamily{phv}\selectfont temp}%
  \includegraphics[width=\facA\columnwidth]{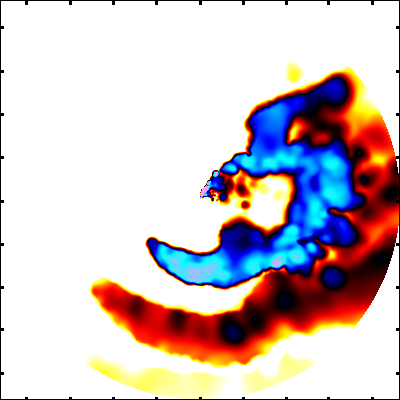}\put(-117,110){\fontfamily{phv}\selectfont X-ray}%
  \includegraphics[width=\facA\columnwidth]{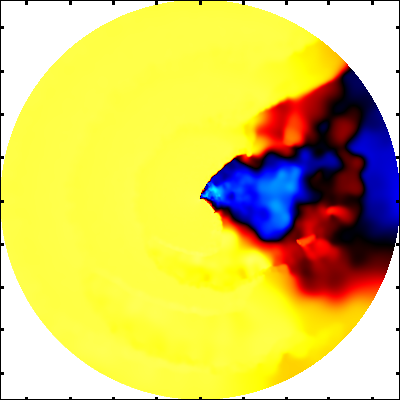}\put(-117,110){\fontfamily{phv}\selectfont vel}

  \includegraphics[width=\facA\columnwidth]{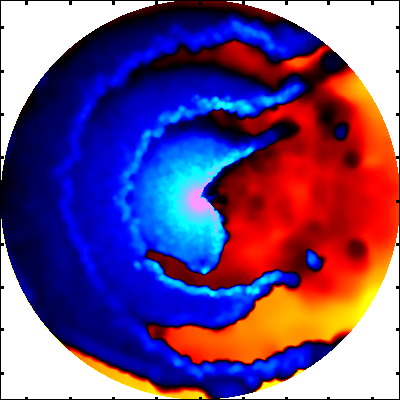}\put(-117,3){\fontfamily{phv}\selectfont SC8.5}%
  \includegraphics[width=\facA\columnwidth]{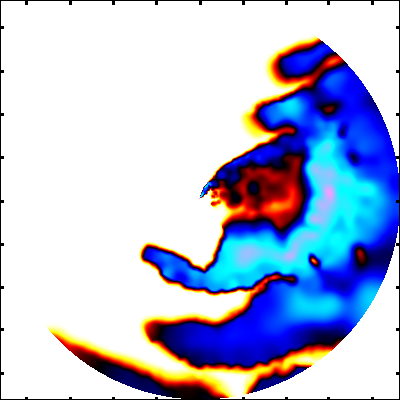}%
  \includegraphics[width=\facA\columnwidth]{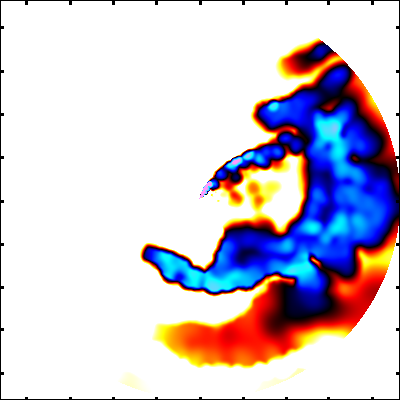}%
  \includegraphics[width=\facA\columnwidth]{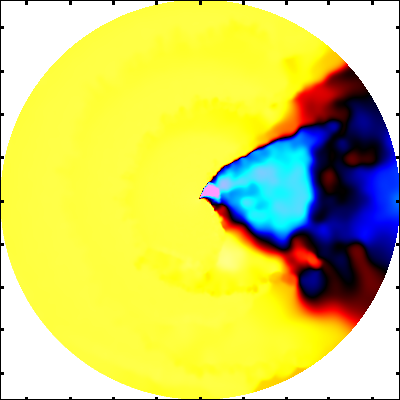}
  \includegraphics[width=\facA\columnwidth]{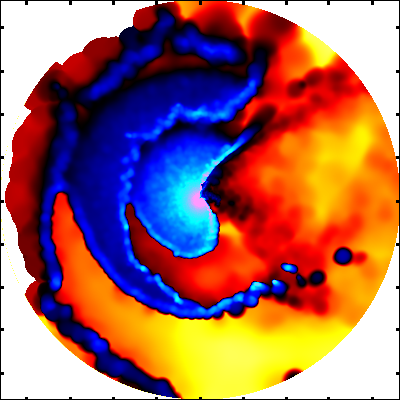}\put(-117,3){\fontfamily{phv}\selectfont SC4.8}%
  \includegraphics[width=\facA\columnwidth]{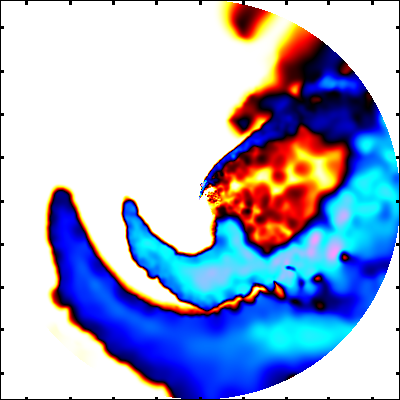}%
  \includegraphics[width=\facA\columnwidth]{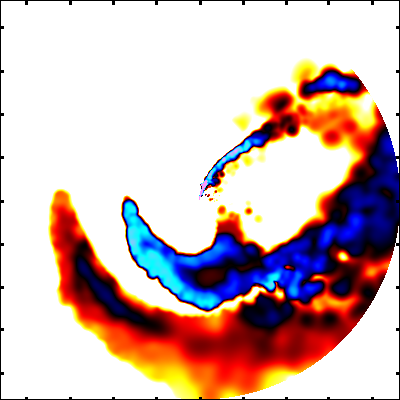}%
  \includegraphics[width=\facA\columnwidth]{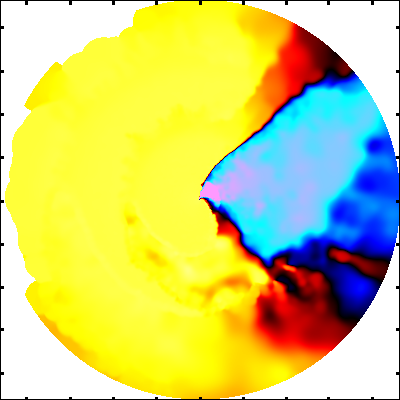}
  \includegraphics[width=\facA\columnwidth]{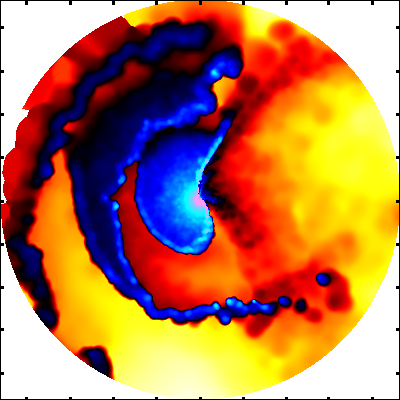}\put(-117,3){\fontfamily{phv}\selectfont SC2.4}%
  \includegraphics[width=\facA\columnwidth]{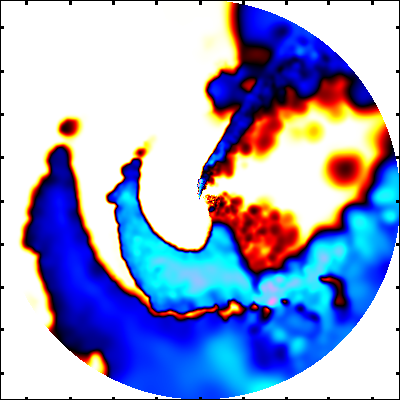}%
  \includegraphics[width=\facA\columnwidth]{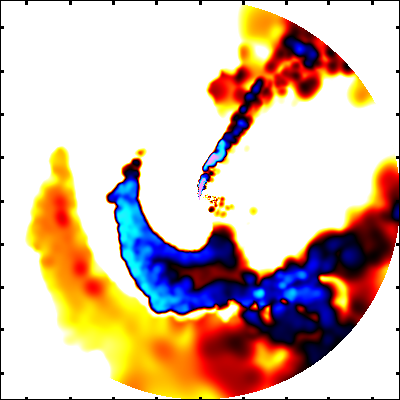}%
  \includegraphics[width=\facA\columnwidth]{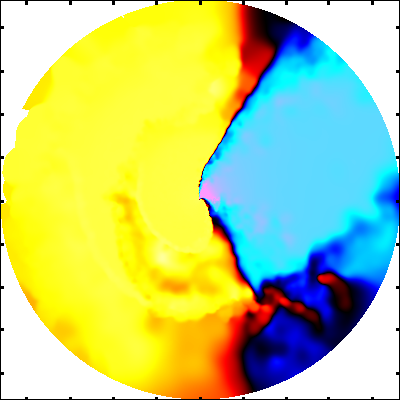}
  \includegraphics[width=\facAB\columnwidth]{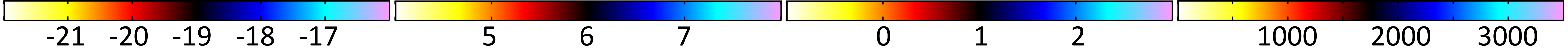}
  \put(-437,-8){\fontfamily{phv}\selectfont log g/cm$^3$}
  \put(-310,-8){\fontfamily{phv}\selectfont log K}
  \put(-215,-8){\fontfamily{phv}\selectfont log erg/s/cm$^2$/keV/sr}
  \put( -68,-8){\fontfamily{phv}\selectfont km/s}

  \caption{Density, temperature, 1 keV X-ray surface brightness, and speed (left to right) in the orbital plane of the SPH simulations for NC8.5, SC8.5, SC4.8, and SC2.4 (top to bottom).  The plots span $\pm$100a, and the tick marks occur every 5$\times$10$^{15}$~cm.}
  \label{fi:SPH100}
\end{figure*}
\newcommand*{\fac}{0.3}
\newcommand*{\facB}{1.205}
\begin{figure*}
  \includegraphics[width=\fac\columnwidth]{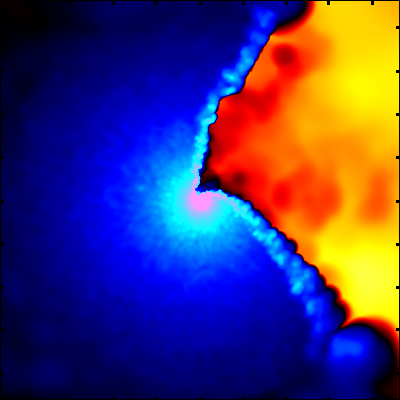}\put(-68,62){\fontfamily{phv}\selectfont \textcolor{white}{dens}}\put(-68,4){\fontfamily{phv}\selectfont \textcolor{white}{NC8.5}}%
  \includegraphics[width=\fac\columnwidth]{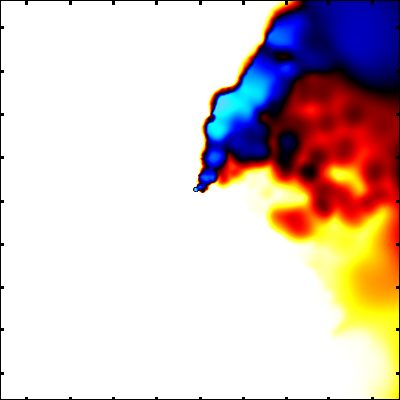}\put(-68,62){\fontfamily{phv}\selectfont temp}%
  \includegraphics[width=\fac\columnwidth]{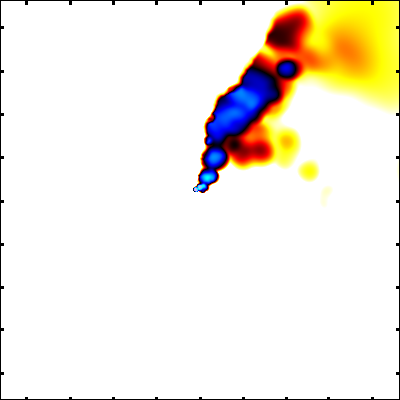}\put(-68,62){\fontfamily{phv}\selectfont X-ray}%
  \includegraphics[width=\fac\columnwidth]{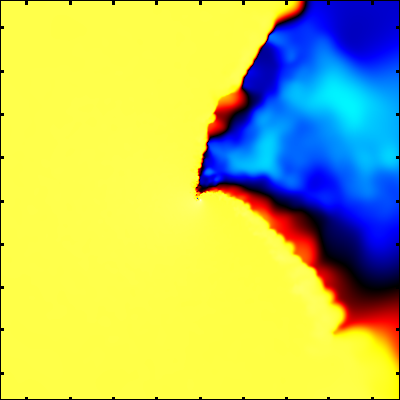}\put(-68,62){\fontfamily{phv}\selectfont vel}

  \includegraphics[width=\fac\columnwidth]{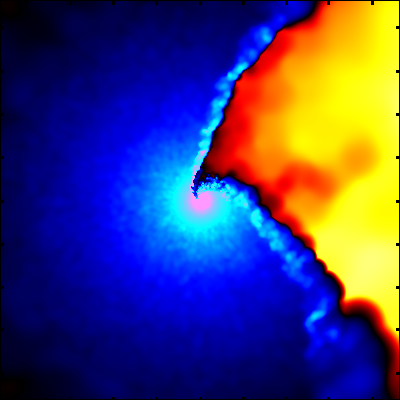}\put(-68,4){\fontfamily{phv}\selectfont \textcolor{white}{SC8.5}}%
  \includegraphics[width=\fac\columnwidth]{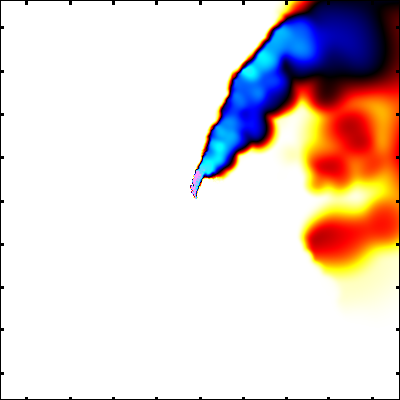}%
  \includegraphics[width=\fac\columnwidth]{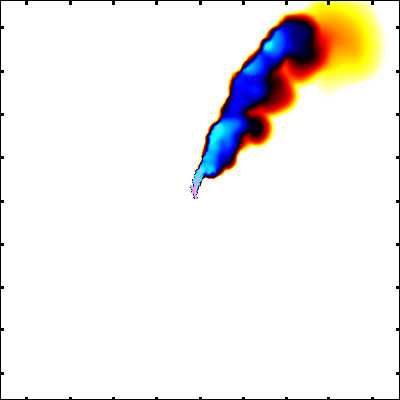}%
  \includegraphics[width=\fac\columnwidth]{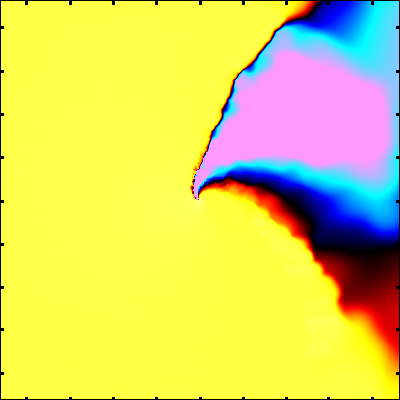}
  \includegraphics[width=\fac\columnwidth]{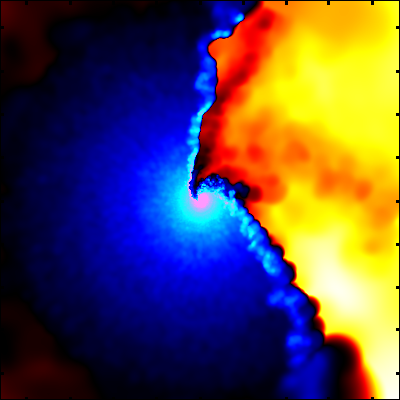}\put(-68,4){\fontfamily{phv}\selectfont \textcolor{white}{SC4.8}}%
  \includegraphics[width=\fac\columnwidth]{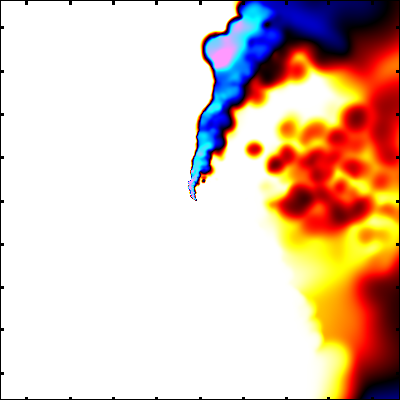}%
  \includegraphics[width=\fac\columnwidth]{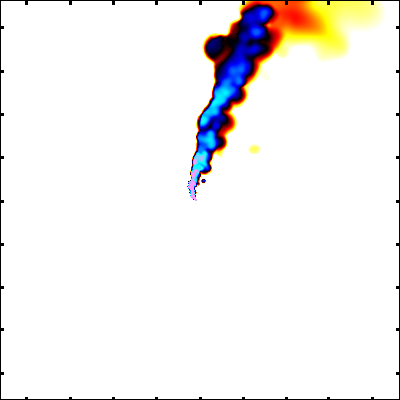}%
  \includegraphics[width=\fac\columnwidth]{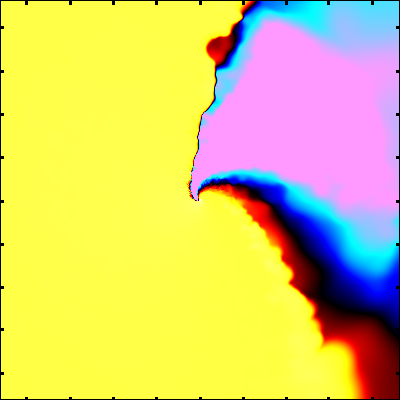}
  \includegraphics[width=\fac\columnwidth]{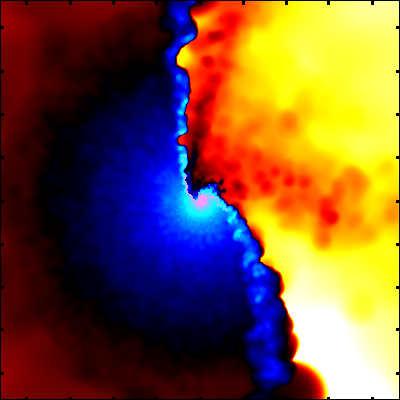}\put(-68,4){\fontfamily{phv}\selectfont \textcolor{white}{SC2.4}}%
  \includegraphics[width=\fac\columnwidth]{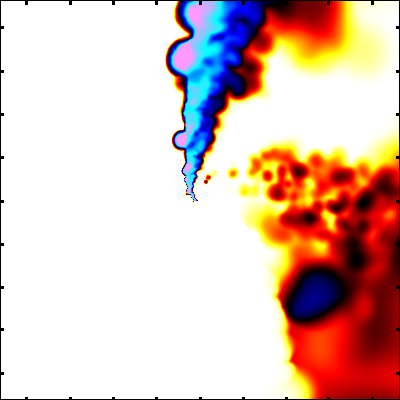}%
  \includegraphics[width=\fac\columnwidth]{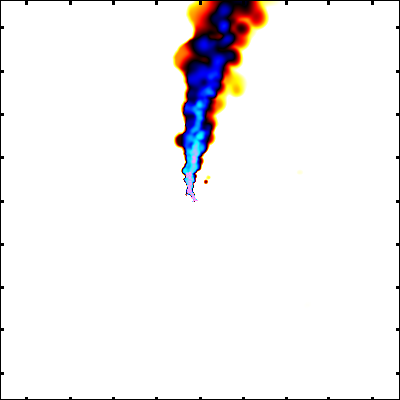}%
  \includegraphics[width=\fac\columnwidth]{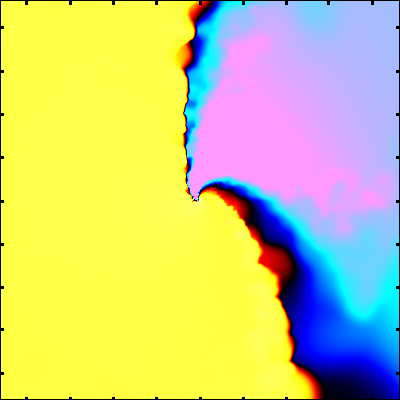}
  \includegraphics[width=\facB\columnwidth]{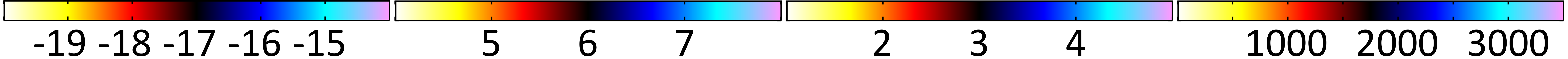}

  \caption{Same as Fig.~\ref{fi:SPH100}, except the plots span $\pm$10a.  The tick marks occur every 5$\times$10$^{14}$~cm.}
  \label{fi:SPH010}
\end{figure*}
\begin{figure*}
  \includegraphics[width=\fac\columnwidth]{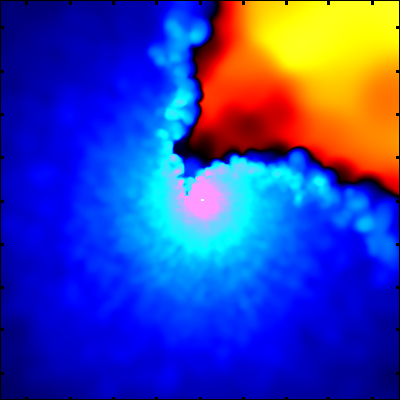}\put(-68,62){\fontfamily{phv}\selectfont \textcolor{white}{dens}}\put(-68,4){\fontfamily{phv}\selectfont \textcolor{white}{NC8.5}}%
  \includegraphics[width=\fac\columnwidth]{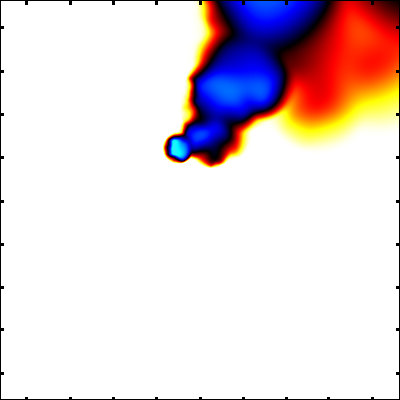}\put(-68,62){\fontfamily{phv}\selectfont temp}%
  \includegraphics[width=\fac\columnwidth]{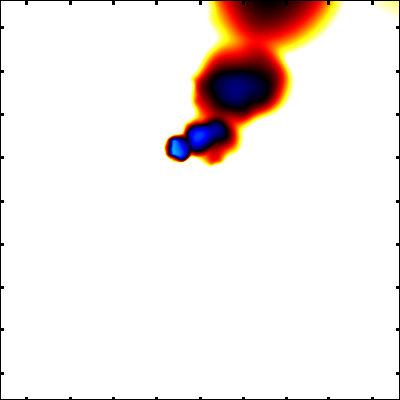}\put(-68,62){\fontfamily{phv}\selectfont X-ray}%
  \includegraphics[width=\fac\columnwidth]{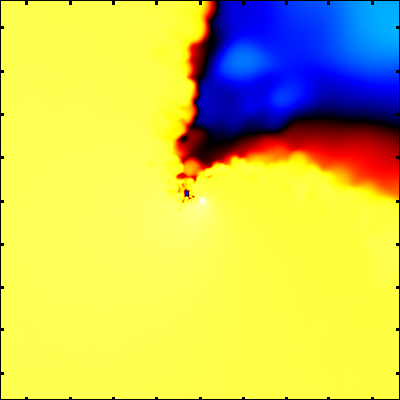}\put(-68,62){\fontfamily{phv}\selectfont vel}

  \includegraphics[width=\fac\columnwidth]{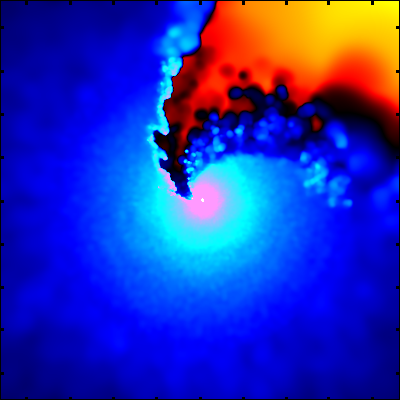}\put(-68,4){\fontfamily{phv}\selectfont \textcolor{white}{SC8.5}}%
  \includegraphics[width=\fac\columnwidth]{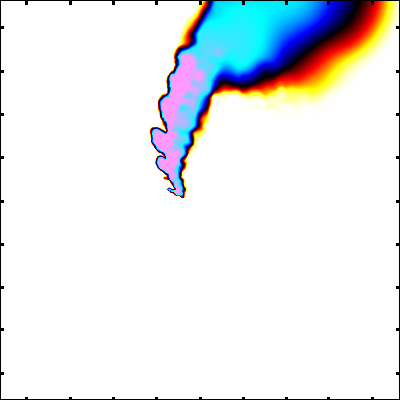}%
  \includegraphics[width=\fac\columnwidth]{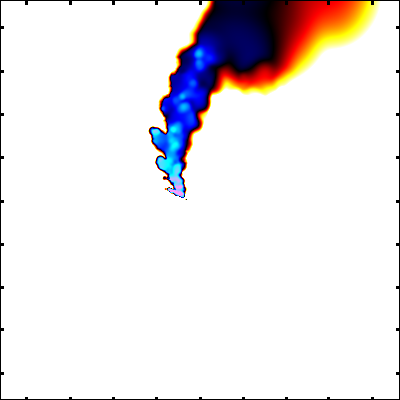}%
  \includegraphics[width=\fac\columnwidth]{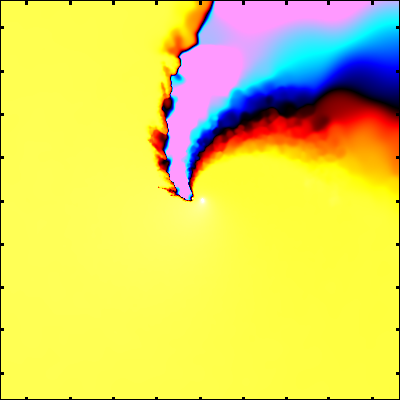}
  \includegraphics[width=\fac\columnwidth]{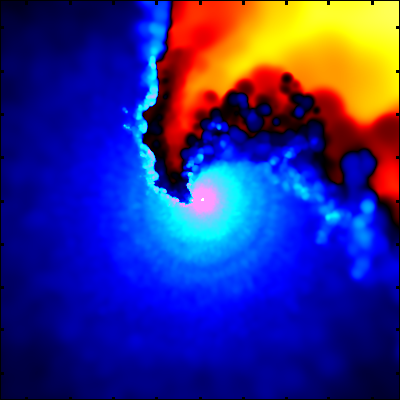}\put(-68,4){\fontfamily{phv}\selectfont \textcolor{white}{SC4.8}}%
  \includegraphics[width=\fac\columnwidth]{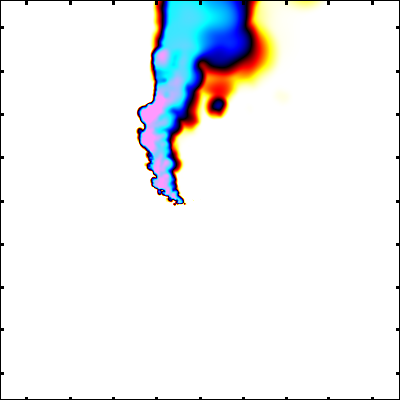}%
  \includegraphics[width=\fac\columnwidth]{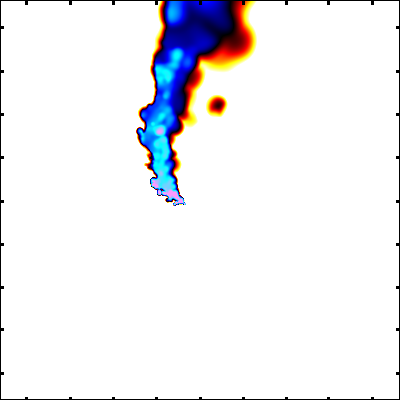}%
  \includegraphics[width=\fac\columnwidth]{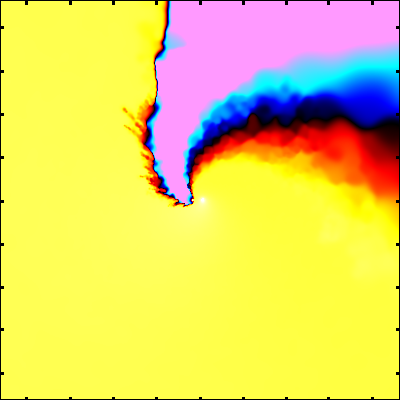}
  \includegraphics[width=\fac\columnwidth]{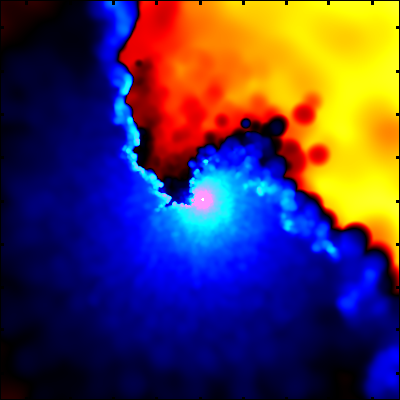}\put(-68,4){\fontfamily{phv}\selectfont \textcolor{white}{SC2.4}}%
  \includegraphics[width=\fac\columnwidth]{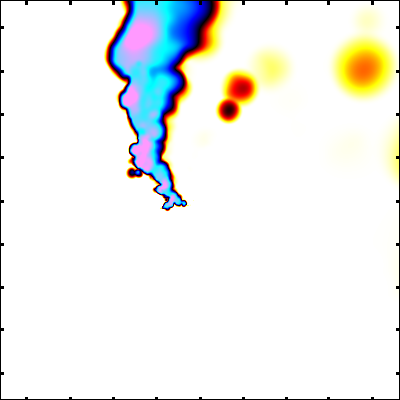}%
  \includegraphics[width=\fac\columnwidth]{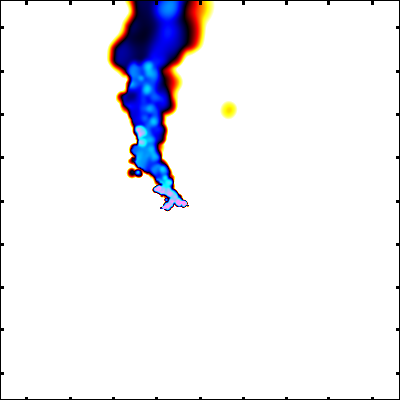}%
  \includegraphics[width=\fac\columnwidth]{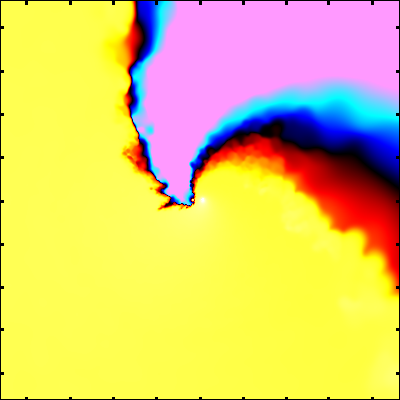}
  \includegraphics[width=\facB\columnwidth]{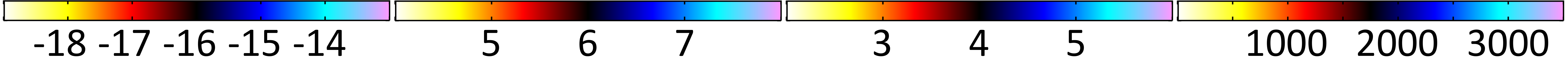}

  \caption{Same as Fig.~\ref{fi:SPH100}, except the plots span $\pm$2a.  The ticks occur every 10$^{14}$~cm.  There are clear differences in the strength of the X-ray emission on this scale due to the different coupling methods chosen (top row versus second row).}
  \label{fi:SPH002}
\end{figure*}

\begin{table}
  \centering
  \caption{Model names and distinguishing parameters of the SPH simulations (both from \citealt{MaduraP13} and new) used in this work.}
  \label{ta:Model}
  \begin{tabular}{lll} 
	\hline
	Model & Coupling of primary radiation & $\dot{M}_A$ \\
	name & to secondary wind & (10$^{-4}$\,M$_\odot$\,yr$^{-1}$) \\
	\hline
	NC8.5 & No Coupling & 8.5 \\
    SC8.5 & Strong Coupling & 8.5 \\
	SC4.8 & Strong Coupling & 4.8 \\
	SC2.4 & Strong Coupling & 2.4 \\
    \hline
  \end{tabular}
\end{table}

In the zoomed-out panels of Fig.~\ref{fi:SPH100}, the effects of the stars speeding by each other at periastron show up as quasi-circular disturbances (`shells' or `walls') in the primary wind on the left of the panels, as well as shells of primary wind material emitted to the right that the secondary wind collides into and gradually destroys.  As expected, the disturbance in the primary wind and the intactness of the primary shell emitted one cycle prior depends on $\dot{M}_A$: NC8.5 and SC8.5 show little primary wind disturbances and visible shells, while the disturbances are large and the shell weak in SC2.4.  For further details of this shell and the time evolution of the simulations, see \citet{MaduraP13}.

The no-coupling simulations (top row) show the largest deviations with the strong-coupling simulations in the speed of the central regions (right-hand panels of Fig.~\ref{fi:SPH002}).  The speed in SC8.5 reaches $>$4000~km\,s$^{-1}$, while the maximum in NC8.5 is slightly under $v_{\infty,B}$, and only obtained in the upper-right portion of the panel.  This is the effect of primary radiation accelerating the secondary wind emitted from the back half of the secondary star (i.e. the half opposite the primary star); the force vectors from both radiation sources co-add to increase the acceleration of these particles, and this extra acceleration becomes the most prominent near periastron when the stars are closest.  The increased acceleration leads to the pre-shock speed of the leading arm being much higher, so the post-shock temperature (centre-left column) and the X-ray emissivity (centre-right column) are also much higher.

\section{Thermal X-ray Radiative Transfer}\label{sec:X}

\subsection{Method}

We perform 3D radiative transfer calculations on the density and temperature structure of the embedded SPH simulations to determine the model thermal X-ray spectrum.  The SPH visualization program \texttt{Splash} \citep{Price07} is the basis for solving the formal solution to radiative transfer
\begin{equation}\label{eq:FSRT}
  I(E,x',y')=\int_{\tau(x',y',-z'_\textrm{max}(x',y'))}^0 S(E,x',y',t)\,e^{-t}\textrm{d}t
\end{equation}
for a grid of rays \{$x',y'$\} through the simulation volume $(-z_\textrm{max}(x',y')$ to $z_\textrm{max}(x',y')$; these values depend on \{x',y'\} since the volume is spherical). The observer is located along the direction $+z'$.

\begin{figure}
  \includegraphics[trim={0.08cm 0 0.08cm 0},clip,width=\columnwidth]{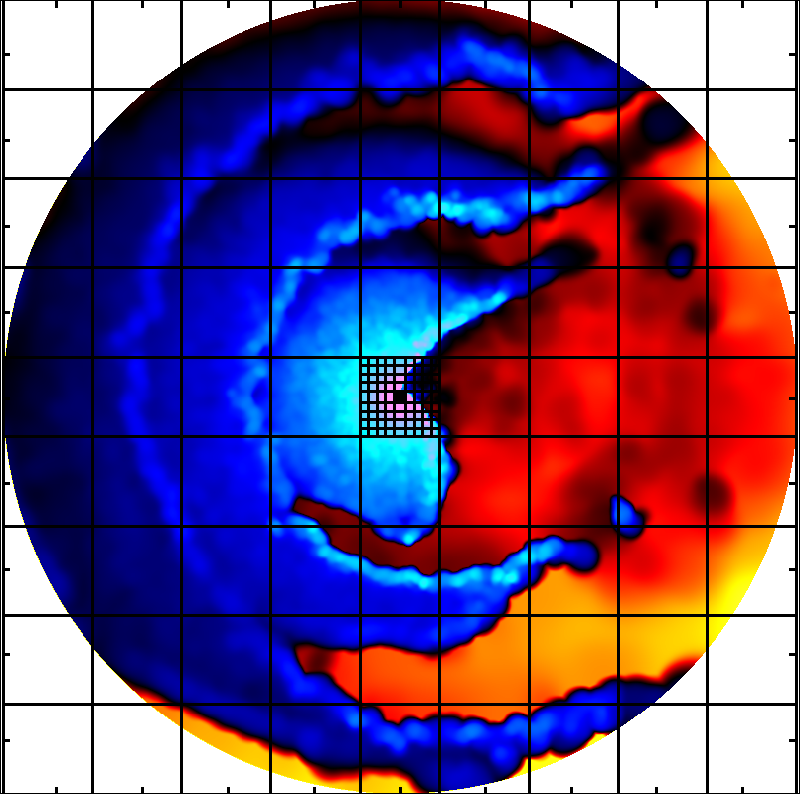}
  \caption{Grid over the density showing the regions where the three levels of spatial resolution for the radiative transfer calculation occur.  The inner region -- a square -- is $\{x,y\}<1.5a$, the middle region -- a square with a square hole in the centre -- is $1.5a<\{x,y\}<10a$, and the outer region -- also a square with a square hole in the centre -- is $10a<\{x,y\}<100a$.  Each of these regions is covered by 400$\times$400 pixels (much less than this number of grids is shown for clarity).}
  \label{fi:XGrid}
\end{figure}

The optical depth is
\begin{equation}\label{eq:OpDe}
  \tau(x',y',z')=\int_{z'}^{z_\textrm{max}(x',y')} \kappa(E) \rho(x',y',z'') \textrm{d}z'',
\end{equation}
so $\tau(z'=z'_\textrm{max})=0$.  The intensity at the boundary of the simulation is $I$, and $S$ is the source function.
This formal solution generates an X-ray map for each energy $E$ that are summed to produce the model spectra, and then folded through an X-ray telescope response function to directly compare with observations.
The radiative transfer is performed at an energy resolution of 800 bins dex$^{-1}$ from 0.3--12 keV (covering the full input range of the \textit{Chandra} ACIS-S response function).  This energy resolution is more than needed for most of the spectra, but is required to properly resolve the Fe--K emission at $\sim$6.7~keV.  Just as embedding the successively smaller outer boundary/higher resolution simulations inside one another obtains the maximum spatial resolution of the density and temperature structure, the \{$x',y'$\} grids of the radiative transfer calculation are also embedded in each other.  The inner region is the square of $\{x',y'\}<\pm1.5a$, the middle region is a square with a hole of the inner-region size at its centre $\pm1.5a<\{x',y'\}<\pm10a$, and the outer region is the same shape $\pm10a<\{x',y'\}<\pm100a$. Each of these regions have 400$\times$400 pixels across them, so their resolutions are 1.73, 11.6, and 116\,$\times10^{12}$cm for the inner, middle, and outer regions, respectively.  Fig.~\ref{fi:XGrid} shows a schematic of the three calculation regions that sum together to generate the final model spectra.

The source function for these thermal X-rays is $S(E)=j(E)/(\kappa(E)\rho)$, where $j(E)=n_\textrm{e}n_\textrm{H}'\Lambda(E,T)$ is the emissivity for electron and hydrogen densities $n_\textrm{e}$ and $n_\textrm{H}'$.  The emission function $\Lambda(E,T)$ is from \texttt{APEC} \citep{SmithP01} using \texttt{AtomDB} version~2.0.2, as implemented in \texttt{XSpec} \citep{Arnaud96} version~12.9.0c.  The circumstellar absorption is from \texttt{windtabs} \citep{LeuteneggerP10}, and the Homunculus/interstellar absorption is from \texttt{TBabs} \citep{WilmsAllenMcCray00}.  Solar abundances \citep{AsplundP09} are used throughout this work; the enhancement of nitrogen at the expense of carbon and oxygen in $\eta$ Carinae's primary spectra will only have a minute effect on the X-ray absorption.

\vspace{\baselineskip}

The line of sight to the binary orbit of $\eta$ Carinae has been the source of much discussion in the literature.  While the consensus is that the binary orbit is inclined $i\sim135^\circ$ to align the orbital axis with the Homunculus axis, there are two widely discrepant azimuthal viewing angles.  The first has the secondary star in front for the majority of the eccentric orbit 
(observer at the $+x$-axis, longitude of periastron $\omega\sim270^\circ$),  while the other is the exact opposite (observer at the $-x$-axis, $\omega\sim90^\circ$).  These place the observer on the right or left side of the orbital-plane images in Figs.~\ref{fi:SPH100}-\ref{fi:SPH002}, respectively.  A subset of the recent work supporting the first is
\citet{OkazakiP08}, \citet{ParkinP09}, \citet{ParkinP11}, \citet{MaduraP12}, \citet{Russell13}, \citet{HamaguchiP14}, \citet{ClementelP15a}, and \citet{ClementelP15b},
while the second is supported by
\citet{SokerBehar06}, \citet{KashiSoker08}, \citet{KashiSoker09c}, \citet{FalcetaGoncalvesAbraham09}, \citet{AbrahamFalcetaGoncalves10}, and \citet{KashiSokerAkashi11}.
The most constraining work for determining the line of sight to $\eta$ Carinae is \citet{MaduraP12} since, as opposed to the point-source nature of the other diagnostics, they used spatially resolved [Fe] emission to constrain all three viewing angles -- inclination, azimuth, and position angle (PA) -- to $i\sim135^\circ$, $\omega\sim252^\circ$, and PA\,$\sim40^\circ$, which aligns the orbital axis with the Homunculus axis.

\begin{figure}
  \frame{\includegraphics[clip=false,width=0.5\columnwidth]{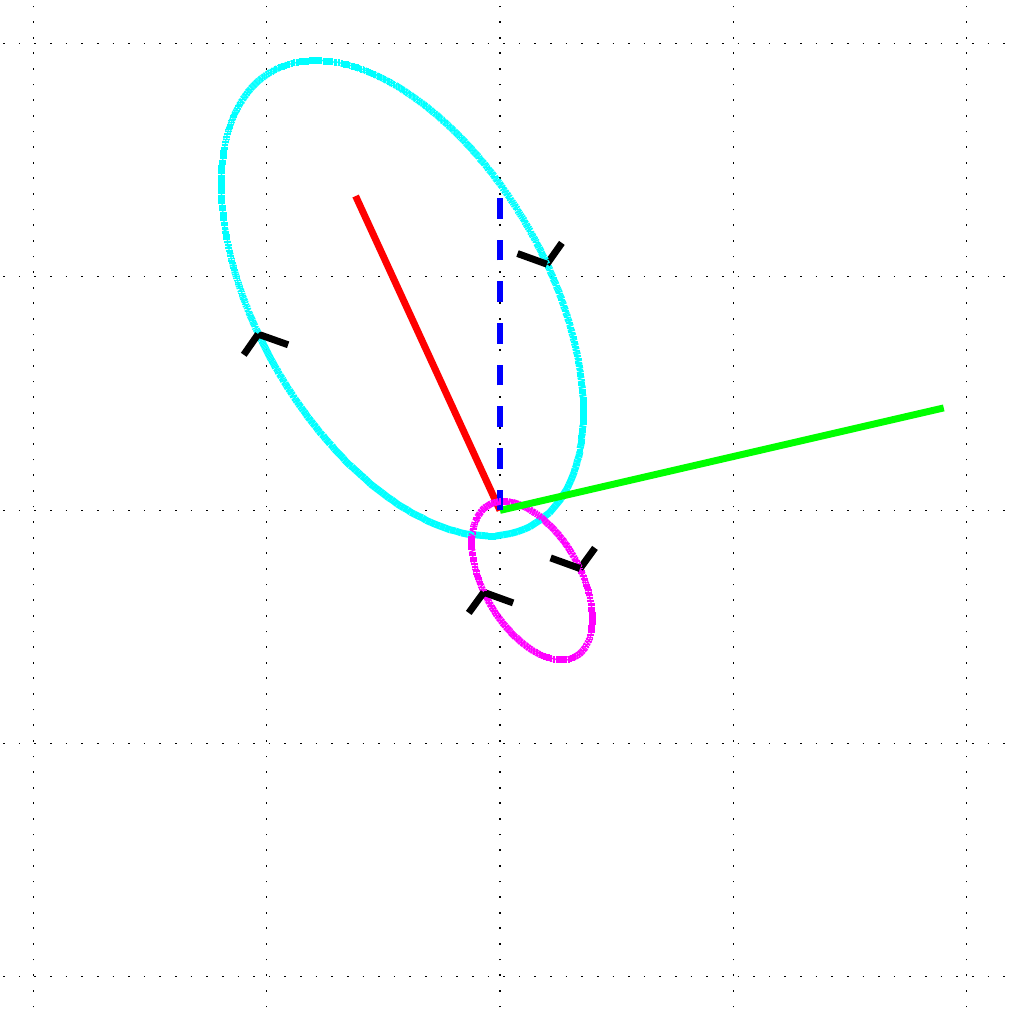}}\put(-30,105){\scriptsize \fontfamily{phv}\selectfont $\omega$\,=\,252$^\circ$}%
  \frame{\includegraphics[clip=false,width=0.5\columnwidth]{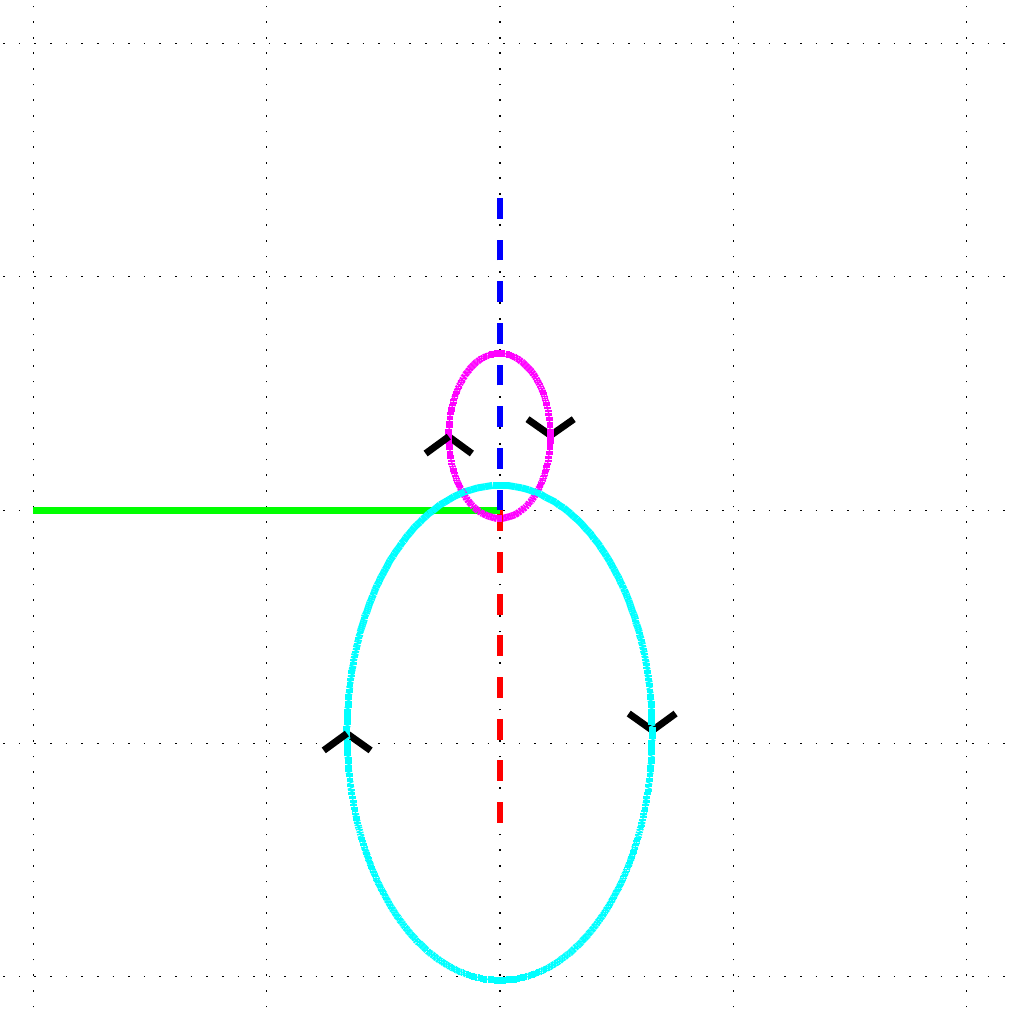}}\put(-30,105){\scriptsize \fontfamily{phv}\selectfont $\omega$\,=\,90$^\circ$}

  \caption{Axes and the orbits in the lab frame of the two lines of sight explored in this work: $\omega=252^\circ$, $i=135^\circ$ (left) and $\omega=90^\circ$, $i=135^\circ$ (right).  Red: major ($x$) axis, green: minor ($y$) axis, blue: orbital ($z$) axis, magenta: primary orbit, cyan: secondary orbit.  These lines of sight only rotate through $\omega$ and $i$, which results in the projected orbital axis (blue) pointing up.  To compare with the images of $\eta$ Carinae on the sky, rotate through a position angle of $\textrm{PA}=40^\circ$ clockwise to align the orbital axis with the Homunculus axis, which is found to be the case by \citet{MaduraP12}.  Dashed axes are oriented into the page.  The solid axes on the left are out of the page, while the solid axis on the right is parallel to the page.  The secondary is in front at apastron for $\omega=252^\circ$ (left) and behind at apastron for $\omega=90^\circ$ (right). Each axes length, prior to projection, is 1$a$.  The black arrows indicate the directions of the orbits: clockwise (due to $i>90^\circ$).}
  \label{fi:LoS}
\end{figure}

Since the X-ray optical depths should differ noticeably from one line of sight to the other, we perform the radiative transfer with the observer at both viewing locations: $\omega=252^\circ$ and $\omega=90^\circ$.  Fig.~\ref{fi:LoS} shows the projections of these viewpoints on the sky without the PA rotation.  Because the point-like CCE X-rays are insensitive to the PA rotation, and the vertical is a better reference line than 40$^\circ$ clockwise of north, the X-ray images presented subsequently do not include the PA rotation.

\subsection{Results}\label{sec:Xres}

Fig.~\ref{fi:XPix} shows, with the same colour scale for all panels, X-ray surface brightness maps of the emission at 1 and 10~keV for both lines of sight.  The 10~keV flux (right half) is only mildly susceptible to circumstellar absorption (i.e.\ absorption within the $r_\textrm{max}=100a$ simulation volume) due to the low opacity at this energy ($\kappa(10~\textrm{keV})=0.36$~cm$^2$\,g$^{-1}$), so it is a good proxy for the location of the intrinsic emission, while the 1~keV flux (left half) is highly influenced by this absorption ($\kappa(1~\textrm{keV})=5.9$~cm$^2$\,g$^{-1}$).  Furthermore, the $\omega=90^\circ$ line of sight (bottom half) suffers much more absorption than its counterpart since it views the system through the denser primary wind, while $\omega=252^\circ$ (top half) looks predominantly through secondary wind material.
The coupling choice also strongly influences the inner regions (right column of each quadrant).  The strong-coupling simulations (bottom three rows of each quadrant) have significant emission at this scale, while the no-coupling model (top row of each quadrant) does not.  In the large outer boundary images (left column of each quadrant), the emission from the comoving shock of the secondary wind catching up with the primary wind shell is visible; it is much lower per solid angle, but it also occupies a larger volume and is therefore necessary for the CCE spectral matching.  We therefore state that the large outer boundary of the simulations is a requirement to match the CCE observations.

\newcommand*{\fact}{0.3}
\newcommand*{\factdouble}{0.6}
\newcommand*{\panspac}{0.1}
\begin{figure*}
\begin{center}
  \includegraphics[width=\fact\columnwidth]{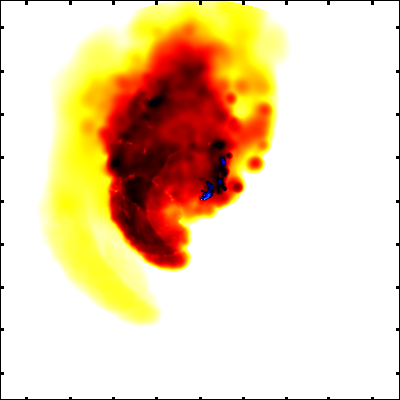}\put(-68,62){\fontfamily{phv}\selectfont $\pm$100$a$}\put(-68,4){\fontfamily{phv}\selectfont NC8.5}\put(25,75){\fontfamily{phv}\selectfont 1\,keV}%
  \includegraphics[width=\fact\columnwidth]{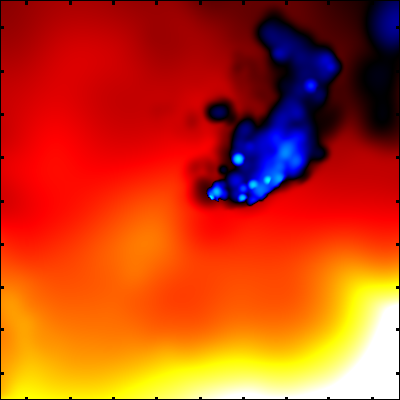}\put(-68,62){\fontfamily{phv}\selectfont $\pm$10$a$}%
  \includegraphics[width=\fact\columnwidth]{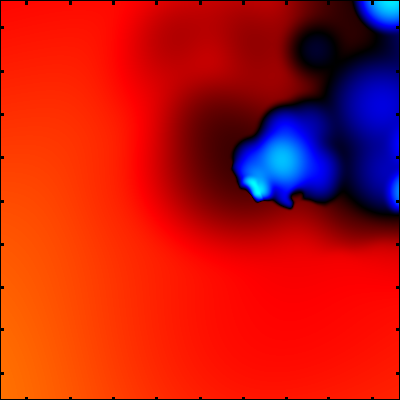}\put(-68,62){\fontfamily{phv}\selectfont $\pm$2$a$}%
  \hspace{\panspac cm}
  \vrule
  \hspace{\panspac cm}
  \includegraphics[width=\fact\columnwidth]{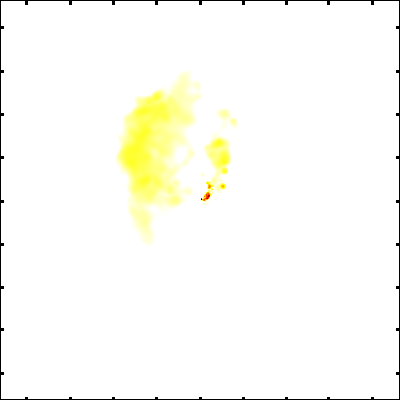}\put(-68,62){\fontfamily{phv}\selectfont $\pm$100$a$}\put(24,75){\fontfamily{phv}\selectfont 10\,keV}%
  \includegraphics[width=\fact\columnwidth]{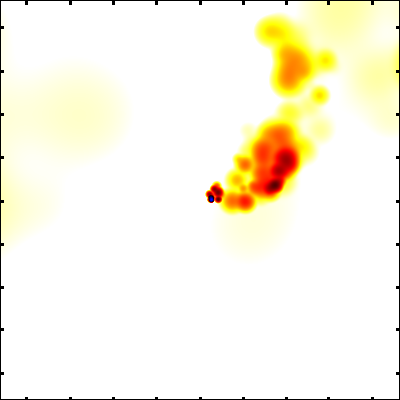}\put(-68,62){\fontfamily{phv}\selectfont $\pm$10$a$}%
  \includegraphics[width=\fact\columnwidth]{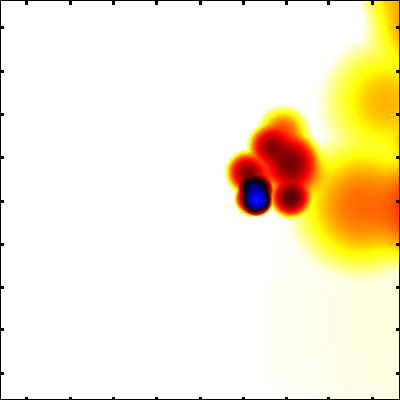}\put(-68,62){\fontfamily{phv}\selectfont $\pm$2$a$}

  \includegraphics[width=\fact\columnwidth]{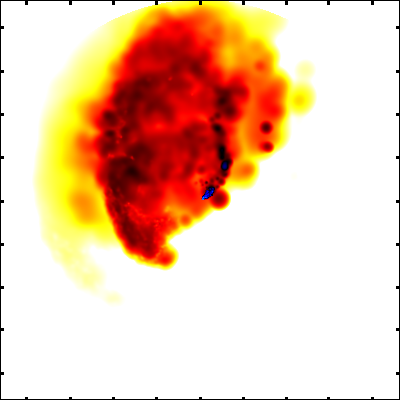}\put(-68,4){\fontfamily{phv}\selectfont SC8.5}\put(-82,0){\fontfamily{phv}\selectfont \rotatebox{90}{252$^\circ$}}%
  \includegraphics[width=\fact\columnwidth]{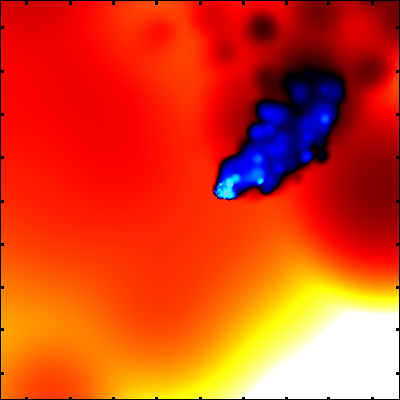}%
  \includegraphics[width=\fact\columnwidth]{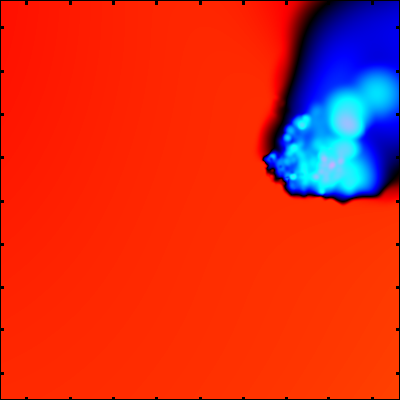}%
  \hspace{\panspac cm}
  \vrule
  \hspace{\panspac cm}
  \includegraphics[width=\fact\columnwidth]{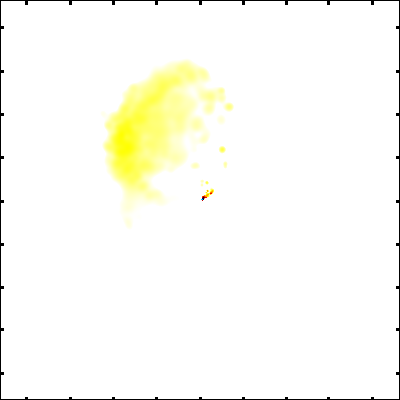}%
  \includegraphics[width=\fact\columnwidth]{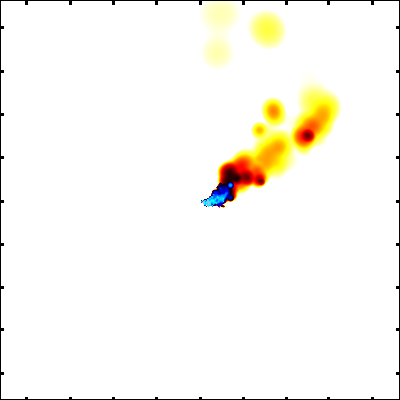}%
  \includegraphics[width=\fact\columnwidth]{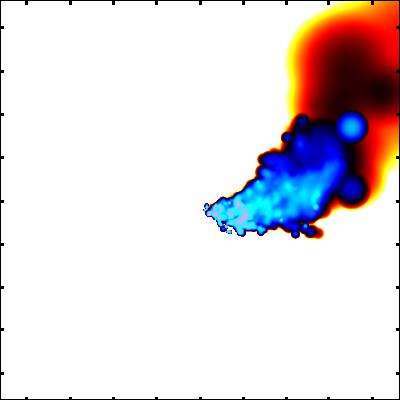}

  \includegraphics[width=\fact\columnwidth]{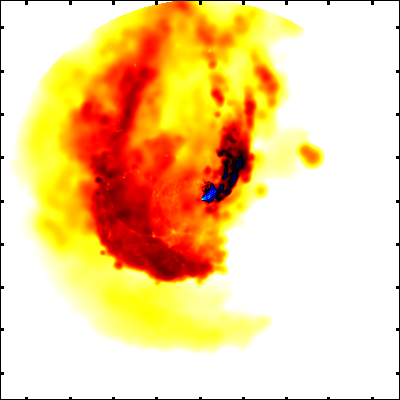}\put(-68,4){\fontfamily{phv}\selectfont SC4.8}\put(-80,58){\fontfamily{phv}\selectfont \rotatebox{90}{$\omega=$}}%
  \includegraphics[width=\fact\columnwidth]{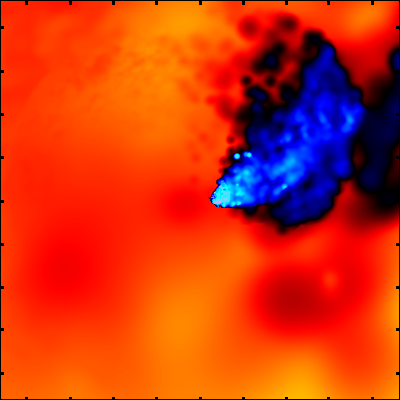}%
  \includegraphics[width=\fact\columnwidth]{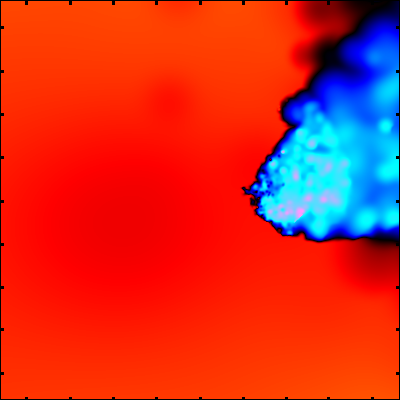}%
  \hspace{\panspac cm}
  \vrule
  \hspace{\panspac cm}
  \includegraphics[width=\fact\columnwidth]{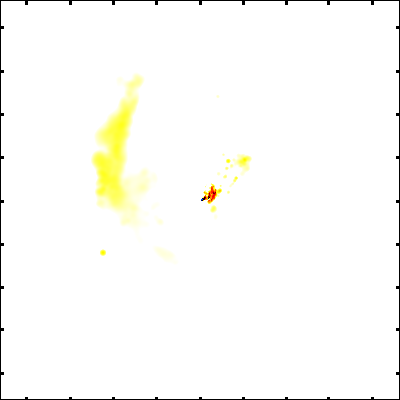}%
  \includegraphics[width=\fact\columnwidth]{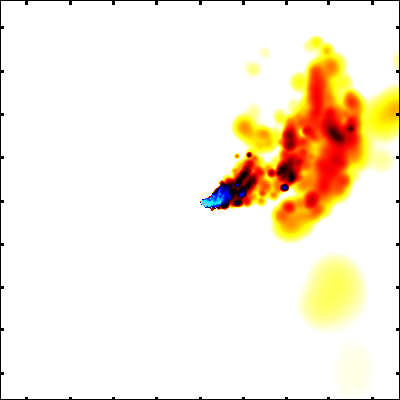}%
  \includegraphics[width=\fact\columnwidth]{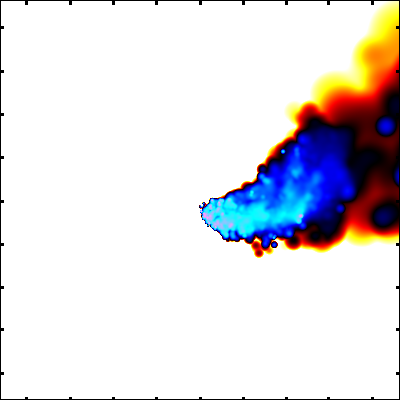}

  \includegraphics[width=\fact\columnwidth]{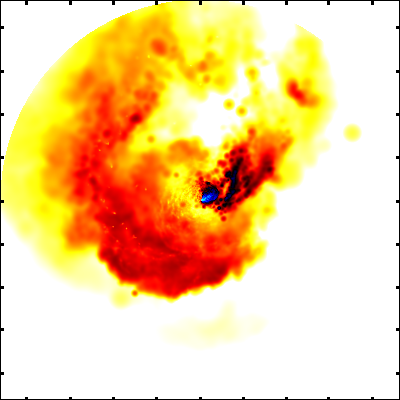}\put(-68,4){\fontfamily{phv}\selectfont SC2.4}%
  \includegraphics[width=\fact\columnwidth]{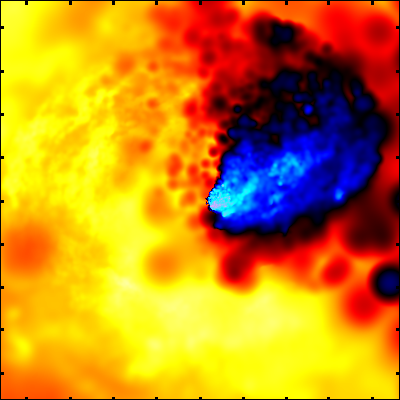}%
  \includegraphics[width=\fact\columnwidth]{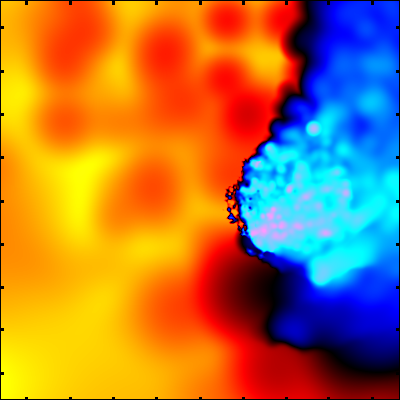}%
  \hspace{\panspac cm}
  \vrule
  \hspace{\panspac cm}
  \includegraphics[width=\fact\columnwidth]{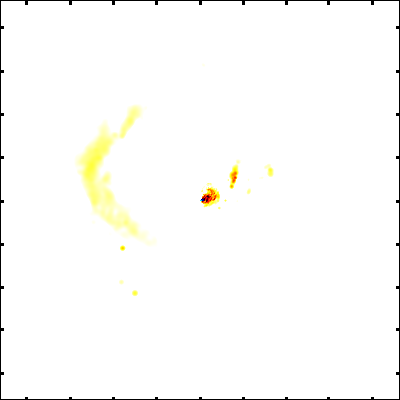}%
  \includegraphics[width=\fact\columnwidth]{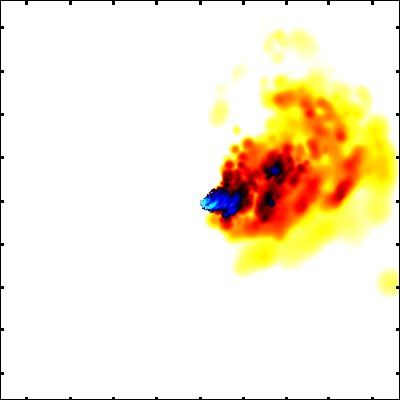}%
  \includegraphics[width=\fact\columnwidth]{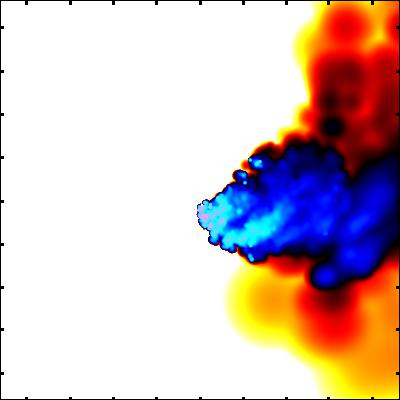}

  \vspace{\panspac cm}
  \hrule
  \vspace{\panspac cm}

  \includegraphics[width=\fact\columnwidth]{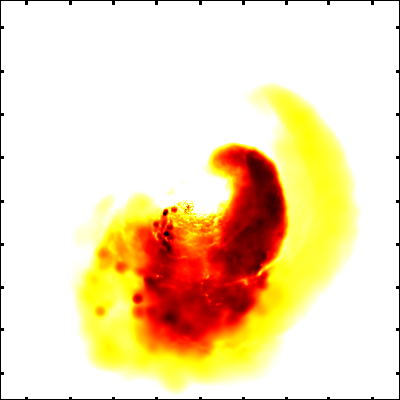}\put(-68,62){\fontfamily{phv}\selectfont NC8.5}%
  \includegraphics[width=\fact\columnwidth]{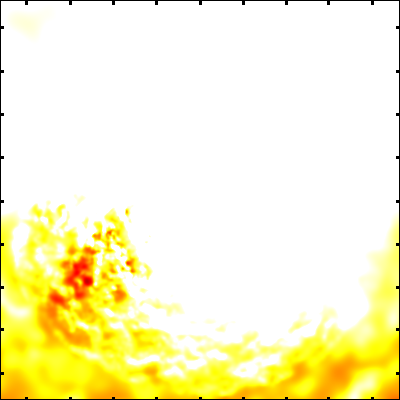}%
  \includegraphics[width=\fact\columnwidth]{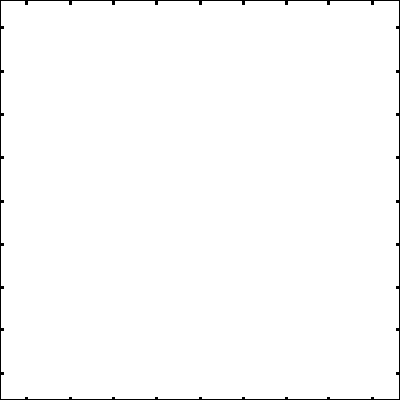}%
  \hspace{\panspac cm}
  \vrule
  \hspace{\panspac cm}
  \includegraphics[width=\fact\columnwidth]{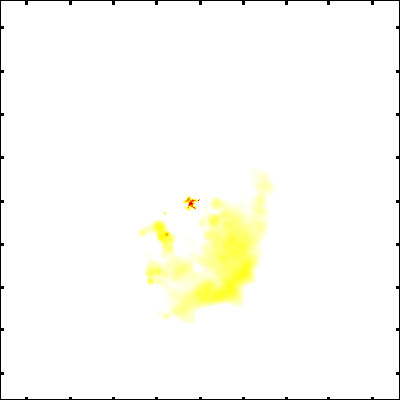}%
  \includegraphics[width=\fact\columnwidth]{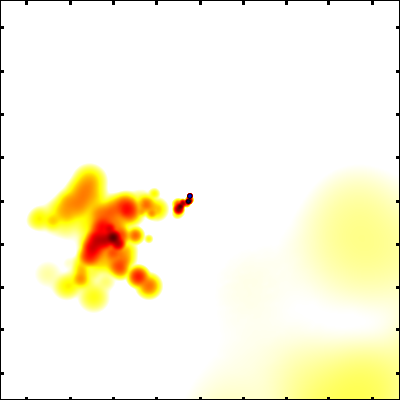}%
  \includegraphics[width=\fact\columnwidth]{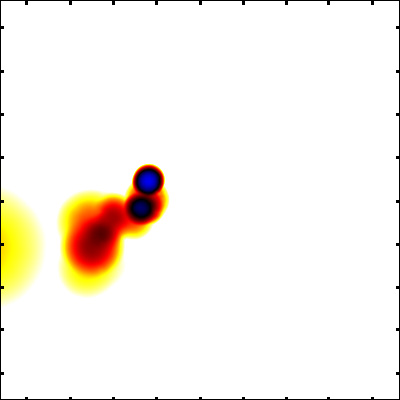}

  \includegraphics[width=\fact\columnwidth]{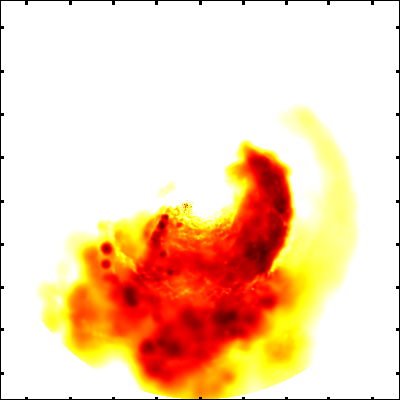}\put(-68,62){\fontfamily{phv}\selectfont SC8.5}\put(-82,0){\fontfamily{phv}\selectfont \rotatebox{90}{90$^\circ$}}%
  \includegraphics[width=\fact\columnwidth]{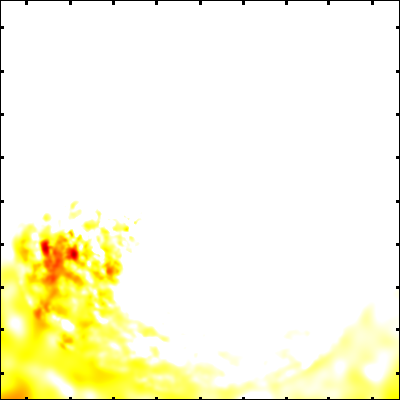}%
  \includegraphics[width=\fact\columnwidth]{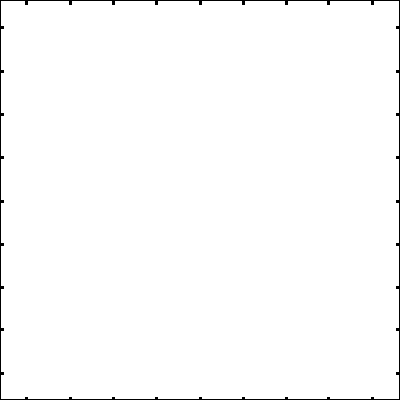}%
  \hspace{\panspac cm}
  \vrule
  \hspace{\panspac cm}
  \includegraphics[width=\fact\columnwidth]{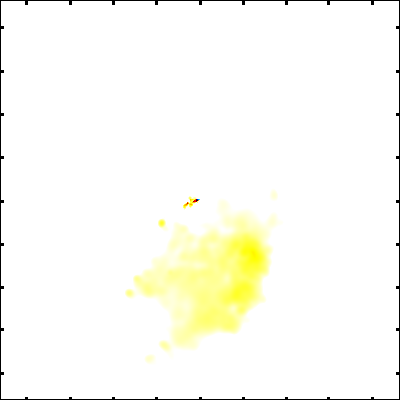}%
  \includegraphics[width=\fact\columnwidth]{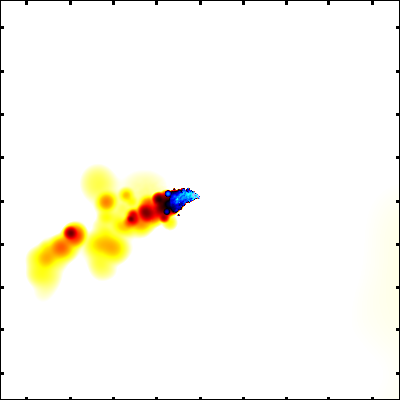}%
  \includegraphics[width=\fact\columnwidth]{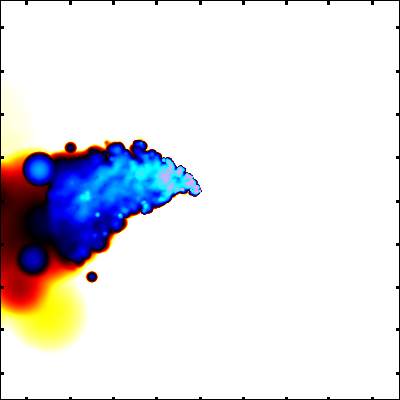}

  \includegraphics[width=\fact\columnwidth]{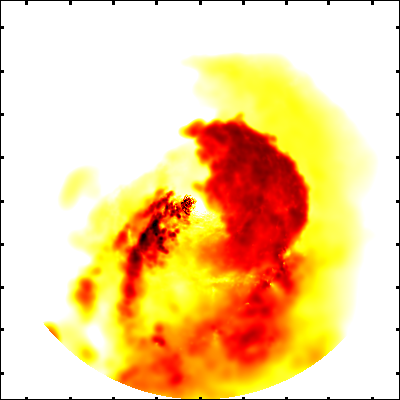}\put(-68,62){\fontfamily{phv}\selectfont SC4.8}\put(-80,58){\fontfamily{phv}\selectfont \rotatebox{90}{$\omega=$}}%
  \includegraphics[width=\fact\columnwidth]{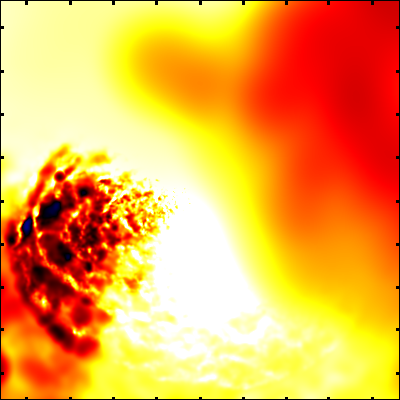}%
  \includegraphics[width=\fact\columnwidth]{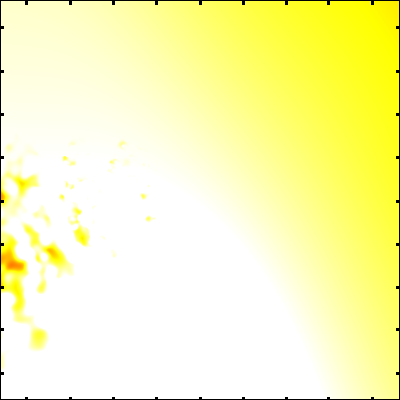}%
  \hspace{\panspac cm}
  \vrule
  \hspace{\panspac cm}
  \includegraphics[width=\fact\columnwidth]{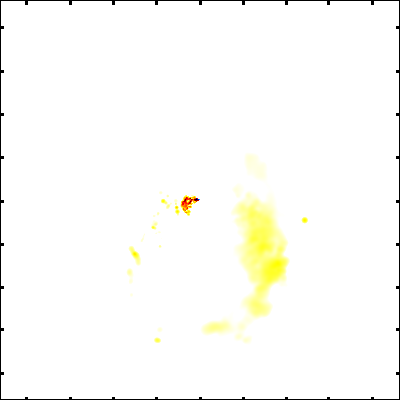}%
  \includegraphics[width=\fact\columnwidth]{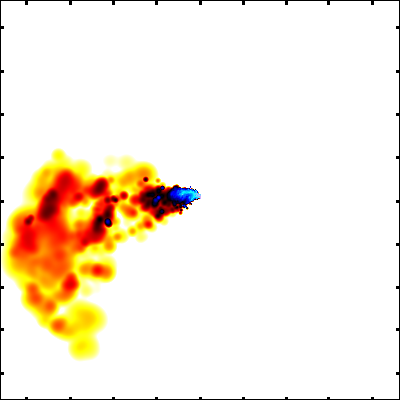}%
  \includegraphics[width=\fact\columnwidth]{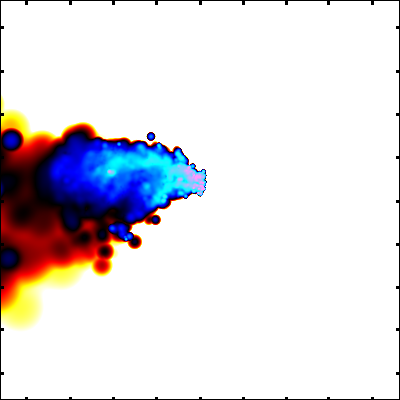}

  \includegraphics[width=\fact\columnwidth]{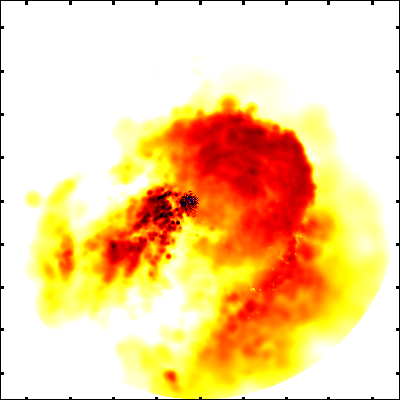}\put(-68,62){\fontfamily{phv}\selectfont SC2.4}%
  \includegraphics[width=\fact\columnwidth]{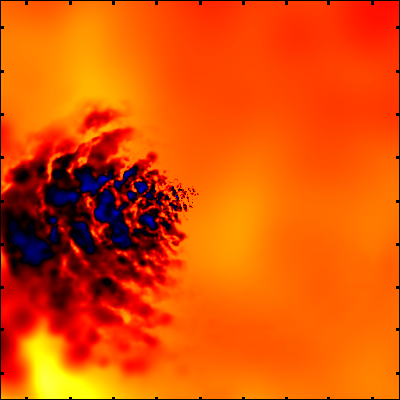}%
  \includegraphics[width=\fact\columnwidth]{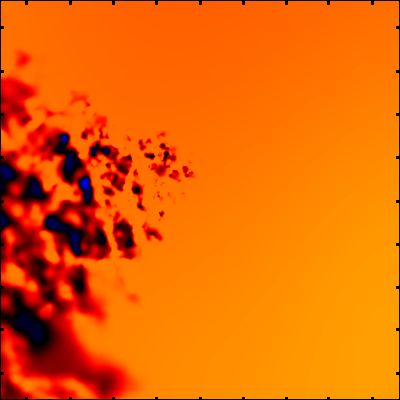}%
  \hspace{\panspac cm}
  \vrule
  \hspace{\panspac cm}
  \includegraphics[width=\fact\columnwidth]{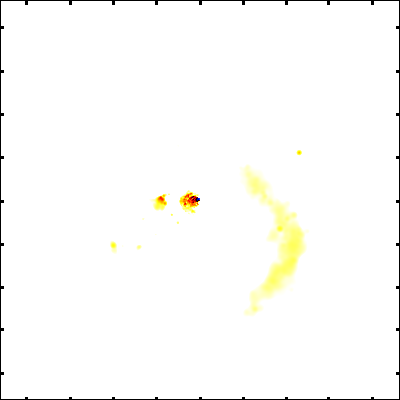}%
  \includegraphics[width=\fact\columnwidth]{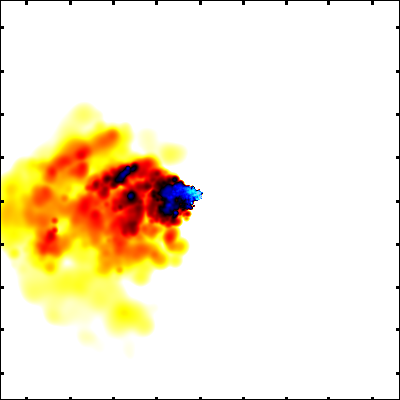}%
  \includegraphics[width=\fact\columnwidth]{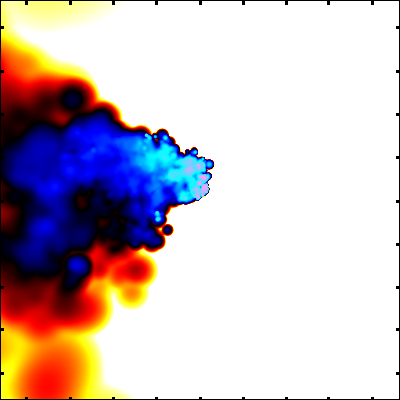}

  \includegraphics[trim={0.1cm 0.5cm 0.1cm 0},clip,width=\factdouble\columnwidth]{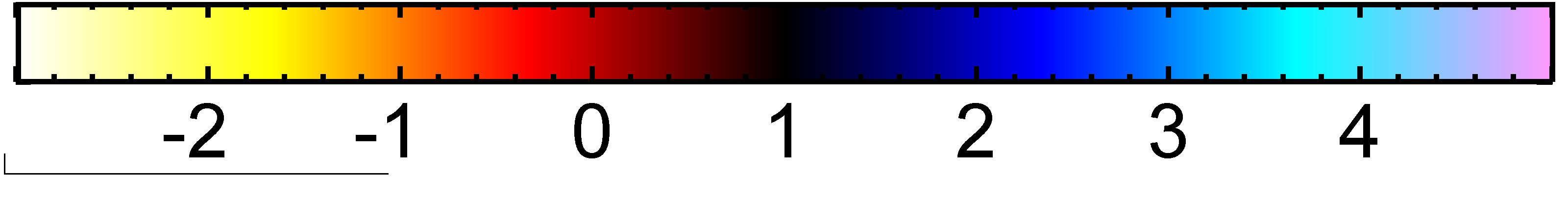}

  \caption{Surface brightness maps of the thermal X-ray emission at 1 keV (left half) and 10 keV (right half) viewed from $i=135^\circ$, $\omega=252^\circ$ (top half) and $i=135^\circ$, $\omega=90^\circ$ (bottom half).  Within each quadrant of 12 panels, left to right are the $\pm$100$a$, $\pm$10$a$, and $\pm$2$a$ regions, and top to bottom are NC8.5, SC8.5, SC4.8, and SC2.4.  The units of the colour scale, which is the same for all 48 panels, is log flux in erg/cm$^2$/s/keV/sr.}
  \label{fi:XPix}
  \end{center}
\end{figure*}

\begin{figure}
  \includegraphics[width=\columnwidth]{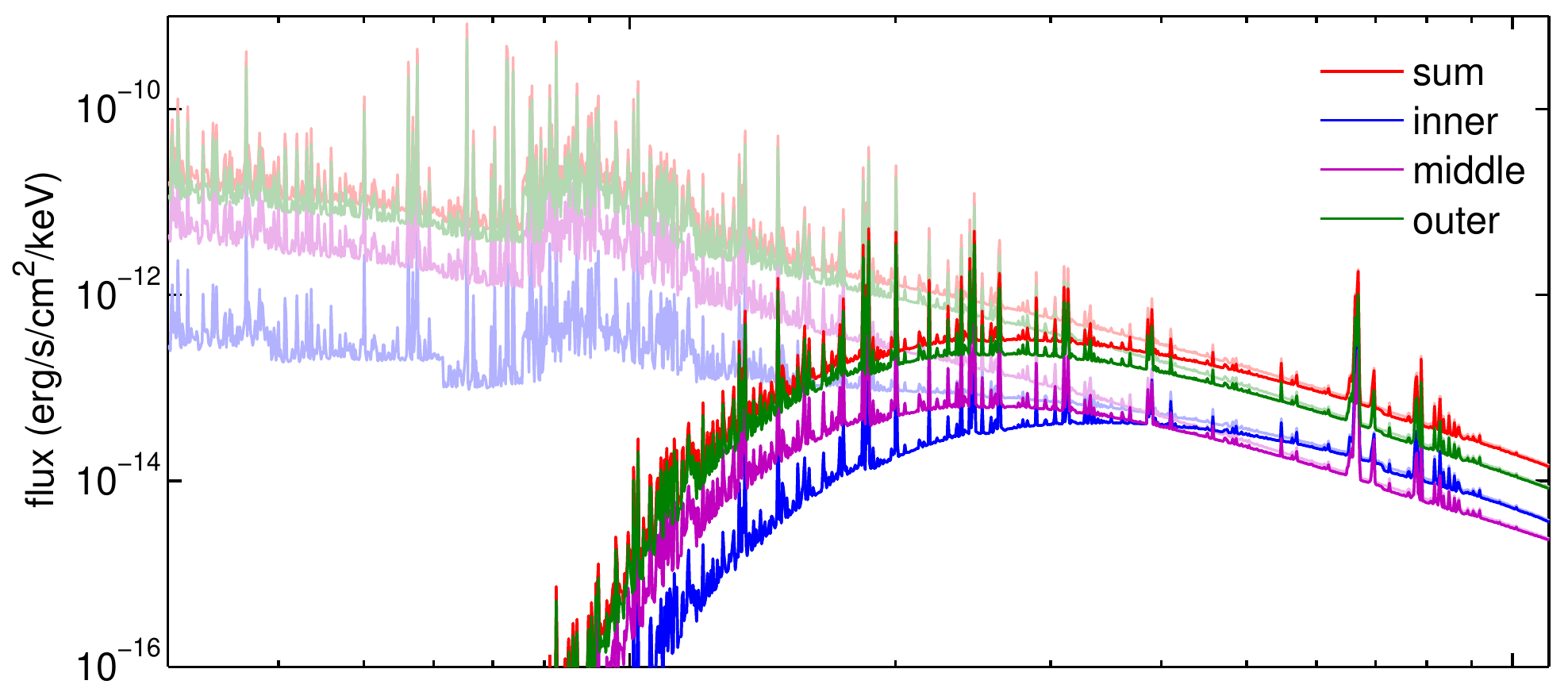}\put(-110,95){\scriptsize \fontfamily{phv}\selectfont NC8.5, $\omega$\,=\,252$^\circ$}

  \includegraphics[width=\columnwidth]{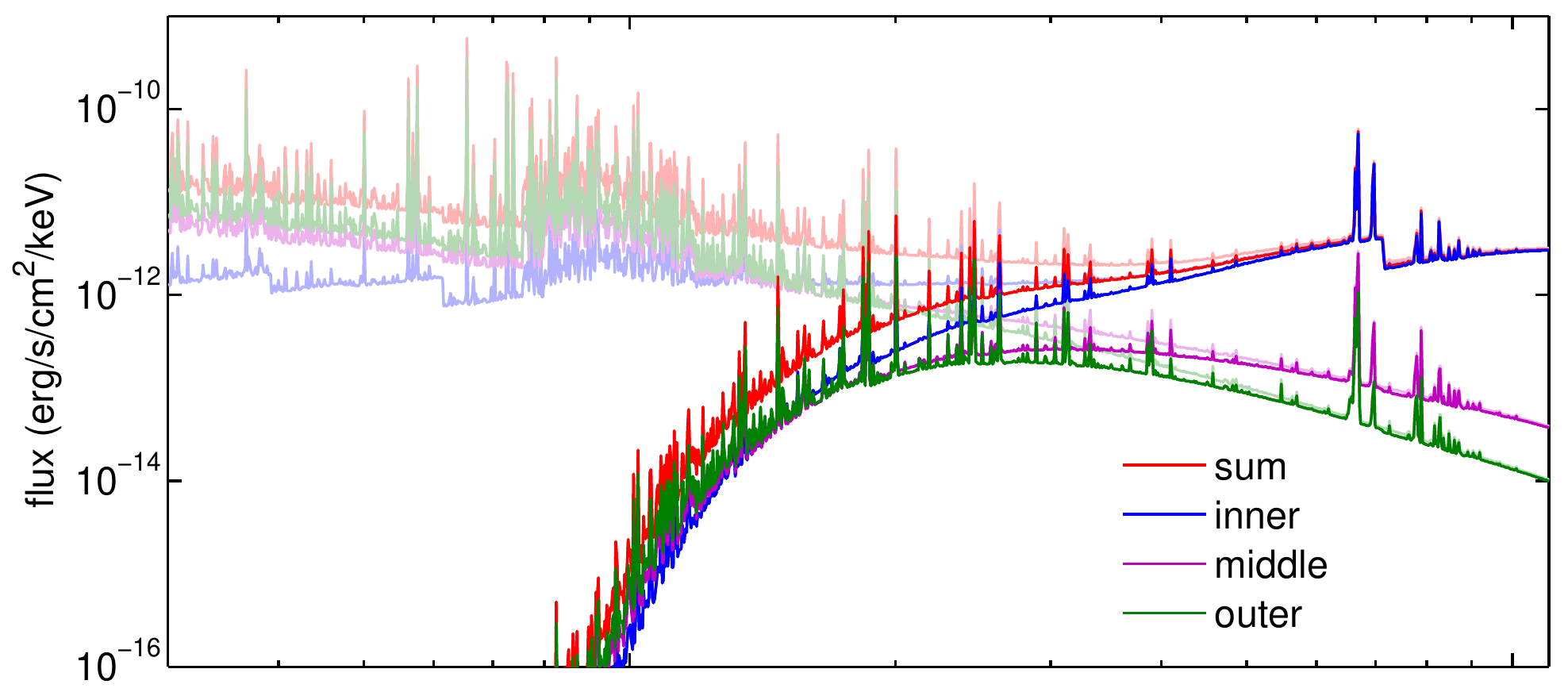}\put(-110,95){\scriptsize \fontfamily{phv}\selectfont SC8.5, $\omega$\,=\,252$^\circ$}

  \includegraphics[width=\columnwidth]{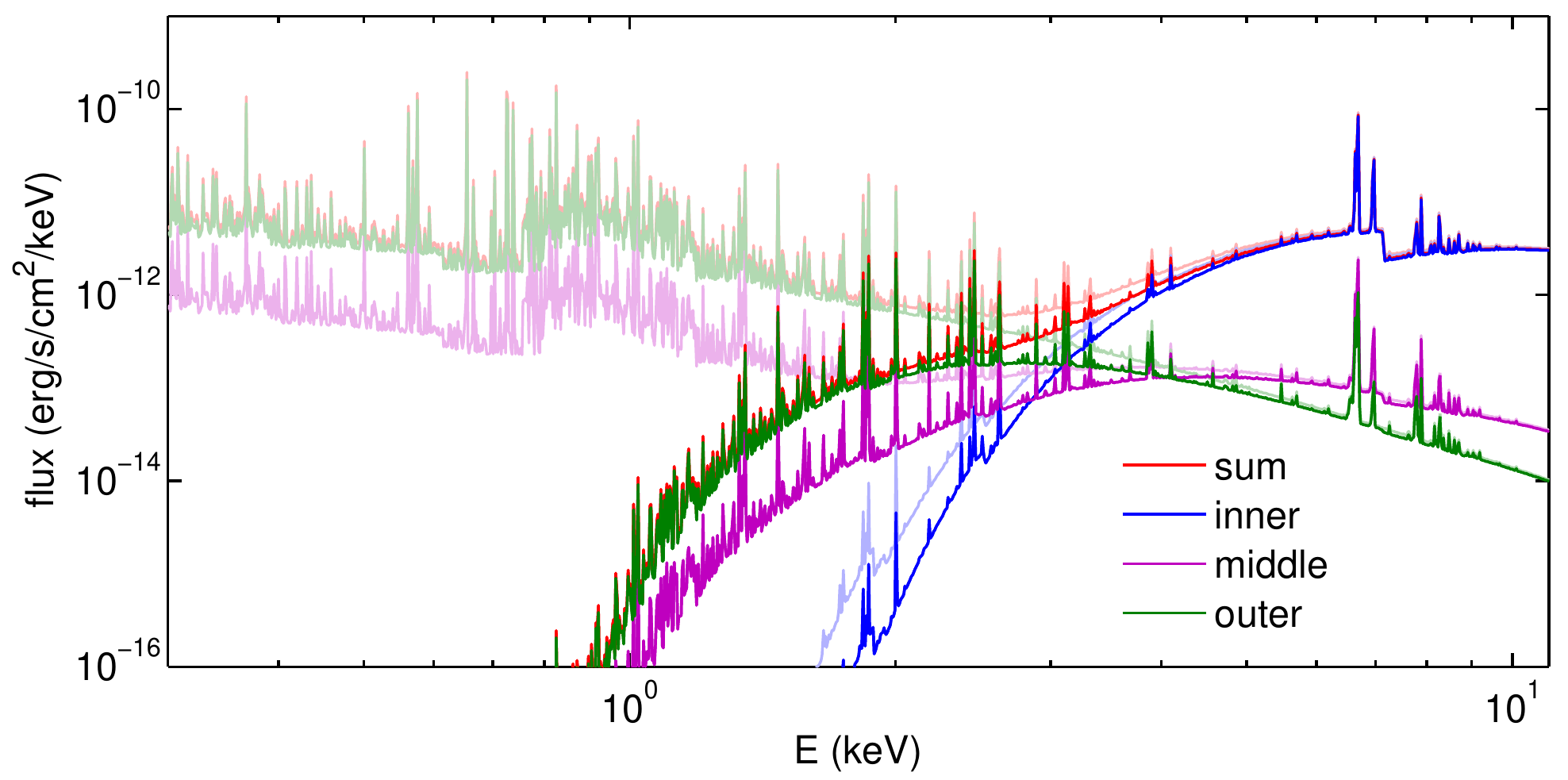}\put(-110,108){\scriptsize \fontfamily{phv}\selectfont SC8.5, $\omega$\,=\,90$^\circ$}

  \caption{Model spectra for NC8.5 with $\omega=252^\circ$ (top), SC8.5 with $\omega=252^\circ$ (middle), and SC8.5 with $\omega=90^\circ$ (bottom).  The dark lines are the spectra accounting for Homunculus and ISM absorption, i.e.\ what reaches the X-ray detector, while the faint lines do not account for these types of absorption, i.e.\ this is what comes from the radiative transfer calculation, so it only includes circumstellar absorption.}
  \label{fi:XSpecMod1}
\end{figure}

\begin{figure}
  \includegraphics[width=\columnwidth]{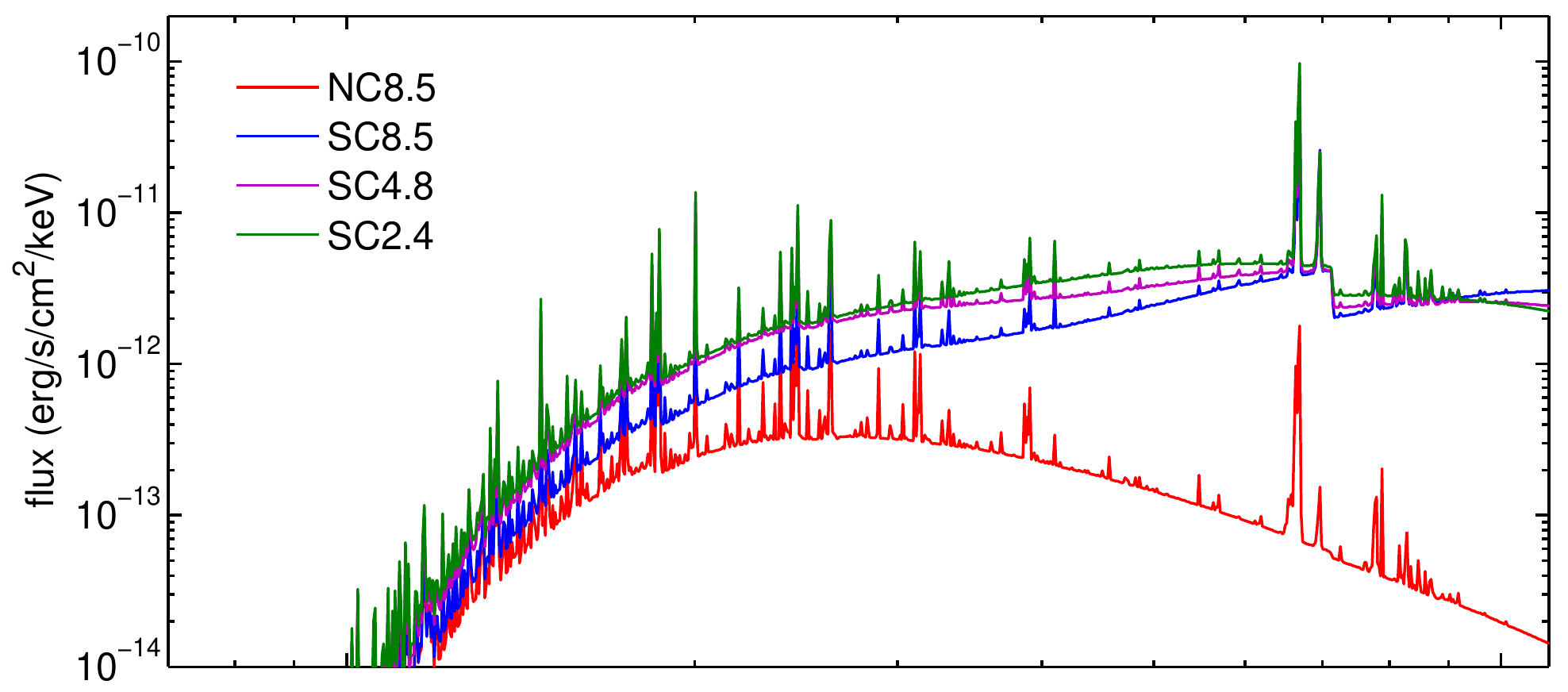}\put(-110,95){\scriptsize \fontfamily{phv}\selectfont $\omega$\,=\,252$^\circ$}

  \includegraphics[width=\columnwidth]{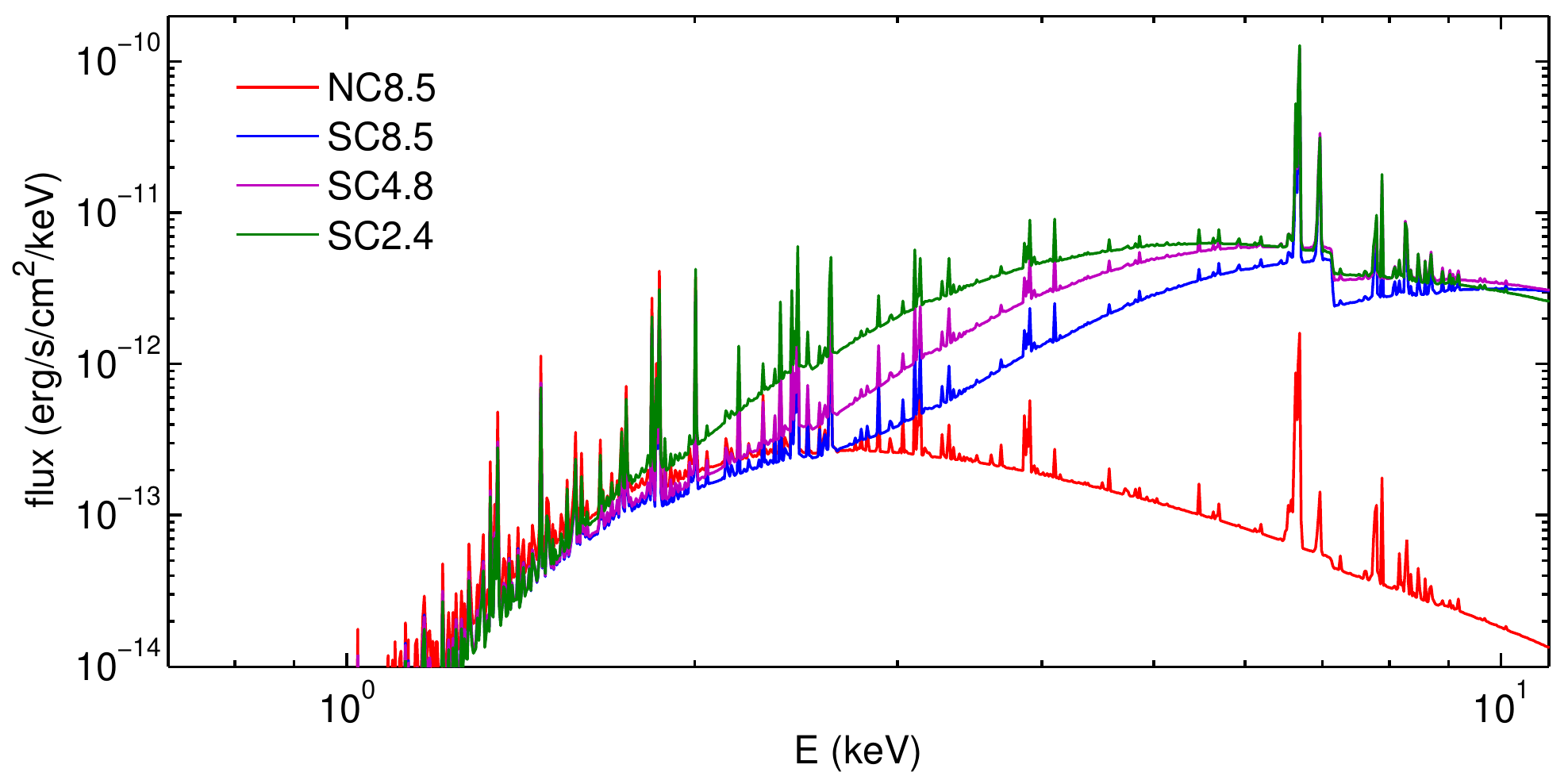}\put(-110,108){\scriptsize \fontfamily{phv}\selectfont $\omega$\,=\,90$^\circ$}

  \caption{Model spectra of the summed three components for $\omega=252^\circ$ (top) and $\omega=90^\circ$ (bottom).}
  \label{fi:XSpecMod2}
\end{figure}

\begin{figure}
  \includegraphics[width=\columnwidth]{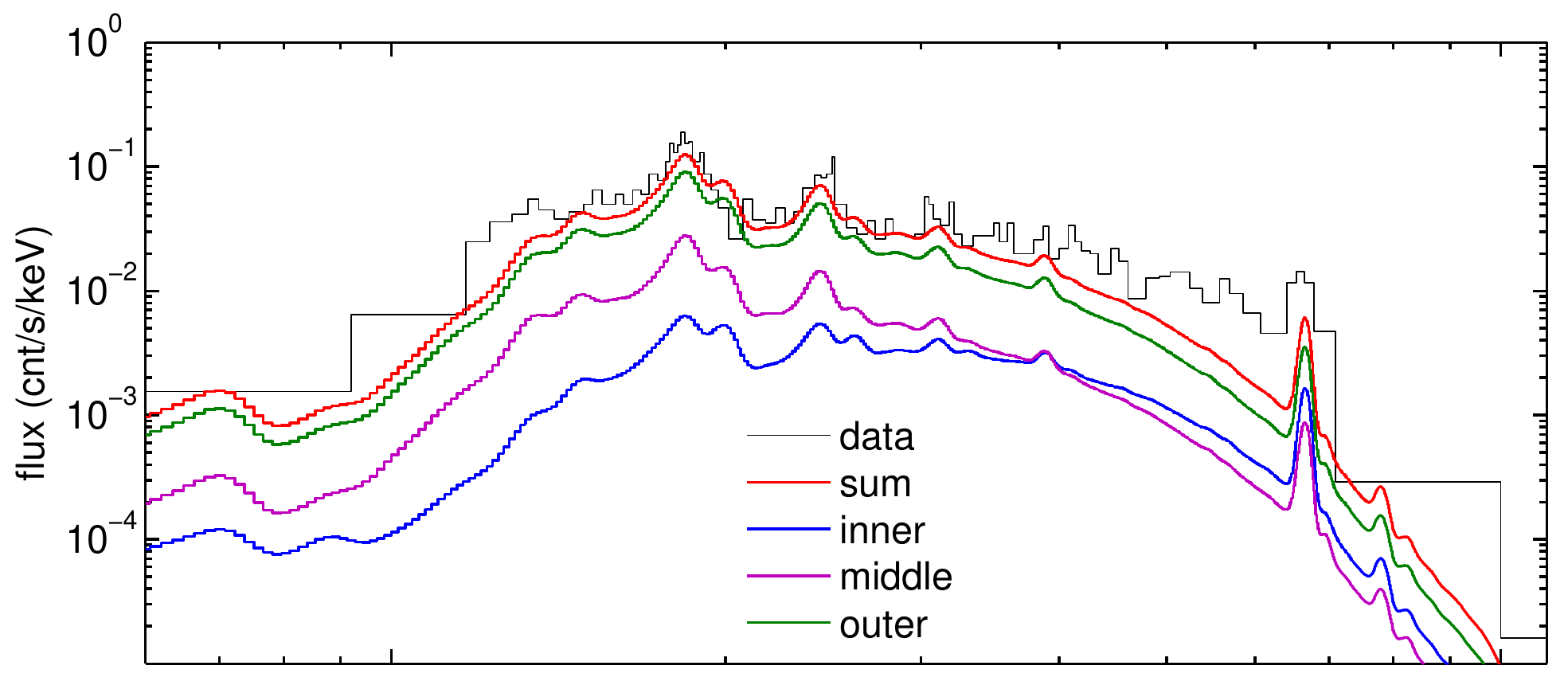}\put(-210,89){\scriptsize \fontfamily{phv}\selectfont NC8.5, $\omega$\,=\,252$^\circ$}

  \includegraphics[width=\columnwidth]{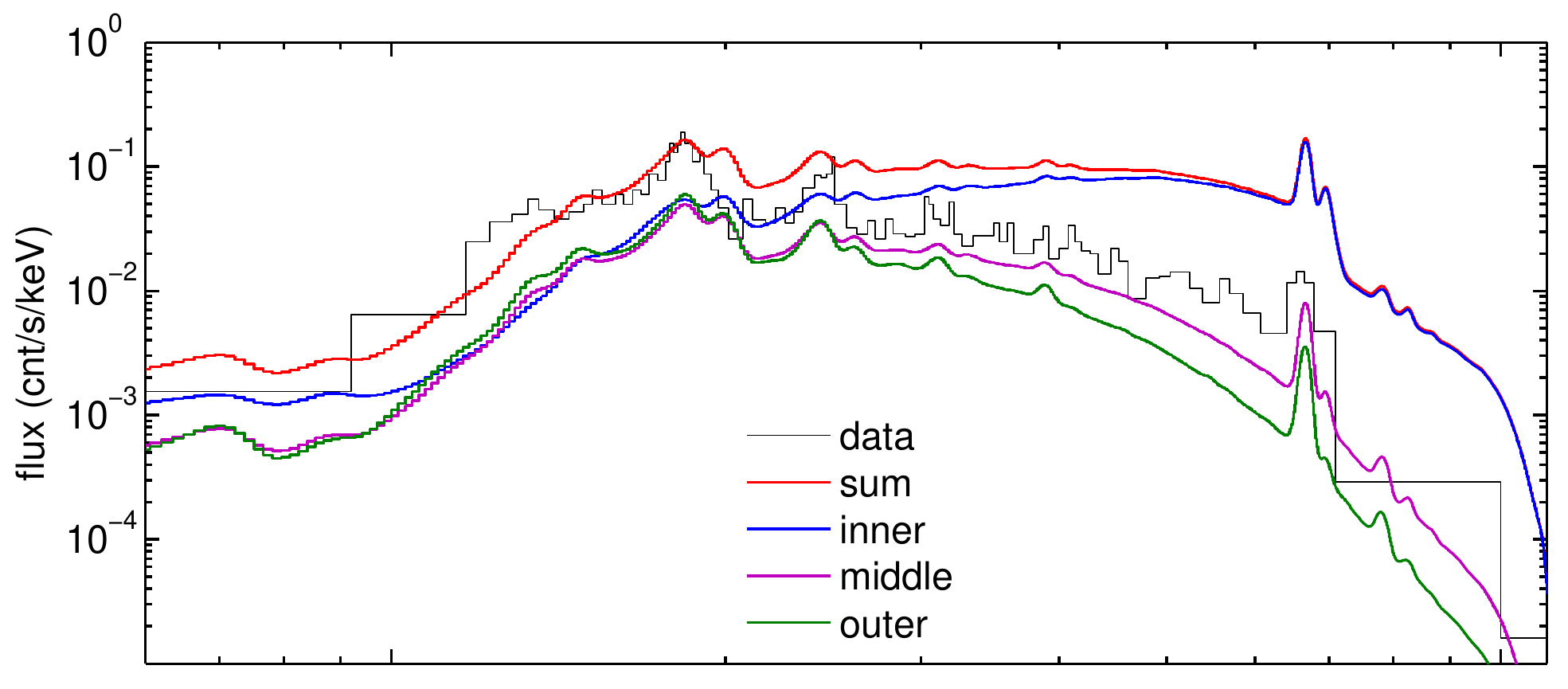}\put(-210,89){\scriptsize \fontfamily{phv}\selectfont SC8.5, $\omega$\,=\,252$^\circ$}

  \includegraphics[width=\columnwidth]{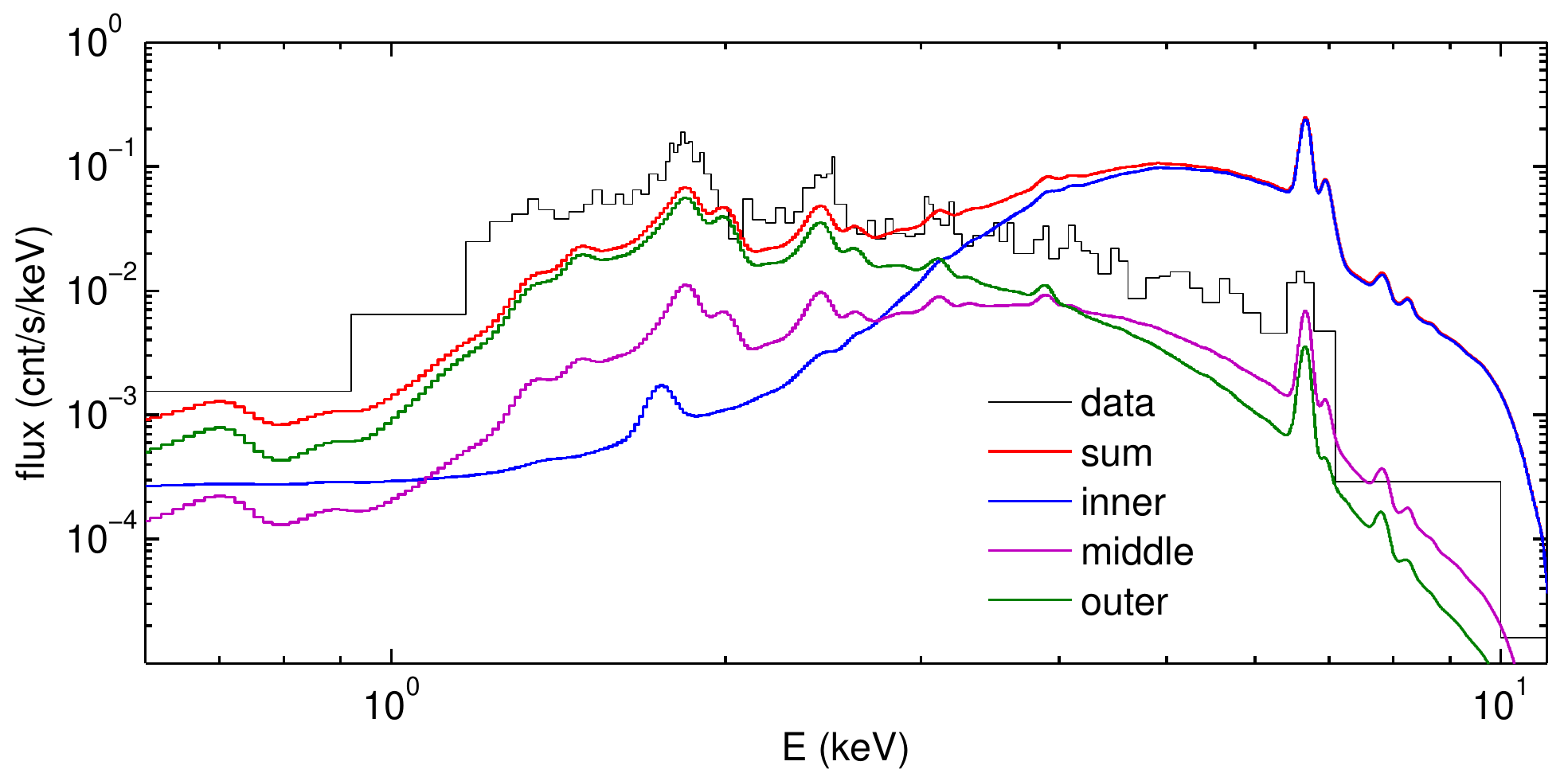}\put(-210,102){\scriptsize \fontfamily{phv}\selectfont SC8.5, $\omega$\,=\,90$^\circ$}

  \caption{\textit{Chandra} spectra for NC8.5 with $\omega=252^\circ$ (top), SC8.5 with $\omega=252^\circ$ (middle), and SC8.5 with $\omega=90^\circ$ (bottom) with the data.}
  \label{fi:XSpec1}
\end{figure}

\begin{figure}
  \includegraphics[width=\columnwidth]{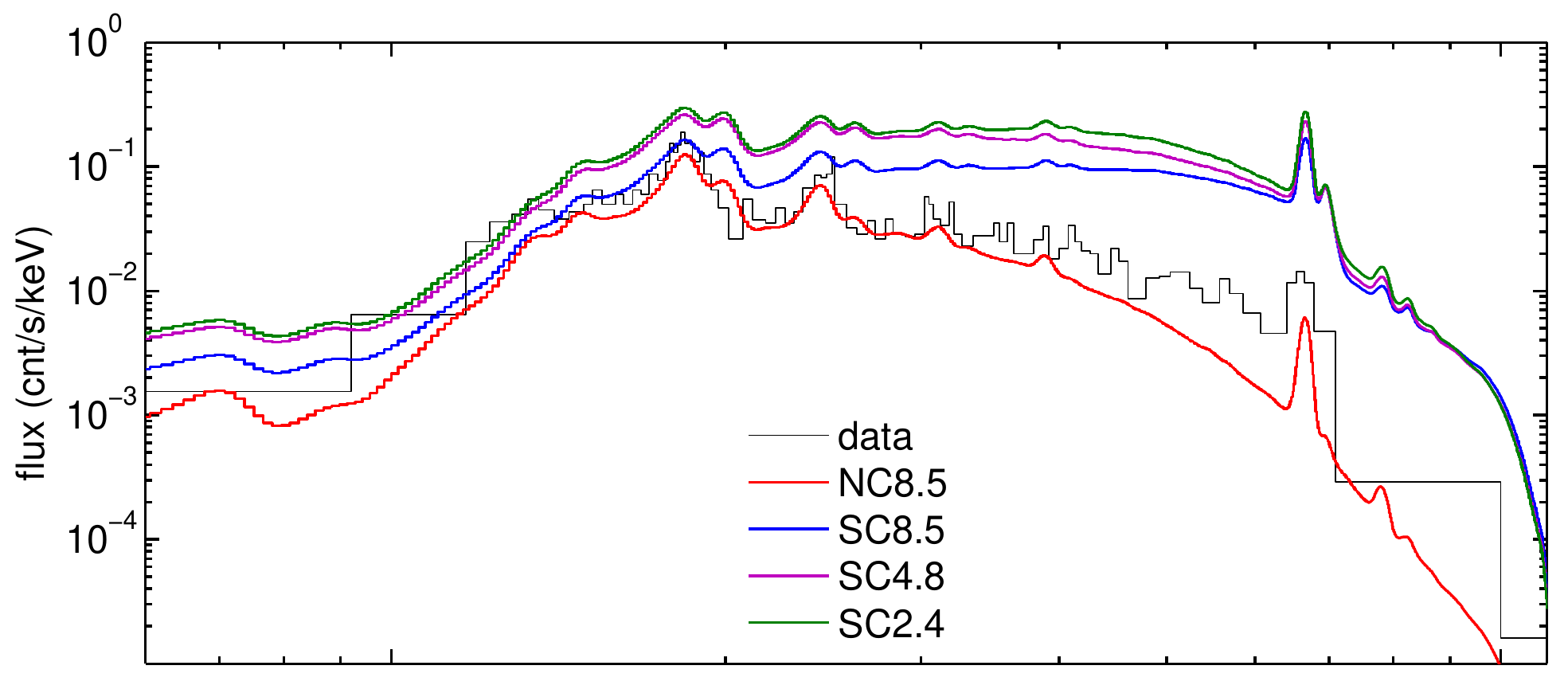}\put(-210,89){\scriptsize \fontfamily{phv}\selectfont $\omega$\,=\,252$^\circ$}

  \includegraphics[width=\columnwidth]{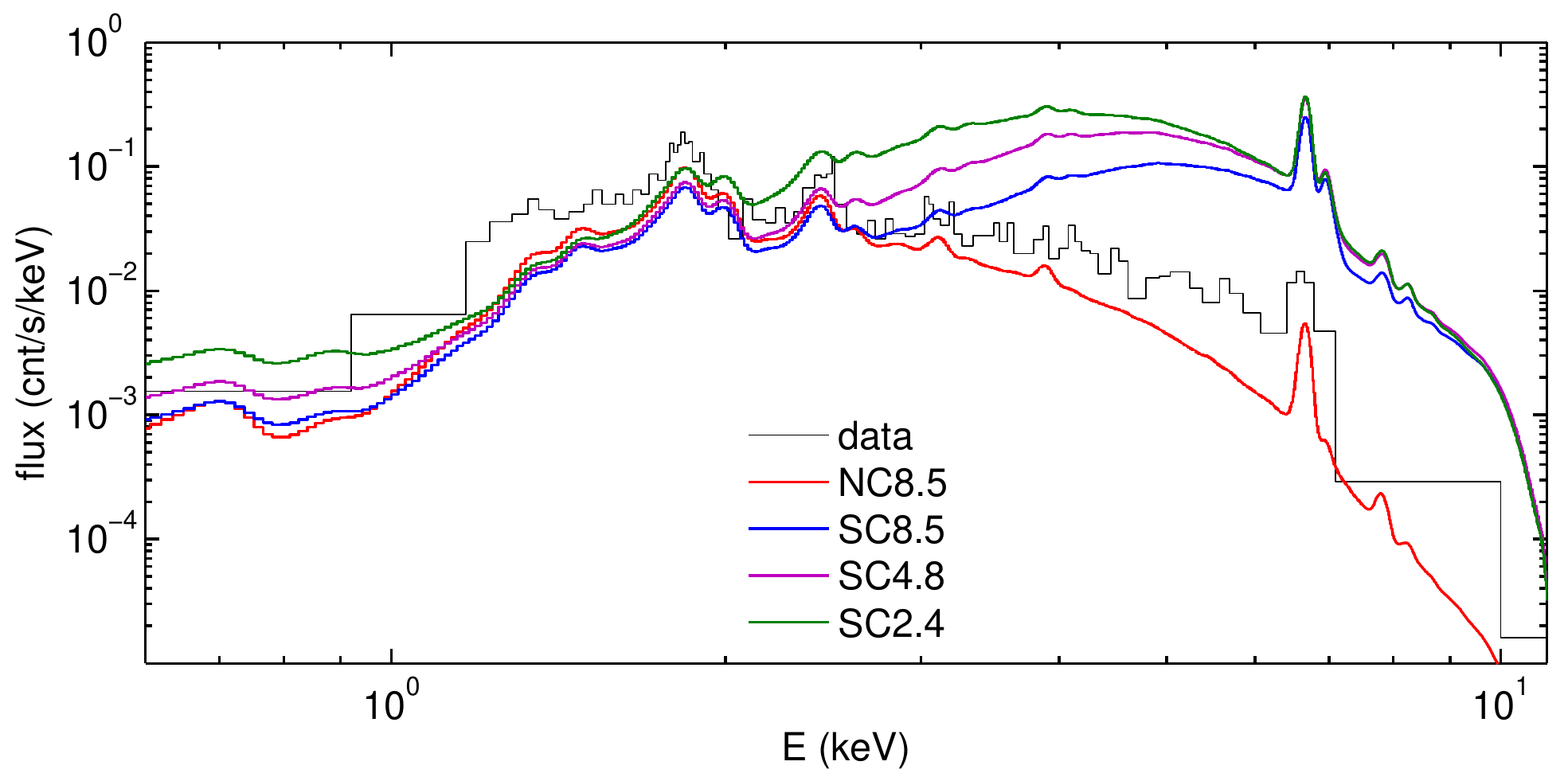}\put(-210,102){\scriptsize \fontfamily{phv}\selectfont $\omega$\,=\,90$^\circ$}

  \caption{\textit{Chandra} spectra of the summed three components for $\omega=252^\circ$ (top) and $\omega=90^\circ$ (bottom) with the data.}
  \label{fi:XSpec2}
\end{figure}

The next step is summing the pixel maps for all energies to generate model spectra for the two lines of sight.  Fig.~\ref{fi:XSpecMod1} shows the inner, middle, outer, and sum model spectra for NC8.5 at $\omega=252^\circ$ (top), SC8.5 at $\omega=252^\circ$ (centre), and SC8.5 at $\omega=90^\circ$ (bottom).  The light lines are before accounting for the absorption of the Homunculus and the interstellar medium (ISM), i.e.\ the fluxes from the radiative transfer calculations at the $r_\textrm{max}=100a$ simulation boundary, while the dark lines include the extra absorption component\footnote{To better compare the spectra among the different models and observing orientations, all absorption values for Figs.~\ref{fi:XSpecMod1}-\ref{fi:XSpec2} are $n_\textrm{H}=3.7\times10^{22}$\,cm$^{-2}$, the optimal value for matching the NC8.5/SC8.5, $\omega=252^\circ$ spectra to the observation.}, i.e.\ the fluxes that get folded through the \textit{Chandra} response function.  As is shown in the images, the coupling strength has a large influence on the emission from the inner region.  The flatter shape in the hard region and the ratio of the $\sim6.7$~keV to $\sim6.9$~keV Fe lines indicate the higher temperature of the X-ray-producing gas in the strong-coupling simulations.  The outer region produces the largest contribution to the emission in NC8.5 (green above blue and purple), again showing the need for such large scale hydrodynamic simulations, but the outer region is the weakest component in SC8.5 (green below blue and purple) since the strong coupling produces much central emission.  The reduced X-ray emission below $\sim$4~keV in $\omega=90^\circ$ (bottom panel) compared to $\omega=252^\circ$ (middle panel) is from the extra circumstellar absorption looking through the higher density primary stellar wind.  The circumstellar absorption is even high enough to be the dominant component for the inner region of $\omega=90^\circ$ since the pre-ISM/pre-Homunculus spectra (faint blue, bottom panel) is already heavily absorbed.

Fig.~\ref{fi:XSpecMod2} shows the summed spectra for all four models and both lines of sight.  Among the SC models for $\omega=252^\circ$ (top panel), the variations due to absorption, the dominant source of which is the shell of primary wind material ejected during the previous periastron cycle, are subtle but noticeable due to the differing mass of the primary-wind shell.  On the other hand, the SC models of $\omega=90^\circ$ (bottom panels) show large variations due to looking through primary winds of varying densities when compared with each other (green to purple to blue) and when compared with their $\omega=252^\circ$ counterparts.  The extra absorption of the NC model is also apparent for $\omega=90^\circ$.

Figs.~\ref{fi:XSpec1} \& \ref{fi:XSpec2} show the same spectra as Figs.~\ref{fi:XSpecMod1} \& \ref{fi:XSpecMod2} now folded through the \textit{Chandra} ACIS-S response function.  The CCE observation (see below) is also plotted (black).  All the trends of the previous plots are shown, but the shape of the detector response make the changes in the soft part of the band due to absorption changes harder to see.

The CCE spectrum is created by taking the minimum flux of five (six) spectra at each energy band across the 2009.0 (2014.6) minimum.
Rather than repeat this with the model, which would involve doing the embedding of, and the radiative transfer calculations on, the SPH output at those 11 phases, we simply choose the observation from 2009 January 22 \citep{HamaguchiP14} as representative of the CCE spectrum.  This observation, which occurs $\sim$0.003 in phase (6 d) after X-ray phase 0.0 (according to the \textit{RXTE} light curve observations; \citealt{CorcoranP10}), is the closest to the CCE spectra; the 2009 January 22 observation only has slightly more soft flux than the CCE model.  We choose to match this observation with exactly periastron (orbital phase 0.0) of the SPH simulations.  There can be a shift between X-ray phase 0.0 and periastron, but since the spectra that are used to generate the CCE only vary slightly across the $\sim$0.015 in phase of the deep minimum, modelling the CCE can not strongly constrain this phase shift.

\begin{figure}
  \includegraphics[width=\columnwidth]{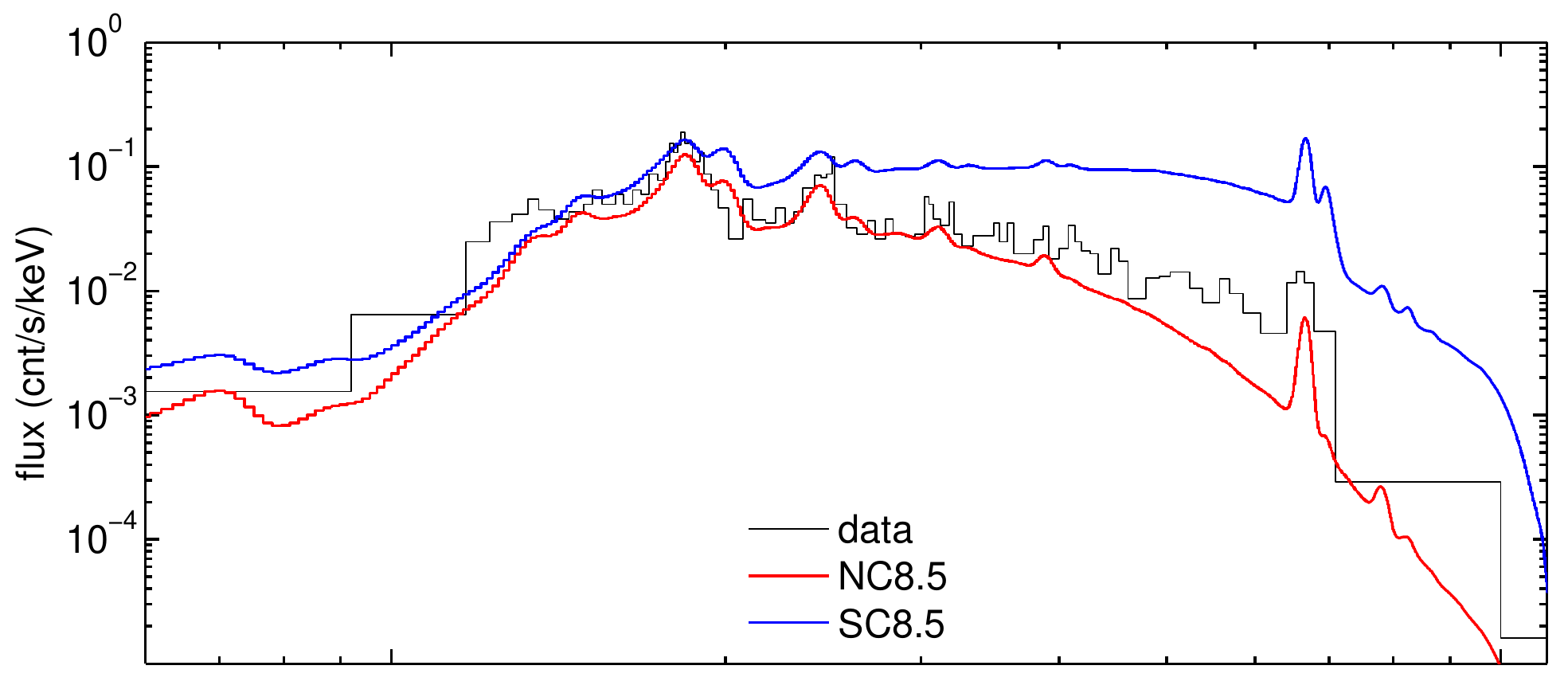}\put(-210,89){\scriptsize \fontfamily{phv}\selectfont $\omega$\,=\,252$^\circ$, ${\mathsfit n_{\mathsfit H}}=3.7\times10^{22}$\,cm$^{-2}$}

  \includegraphics[width=\columnwidth]{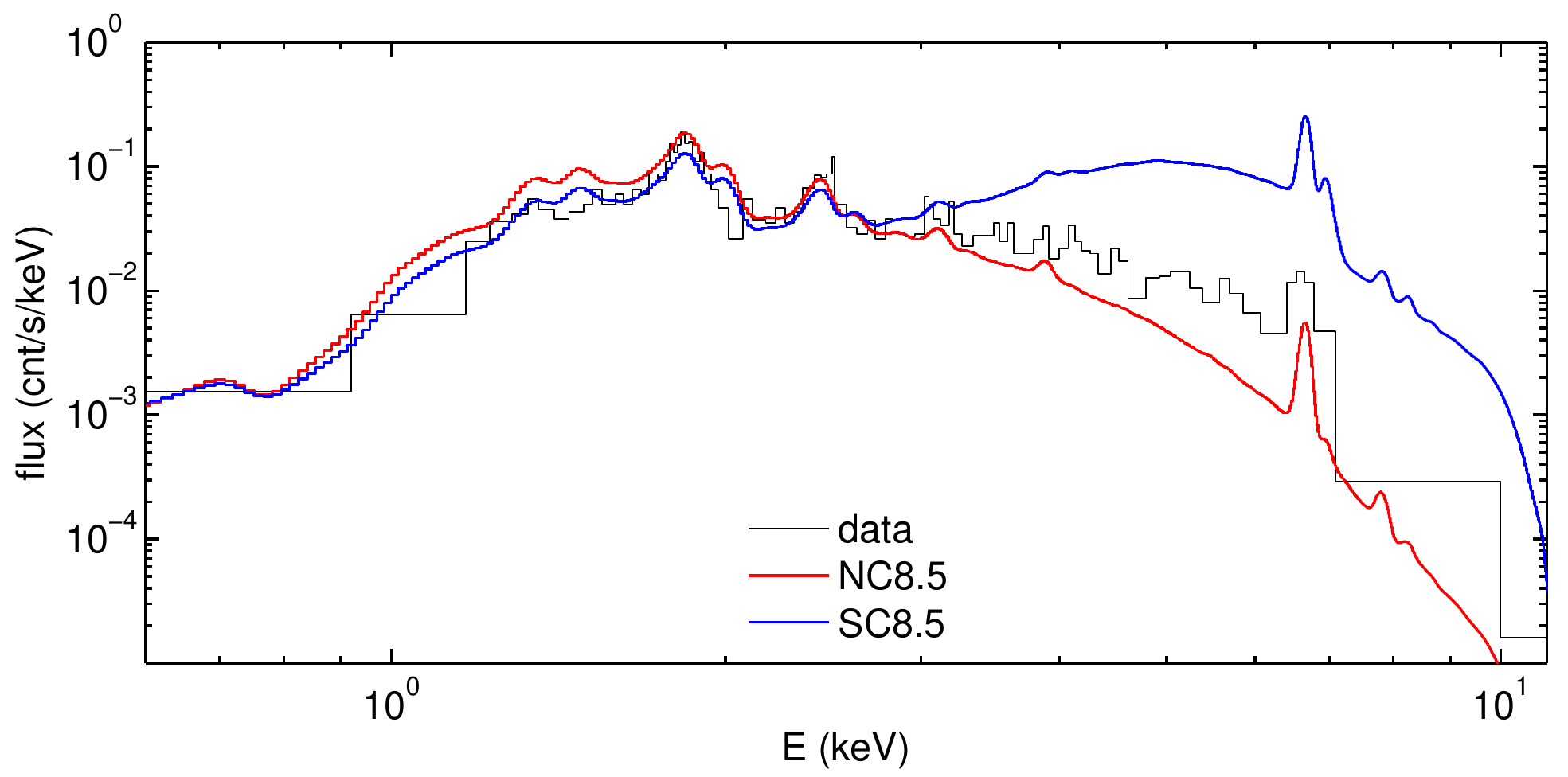}\put(-210,102){\scriptsize \fontfamily{phv}\selectfont $\omega$\,=\,90$^\circ$, ${\mathsfit n_{\mathsfit H}}=2.5\times10^{22}$\,cm$^{-2}$}

  \caption{\textit{Chandra} spectra of the summed three components of NC8.5 and SC8.5 for $\omega=252^\circ$ (top) and $\omega=90^\circ$ (bottom) with the data.}
  \label{fi:XSpec3}
\end{figure}

Fig.~\ref{fi:XSpec3} summarizes the main result of this work.  For the preferred mass-loss rate of $\dot{M}_A=8.5\times10^{-4}$\,M$_\odot$\,yr$^{-1}$, the model spectra summed over the entire $r_\textrm{max}=100a$ simulation volume reproduce the observed CCE spectrum; the strong- and no-coupling spectra well bound the observation for $\omega=252^\circ$(top panel), while they bound the hard portion and converge on the soft portion for $\omega=90^\circ$ (bottom panel) using a lower ISM/Homunculus absorption value. Therefore, the CCE emission arises from the current secondary wind colliding with primary wind ejected during the previous periastron passage, as well as emission on a smaller scale from the downstream portion of the leading-arm of the wind--wind collision region between the stars.  Additionally, the result suggests an intermediate amount of coupling between the primary radiation and secondary wind in $\eta$ Carinae.

\section{Discussion}\label{sec:D}

\subsection{Absorbing column and line of sight}

Each line of sight can reproduce the observations at the same level through employing slightly different ISM/Homunculus absorbing columns.  The line of sight plays a significant role in the inner spectra since the impact parameters to the stars are lower and thus the column densities are higher, but this effect is mitigated in the middle and outer regions as the impact parameters are larger and column densities lower (see Fig.~\ref{fi:XSpecMod1}).  Therefore, the full model CCE spectra is only moderately affected by the line of sight, which can be counteracted by different $n_\textrm{H}$ values.  By matching the spectral set (strong coupling and no coupling) for each line of sight independently, the optimal values are  $n_\textrm{H}=\{3.7,2.5\}\times10^{22}$\,cm$^{-2}$ for $\omega=\{252,90\}^\circ$.  Therefore, the CCE X-ray emission is not a good diagnostic for determining the line of sight to the system.

It is also worth noting that both ISM/Homunculus values are consistent with the literature value of $n_\textrm{H}=5\times10^{22}$\,cm$^{-2}$ \citep{HamaguchiP07}.   This value
accounts for all absorption to the X-ray emitting source, while the present work splits the absorption into a circumstellar component (incorporated into the radiative transfer calculation within the $r_\textrm{max}=100a$ simulation boundary) and an ISM/Homunculus component, so the model values being below the literature value means they are in agreement.

\subsection{Comparisons with thermal X-ray spectral fitting parameters}

Common fitting procedures for thermal X-ray data involve constructing a model of 2-3 sources of emission that have different temperatures and possibly different absorbing columns.  This is obviously a simplification since an X-ray source will typically have an assortment of material at a range of temperatures and absorbing columns, but the simplified two-temperature or three-temperature models have been able to well reproduce X-ray data, including that of $\eta$ Carinae \citep[e.g.][]{HamaguchiP14}.  We compute the temperature distribution and absorbing columns of the gas in the hydrodynamic simulations to compare with the simplistic fits.

\begin{figure}
  \includegraphics[width=\columnwidth]{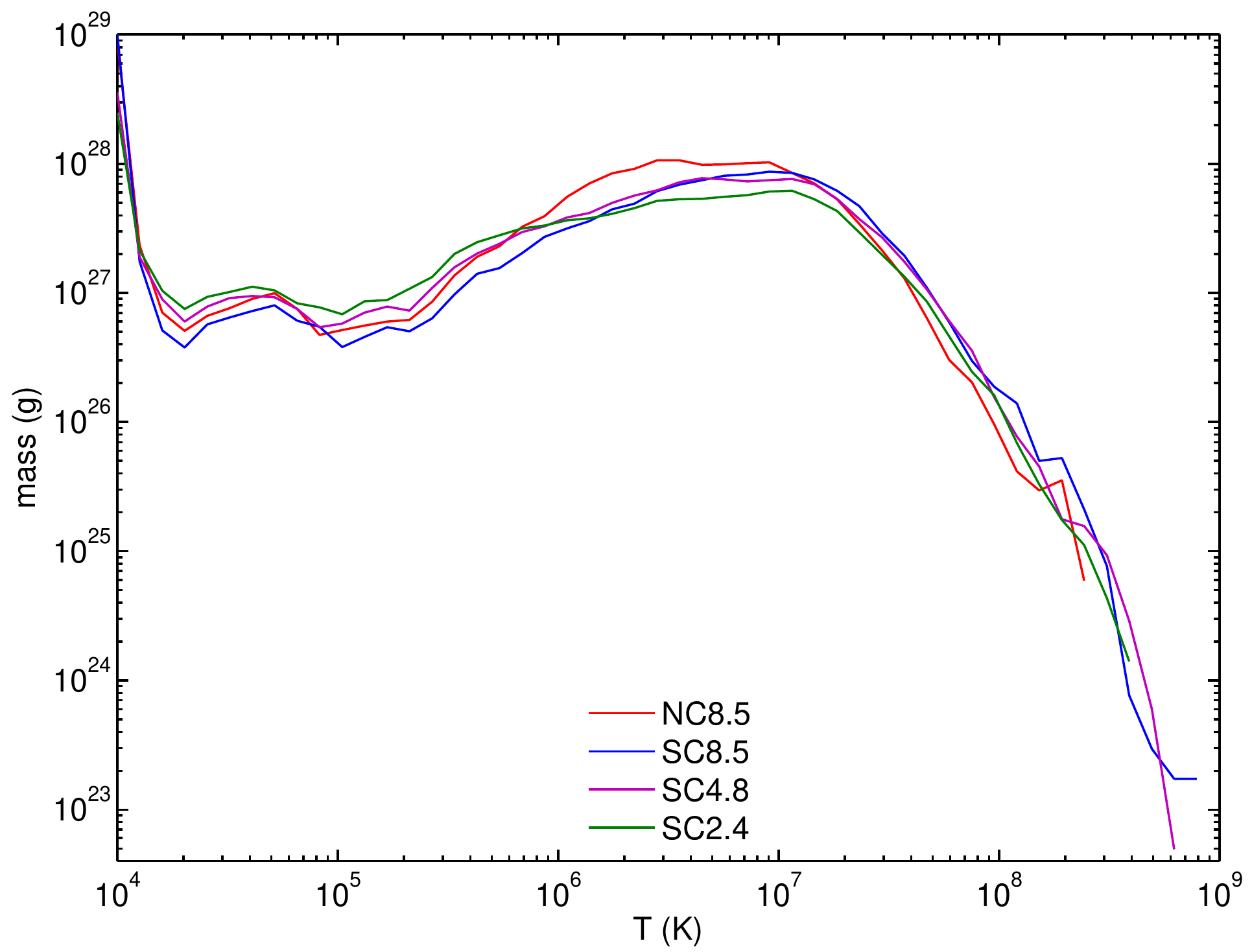}

  \caption{Histogram of mass as a function of temperature for secondary wind particles in all model simulations.}
  \label{fi:EM}
\end{figure}

Fig.~\ref{fi:EM} shows the mass--temperature distribution of the secondary wind for all four models.  (Recall that X-ray emission scales $\sim$$\rho^2$, so the X-ray deviations between models in orders of magnitude is twice that of this plot.)  In log-spaced bins, the peak for NC8.5 is $\sim3\times10^6$\,K, while the SC models peak closer to $\sim10^7$\,K.  This latter value is consistent with the one-temperature fitting parameter of 1.05\,keV $\rightarrow$ $1.22\times10^7\,$K that \citet{HamaguchiP07} found from fitting the 2003.5 CCE observation, even though the strong-coupling spectra from the hydrodynamic modelling are harder than the observation.  The updated fitting of \citet{HamaguchiP14,HamaguchiP15} indicates material at $\sim$5\,keV $\rightarrow$ $5.8\times10^7\,$K, which is also seen in the models, though in less quantity than the 1.05\,keV component.  The ratio of the amount of material in NC8.5 to SC8.5 at log $T$(K) = 6, 7, and 8 is $\sim$2, $\sim$1, and $\sim$0.5, respectively, which further explains why SC8.5 has a harder spectrum.  The figure also shows that the maximum temperature of material produced in each of the simulations, which decreases from SC8.5 to SC4.8 to SC2.4 to NC8.5, follows the trend of the coupling strength\footnote{Within the set of strong-coupling computations, the secondary luminosity increases as $\dot{M}_A$ increases to preserve the predicted mass-loss rate scaling of CAK theory, so the opacity decreases \citep[see Appendix 1 of ][]{MaduraP13}, thus its product with a consistent primary luminosity among the three strong-coupling models also decreases, causing the coupling to slightly increase with increasing $\dot{M}_A$.}.

To compute $n_\textrm{H}$, we use a method that pixel-by-pixel compares the intrinsic intensity and the absorbed intensity as a function of energy, and then uses the opacity of that energy to determine the absorbing column.  The intrinsic intensity is
\begin{equation}\label{eq:Int}
  I_*(E,x',y')=\int_{-z_\textrm{max}(x',y')}^{z_\textrm{max}(x',y')} j(E,x',y',z'') \textrm{d}z'',
\end{equation}
so based on the simplistic radiative transfer equation $I=I_*\exp(-\kappa n_\textrm{H} m_\textrm{p})$,
where $m_\textrm{p}$ is the mass of the proton, the absorbing column for each pixel is
\begin{equation}\label{eq:nH}
  n_\textrm{H}(E,x',y')=\frac{1}{\kappa(E)m_\textrm{p}}\ln\left(\frac{I_*(E,x',y')}{I(E,x',y')}\right).
\end{equation}
Note that $n_\textrm{H}$ is a function of energy since the emission locations of gas producing X-rays at different energies can vary.

Fig.~\ref{fi:nH1} shows the $n_\textrm{H}$ values for NC8.5 for 0.1, 1, and 10 keV (left to right) for $\omega=252^\circ$ (top rows) and $\omega=90^\circ$ (bottom rows) on scales of $\pm$100$a$ and $\pm$20$a$.  The absorbing column is higher for $\omega=90^\circ$, and the absorbing column in the centre increases with energy, indicating the highest energy X-rays are closest to the stars.  The upper portion (red and yellow) of the $\omega=252^\circ$ plots are column densities through secondary wind, with the clumps caused by the secondary wind pummelling the primary shell visible (black).  There is also a portion of the lower-left quadrant that is viewed through primary wind (blue) as the secondary approaches periastron on the right side of centre (see Fig.~\ref{fi:LoS}), so the primary wind fills in left of centre.  This material is the beginning of the next cycle's primary wind shell.

\newcommand*{\facto}{0.31}
\newcommand*{\factoD}{0.6}
\begin{figure}
\begin{center}
  \includegraphics[width=\facto\columnwidth]{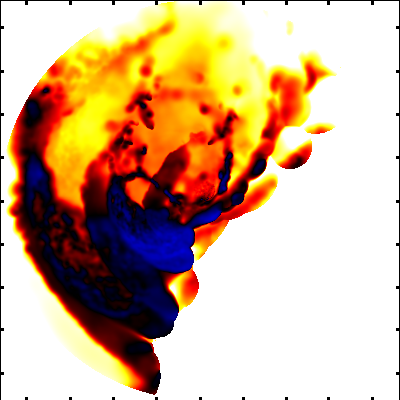}\put(-47,77){\fontfamily{phv}\selectfont 0.1\,keV}\put(-72,67){\fontfamily{phv}\selectfont $\pm$100$a$}\put(-82,0){\fontfamily{phv}\selectfont \rotatebox{90}{252$^\circ$}}%
  \includegraphics[width=\facto\columnwidth]{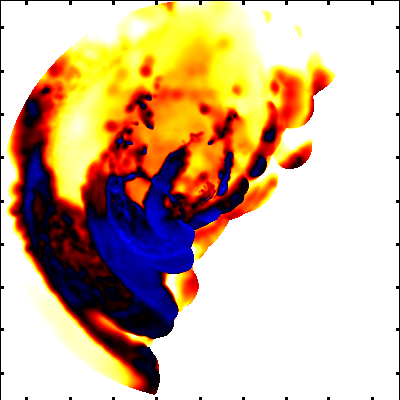}\put(-47,77){\fontfamily{phv}\selectfont 1\,keV}%
  \includegraphics[width=\facto\columnwidth]{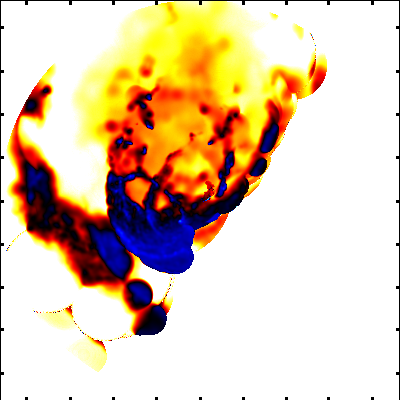}\put(-47,77){\fontfamily{phv}\selectfont 10\,keV}

  \includegraphics[width=\facto\columnwidth]{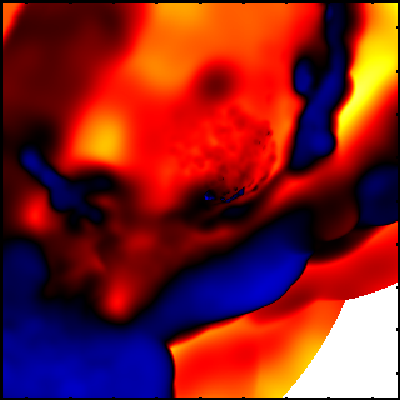}\put(-72,67){\fontfamily{phv}\selectfont \textcolor{white}{$\pm$20$a$}}\put(-80,59){\fontfamily{phv}\selectfont \rotatebox{90}{$\omega=$}}%
  \includegraphics[width=\facto\columnwidth]{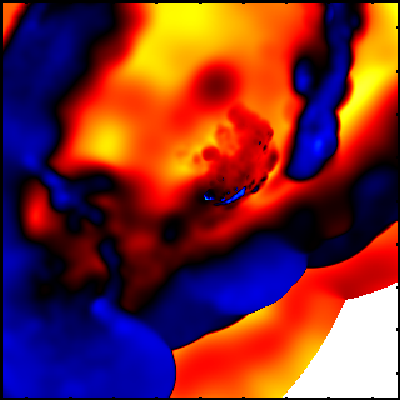}%
  \includegraphics[width=\facto\columnwidth]{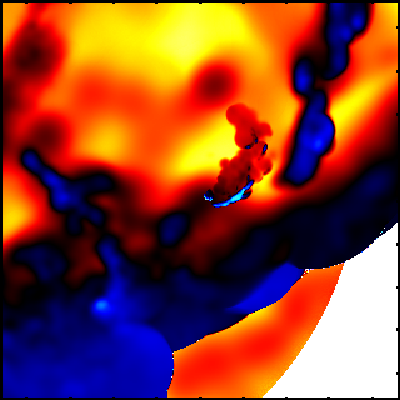}

  \includegraphics[width=\facto\columnwidth]{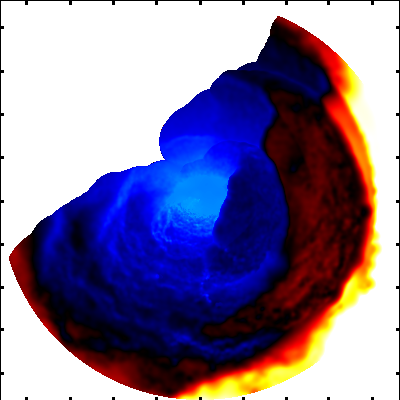}\put(-72,67){\fontfamily{phv}\selectfont $\pm$100$a$}\put(-82,0){\fontfamily{phv}\selectfont \rotatebox{90}{90$^\circ$}}%
  \includegraphics[width=\facto\columnwidth]{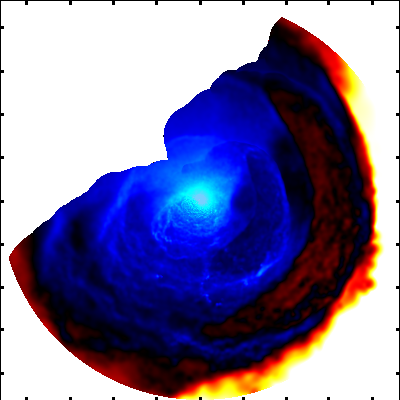}%
  \includegraphics[width=\facto\columnwidth]{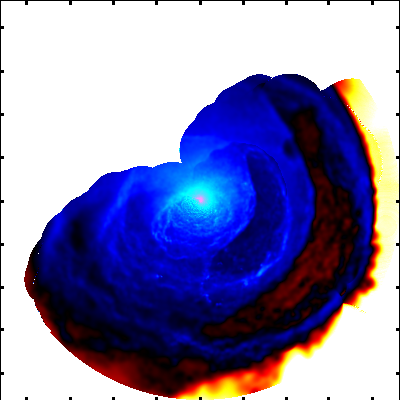}

  \includegraphics[width=\facto\columnwidth]{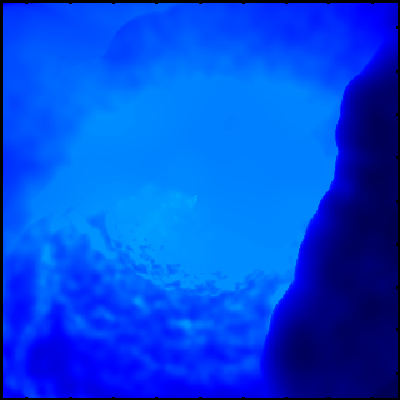}\put(-72,67){\fontfamily{phv}\selectfont $\pm$20$a$}\put(-80,59){\fontfamily{phv}\selectfont \rotatebox{90}{$\omega=$}}%
  \includegraphics[width=\facto\columnwidth]{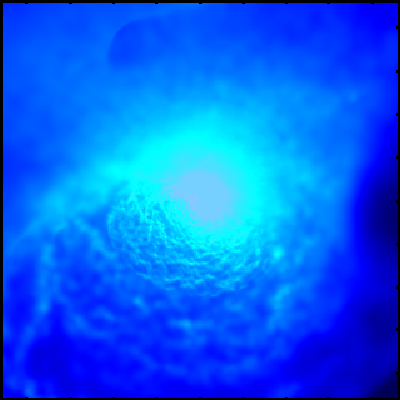}%
  \includegraphics[width=\facto\columnwidth]{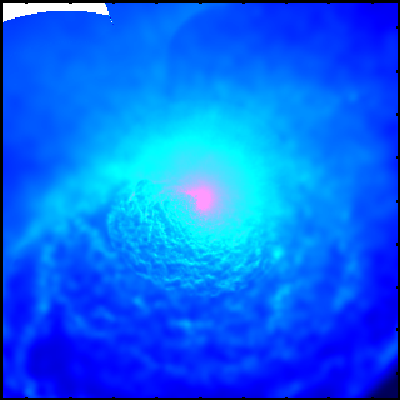}

  \includegraphics[trim={0.1cm 0.5cm 0.1cm 0},clip,width=\factoD\columnwidth]{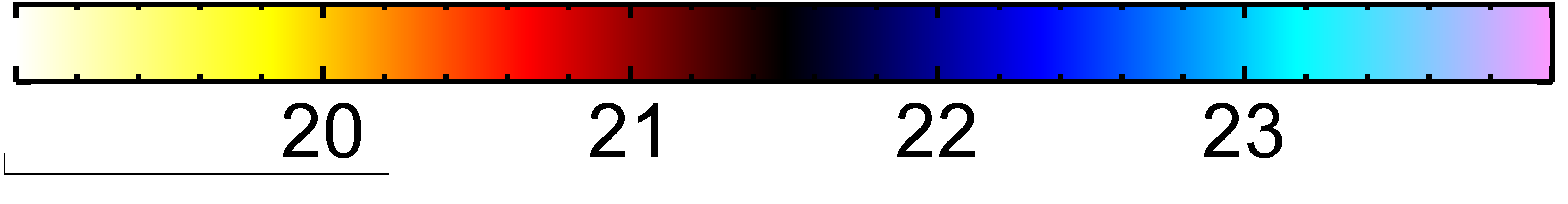}

  \caption{Spatial variation of the absorbing columns computed with equation~(\ref{eq:nH}) for NC8.5 at 0.1, 1, and 10~keV  (left to right) with $\omega=252^\circ$ (top two rows) and $\omega=90^\circ$ (bottom two rows).  The first and third rows show $\pm$100$a$ with tick marks every 5$\times$10$^{15}$\,cm, while the second and fourth rows show $\pm$20$a$ with tick marks every 10$^{15}$\,cm.  The colour scale units are log cm$^{-2}$.}
  \label{fi:nH1}
\end{center}
\end{figure}

Fig.~\ref{fi:nH2} shows the 1\,keV absorbing columns for all four models.  The $\omega=90^\circ$ images (bottom rows) show the decreasing absorbing column as $\dot{M}_A$ decreases, as well as the larger deviations in the primary wind from periastron passage (see Fig.~\ref{fi:SPH100}).  For $\omega=252^\circ$, the projected area of X-rays seen through the primary wind (blue in the left-hand panels, blue to black to red in the right-hand panels) increases with decreasing $\dot{M}_A$ since the opening angle of the shock cone is wider.  The absorption from the primary shell in the upper portion of the plots also decreases with decreasing $\dot{M}_A$.  Comparing the no-coupling model (left column) to the strong-coupling models (right three columns) also shows that the $n_\textrm{H}$ at the very centre of the image (best seen in the second row) is much larger for the strong coupling, again indicating X-rays are coming from near the centre, while the no-coupling X-rays are coming from farther out (see also Fig.~\ref{fi:SPH002}, third row).

\newcommand*{\factB}{0.23}
\begin{figure}
\begin{center}
  \includegraphics[width=\factB\columnwidth]{CCEnHImage_eC7_8p5_1p0keV.png}\put(-40,58){\fontfamily{phv}\selectfont NC8.5}\put(-54,48){\fontfamily{phv}\selectfont $\pm$100$a$}\put(-63,0){\fontfamily{phv}\selectfont \rotatebox{90}{252$^\circ$}}%
  \includegraphics[width=\factB\columnwidth]{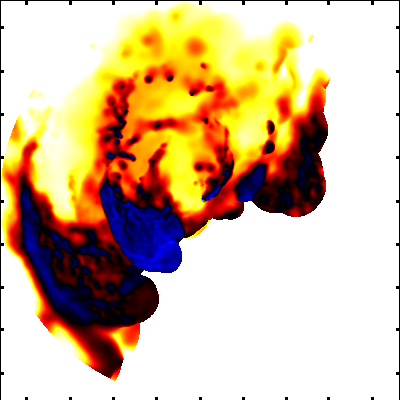}\put(-40,58){\fontfamily{phv}\selectfont SC8.5}%
  \includegraphics[width=\factB\columnwidth]{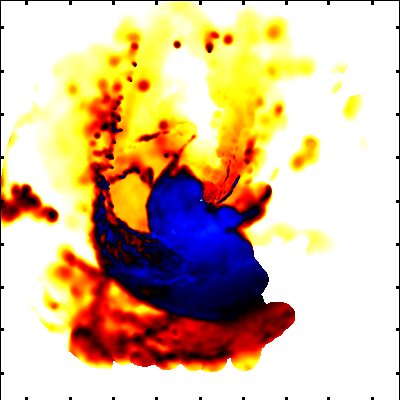}\put(-40,58){\fontfamily{phv}\selectfont SC4.8}%
  \includegraphics[width=\factB\columnwidth]{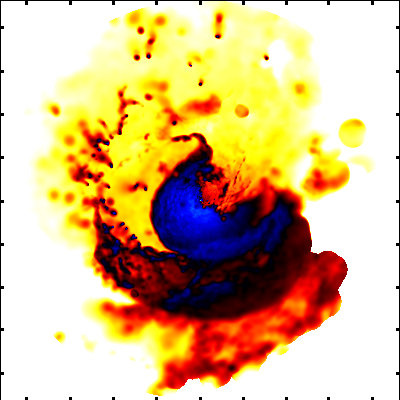}\put(-40,58){\fontfamily{phv}\selectfont SC2.4}

  \includegraphics[width=\factB\columnwidth]{CCEnHImage_eC7_8p5_1p0keV_20a.png}\put(-54,48){\fontfamily{phv}\selectfont \textcolor{white}{$\pm$20$a$}}\put(-61,41){\fontfamily{phv}\selectfont \rotatebox{90}{$\omega=$}}%
  \includegraphics[width=\factB\columnwidth]{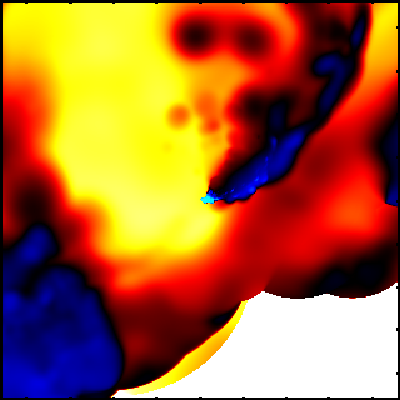}%
  \includegraphics[width=\factB\columnwidth]{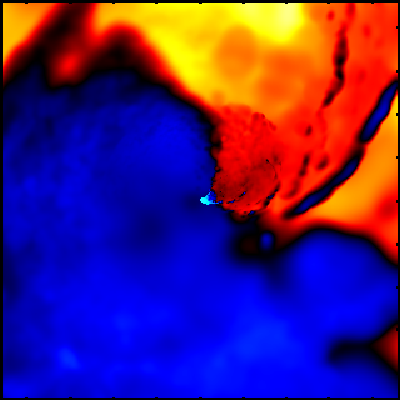}%
  \includegraphics[width=\factB\columnwidth]{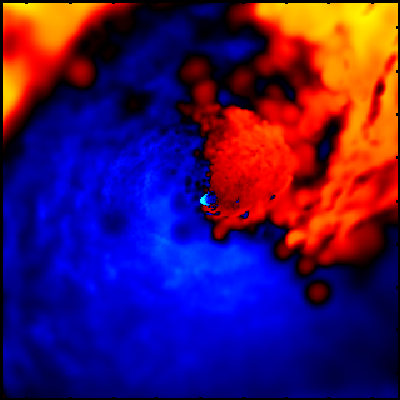}

  \includegraphics[width=\factB\columnwidth]{CCEnHImage_eC7_8p5_1p0keV_SA.png}\put(-54,48){\fontfamily{phv}\selectfont $\pm$100$a$}\put(-63,0){\fontfamily{phv}\selectfont \rotatebox{90}{90$^\circ$}}%
  \includegraphics[width=\factB\columnwidth]{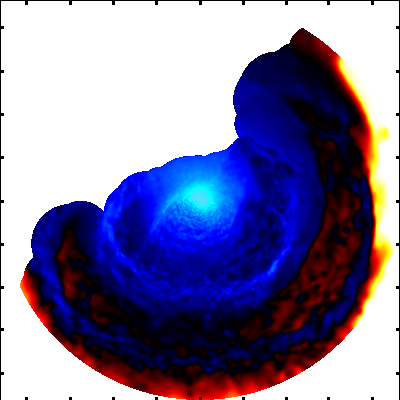}%
  \includegraphics[width=\factB\columnwidth]{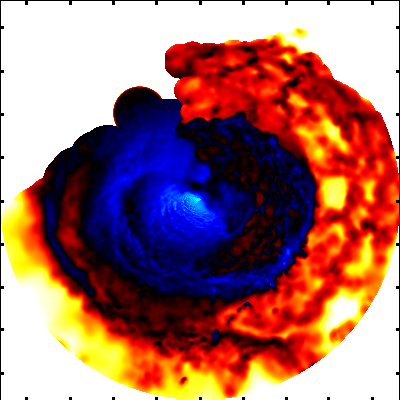}%
  \includegraphics[width=\factB\columnwidth]{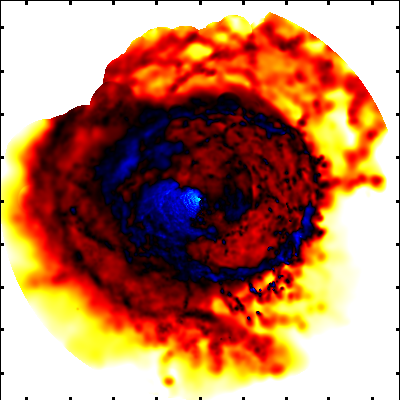}

  \includegraphics[width=\factB\columnwidth]{CCEnHImage_eC7_8p5_1p0keV_SA_20a.png}\put(-54,48){\fontfamily{phv}\selectfont $\pm$20$a$}\put(-61,41){\fontfamily{phv}\selectfont \rotatebox{90}{$\omega=$}}%
  \includegraphics[width=\factB\columnwidth]{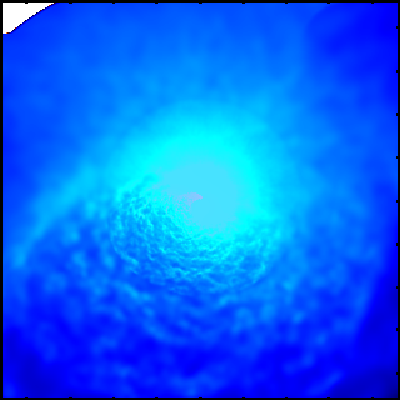}%
  \includegraphics[width=\factB\columnwidth]{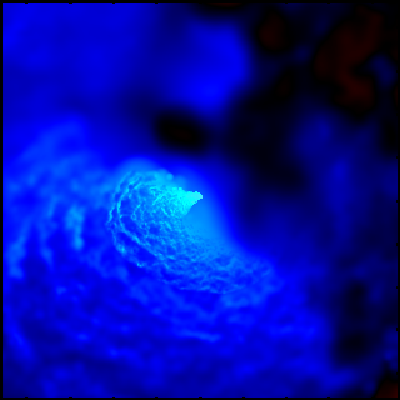}%
  \includegraphics[width=\factB\columnwidth]{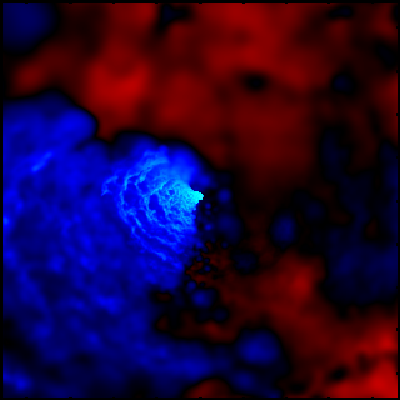}

  \includegraphics[trim={0.1cm 0.5cm 0.1cm 0},clip,width=\factoD\columnwidth]{CCEnH_Colorbar.png}

  \caption{Same as Fig.~\ref{fi:nH1}, but showing the absorbing columns at 1 keV for NC8.5, SC8.5, SC4.8, and SC2.4 (left to right).}
  \label{fi:nH2}
\end{center}
\end{figure}

\begin{figure}
\begin{center}
  \includegraphics[width=\columnwidth]{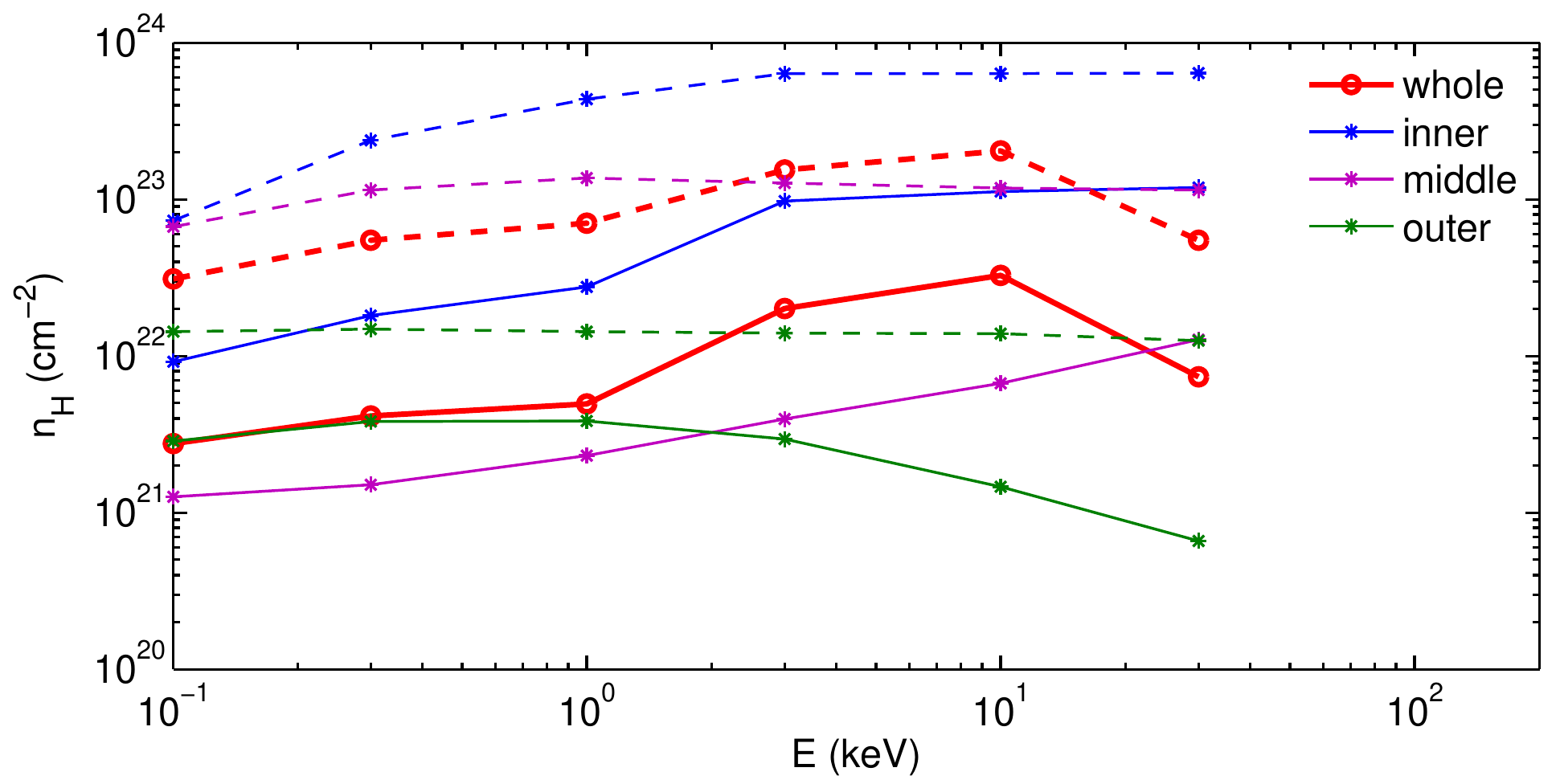}
  \includegraphics[width=\columnwidth]{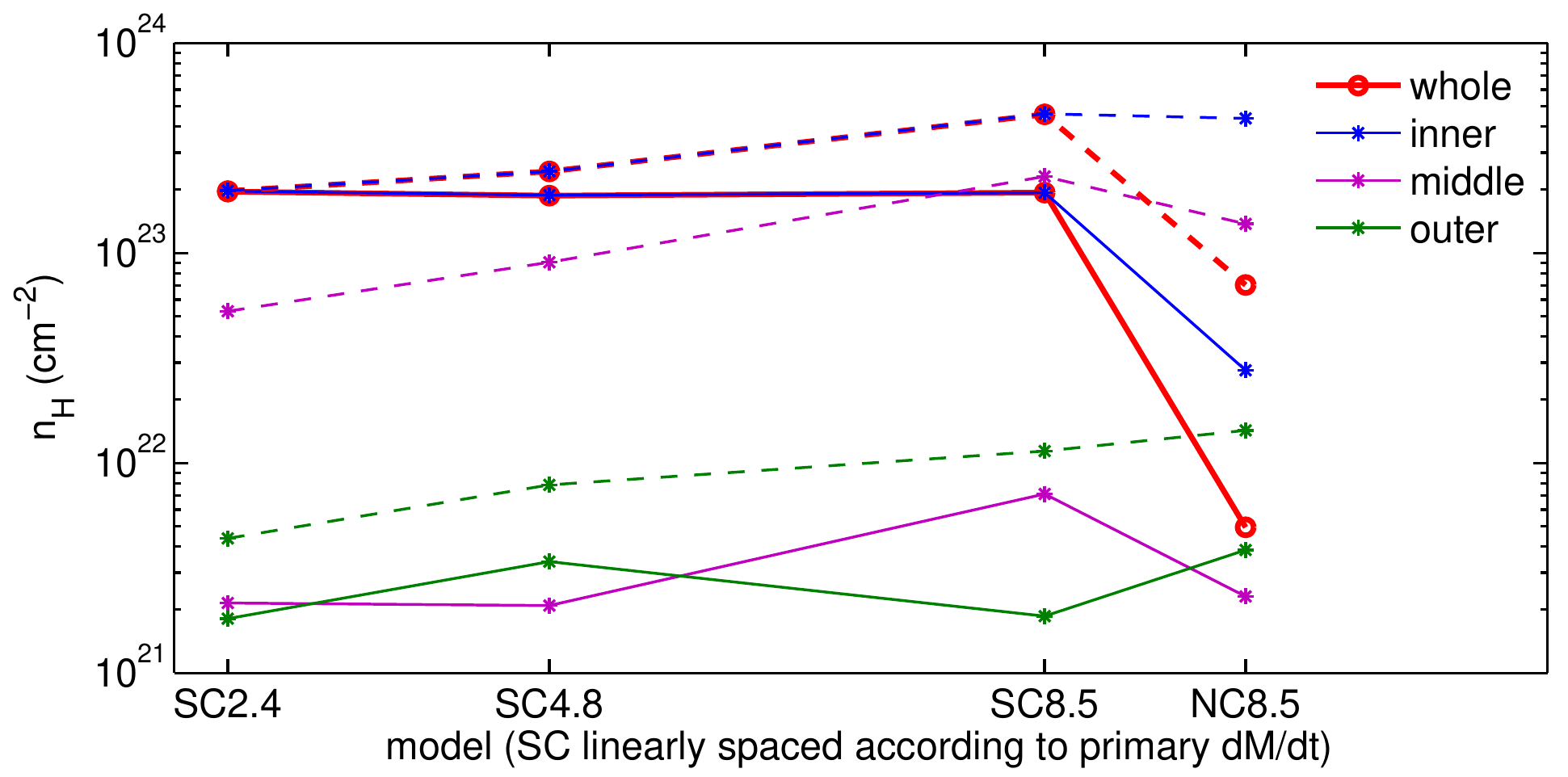}

  \caption{Absorption column versus energy for NC8.5 (top panel), and absorption column at 1\,keV for all models (bottom).  Both panels contain $\omega=252^\circ$ (solid) and $\omega=90^\circ$ (dashed).  The $x$-axis in the bottom panel is linearly spaced in $\dot{M}_A$ for the strong-coupling models, while the no-coupling model is offset to the right for clarity.}
  \label{fi:nH3}
\end{center}
\end{figure}

While the pixel maps are useful for seeing the range of absorption values, the typical data fitting results only produce a single value of absorption per emission location.  Therefore, we convert these absorption images to a single value by weighting it with the intrinsic intensity per pixel to determine the single absorption value for that energy,
\begin{equation}\label{eq:nHw}
  n_\textrm{H}(E)=\frac{\sum n_\textrm{H}(E,x',y')\; I_*(E,x',y')}{\sum I_*(E,x',y')}.
\end{equation}
The top panel of Fig.~\ref{fi:nH3} shows $n_\textrm{H}(E)$ for SC8.5 split into its components for $\omega=252^\circ$ (solid) and $\omega=90^\circ$ (dashed).  (Note that since each component is a weighted average, the average of the inner, middle, and outer does not equate to the value of the whole area.)  The majority of lines increase with increasing $E$, indicating that the higher energy X-rays are produced closer in to the centre.  The exception is the outer emission from $\omega=252^\circ$, which the top row of Fig.~\ref{fi:nH2} shows is from fewer X-rays being produced along rays through the primary wind as $E$ increases.  Consequently, more hard X-rays pass through the lower density secondary wind in the outer region, and thus have lower $n_\textrm{H}$'s as $E$ increases.  The bottom panel of Fig.~\ref{fi:nH3} show the absorption variation at 1\,keV for the four models.  Since the strong-coupling models are dominated by the X-ray emission from the centre, the whole and inner absorbing columns (blue and red) are identical for both lines of sight.  On the other hand, the no-coupling whole values have meaningful contributions from each region.  The strong-coupling models for $\omega=252^\circ$ produce approximately the same $n_\textrm{H}$ since they are dominated by the emission near the centre of the system, as shown by the small cyan region in the centre of the right three panels on the second row of Fig.~\ref{fi:nH2}.  The absorptions for $\omega=90^\circ$ increase as $\dot{M}_A$ increases as expected.

\section{Conclusions}\label{sec:C}

We model the CCE X-ray spectra of $\eta$ Carinae from 3D hydrodynamic models and radiative transfer calculations.  Embedding successively higher resolution simulations of $r_\textrm{max}$ = 100$a$, 10$a$, and 1.5$a$ allows for the most detailed density and temperature structure of $\eta$ Carinae out to $r=100a$ to date, and thus provides an excellent basis for calculating the CCE spectra.  The coupling of the primary radiation to the secondary wind is important for determining how the secondary wind, ejected on the side of the secondary star away from the primary, accelerates.  The acceleration components from both stellar radiation fields are additive in this region, so a strong coupling produces wind speeds, post-shock temperatures, and X-ray fluxes greater than that expected for a terminal-speed shock, and certainly greater than if there is no coupling between the primary radiation field and the secondary wind.  The primary mass-loss rate is also an important parameter; there is recent observational evidence that it might have changed \citep{CorcoranP10,MehnerP10}, and the wind-wind collision region is strongly affected by it \citep{MaduraP13}.

The model CCE spectra for $\dot{M}_A=8.5\times10^{-4}$\,M$_\odot$\,yr$^{-1}$ reproduce the properties of the observed CCE spectrum.  For $\omega=252^\circ$ the strong- and no-coupling spectra bound the observation, while for $\omega=90^\circ$ the two model spectra bound the hard component and converge on the soft component of the observed spectra. 
Therefore, $\eta$ Carinae has a moderate coupling between the primary radiation and secondary wind, and the CCE X-ray emission is generated from both the secondary wind colliding with primary wind ejected during the previous periastron passage, and the smaller scale emission downstream of the leading arm of the current wind-wind collision shock.  Additionally, the CCE is not a good diagnostic for distinguishing between the two observer lines of sight.

We also compute the temperature distribution and the absorbing column of X-rays for comparing with these parameters typically derived from fitting X-ray data.  As expected, the model produces a much wider range of both parameters than a several-temperature fit to the data.

\section*{Acknowledgements}

CMPR is supported by an appointment to the NASA Postdoctoral Program at the Goddard Space Flight Center, administered by Oak Ridge Associated Universities through a contract with NASA.  KH is supported by the \textit{Chandra} grant GO4-15019A, the \textit{XMM-Newton} grant NNX15AK62G, and the ADAP grant NNX15AM96G. SPO acknowledges partial support of NASA  Astrophysics Theory Program grant NNX11AC40G, awarded to the University of Delaware. Resources supporting this work were provided by the NASA High-End Computing (HEC) Program through the NASA Advanced Supercomputing (NAS) Division at Ames Research Center.




\bibliographystyle{mnras}
\bibliography{sampleAF,sampleGM,sampleNS,sampleTZ}

\begin{thebibliography}{}
\makeatletter
\relax
\def\mn@urlcharsother{\let\do\@makeother \do\$\do\&\do\#\do\^\do\_\do\%\do\~}
\def\mn@doi{\begingroup\mn@urlcharsother \@ifnextchar [ {\mn@doi@}
  {\mn@doi@[]}}
\def\mn@doi@[#1]#2{\def\@tempa{#1}\ifx\@tempa\@empty \href
  {http://dx.doi.org/#2} {doi:#2}\else \href {http://dx.doi.org/#2} {#1}\fi
  \endgroup}
\def\mn@eprint#1#2{\mn@eprint@#1:#2::\@nil}
\def\mn@eprint@arXiv#1{\href {http://arxiv.org/abs/#1} {{\tt arXiv:#1}}}
\def\mn@eprint@dblp#1{\href {http://dblp.uni-trier.de/rec/bibtex/#1.xml}
  {dblp:#1}}
\def\mn@eprint@#1:#2:#3:#4\@nil{\def\@tempa {#1}\def\@tempb {#2}\def\@tempc
  {#3}\ifx \@tempc \@empty \let \@tempc \@tempb \let \@tempb \@tempa \fi \ifx
  \@tempb \@empty \def\@tempb {arXiv}\fi \@ifundefined
  {mn@eprint@\@tempb}{\@tempb:\@tempc}{\expandafter \expandafter \csname
  mn@eprint@\@tempb\endcsname \expandafter{\@tempc}}}

\bibitem[\protect\citeauthoryear{{Abraham} \& {Falceta-Gon{\c
  c}alves}}{{Abraham} \& {Falceta-Gon{\c
  c}alves}}{2010}]{AbrahamFalcetaGoncalves10}
{Abraham} Z.,  {Falceta-Gon{\c c}alves} D.,  2010, \mn@doi [\mnras]
  {10.1111/j.1365-2966.2009.15692.x}, \href
  {http://adsabs.harvard.edu/abs/2010MNRAS.401..687A} {401, 687}

\bibitem[\protect\citeauthoryear{{Arnaud}}{{Arnaud}}{1996}]{Arnaud96}
{Arnaud} K.~A.,  1996, in {Jacoby} G.~H.,  {Barnes} J.,  eds,  Astronomical
  Society of the Pacific Conference Series Vol. 101, Astronomical Data Analysis
  Software and Systems V. p.~17

\bibitem[\protect\citeauthoryear{{Asplund}, {Grevesse}, {Sauval}  \&
  {Scott}}{{Asplund} et~al.}{2009}]{AsplundP09}
{Asplund} M.,  {Grevesse} N.,  {Sauval} A.~J.,   {Scott} P.,  2009, \mn@doi
  [\araa] {10.1146/annurev.astro.46.060407.145222}, \href
  {http://adsabs.harvard.edu/abs/2009ARA%26A..47..481A} {47, 481}

\bibitem[\protect\citeauthoryear{{Bate}, {Bonnell}  \& {Price}}{{Bate}
  et~al.}{1995}]{BateBonnellPrice95}
{Bate} M.~R.,  {Bonnell} I.~A.,   {Price} N.~M.,  1995, \mnras, \href
  {http://adsabs.harvard.edu/abs/1995MNRAS.277..362B} {277, 362}

\bibitem[\protect\citeauthoryear{{Benz}}{{Benz}}{1990}]{Benz90}
{Benz} W.,  1990, in {Buchler} J.~R.,  ed., Numerical Modelling of Nonlinear
  Stellar Pulsations Problems and Prospects. p.~269

\bibitem[\protect\citeauthoryear{{Castor}, {Abbott}  \& {Klein}}{{Castor}
  et~al.}{1975}]{CastorAbbottKlein75}
{Castor} J.~I.,  {Abbott} D.~C.,   {Klein} R.~I.,  1975, \mn@doi [\apj]
  {10.1086/153315}, \href {http://adsabs.harvard.edu/abs/1975ApJ...195..157C}
  {195, 157}

\bibitem[\protect\citeauthoryear{{Clementel}, {Madura}, {Kruip}, {Paardekooper}
   \& {Gull}}{{Clementel} et~al.}{2015a}]{ClementelP15a}
{Clementel} N.,  {Madura} T.~I.,  {Kruip} C.~J.~H.,  {Paardekooper} J.-P.,
  {Gull} T.~R.,  2015a, \mn@doi [\mnras] {10.1093/mnras/stu2614}, \href
  {http://adsabs.harvard.edu/abs/2015MNRAS.447.2445C} {447, 2445}

\bibitem[\protect\citeauthoryear{{Clementel}, {Madura}, {Kruip}  \&
  {Paardekooper}}{{Clementel} et~al.}{2015b}]{ClementelP15b}
{Clementel} N.,  {Madura} T.~I.,  {Kruip} C.~J.~H.,   {Paardekooper} J.-P.,
  2015b, \mn@doi [\mnras] {10.1093/mnras/stv696}, \href
  {http://adsabs.harvard.edu/abs/2015MNRAS.450.1388C} {450, 1388}

\bibitem[\protect\citeauthoryear{{Corcoran}}{{Corcoran}}{2005}]{Corcoran05}
{Corcoran} M.~F.,  2005, \mn@doi [\aj] {10.1086/428756}, \href
  {http://adsabs.harvard.edu/abs/2005AJ....129.2018C} {129, 2018}

\bibitem[\protect\citeauthoryear{{Corcoran}, {Ishibashi}, {Swank}  \&
  {Petre}}{{Corcoran} et~al.}{2001}]{CorcoranP01}
{Corcoran} M.~F.,  {Ishibashi} K.,  {Swank} J.~H.,   {Petre} R.,  2001, \mn@doi
  [\apj] {10.1086/318416}, \href
  {http://adsabs.harvard.edu/abs/2001ApJ...547.1034C} {547, 1034}

\bibitem[\protect\citeauthoryear{{Corcoran}, {Hamaguchi}, {Pittard}, {Russell},
  {Owocki}, {Parkin}  \& {Okazaki}}{{Corcoran} et~al.}{2010}]{CorcoranP10}
{Corcoran} M.~F.,  {Hamaguchi} K.,  {Pittard} J.~M.,  {Russell} C.~M.~P.,
  {Owocki} S.~P.,  {Parkin} E.~R.,   {Okazaki} A.,  2010, \mn@doi [\apj]
  {10.1088/0004-637X/725/2/1528}, \href
  {http://adsabs.harvard.edu/abs/2010ApJ...725.1528C} {725, 1528}

\bibitem[\protect\citeauthoryear{{Damineli} et~al.,}{{Damineli}
  et~al.}{2008}]{DamineliP08}
{Damineli} A.,  et~al., 2008, \mn@doi [\mnras]
  {10.1111/j.1365-2966.2008.13214.x}, \href
  {http://adsabs.harvard.edu/abs/2008MNRAS.386.2330D} {386, 2330}

\bibitem[\protect\citeauthoryear{{Davidson} \& {Humphreys}}{{Davidson} \&
  {Humphreys}}{1997}]{DavidsonHumphreys97}
{Davidson} K.,  {Humphreys} R.~M.,  1997, \mn@doi [\araa]
  {10.1146/annurev.astro.35.1.1}, \href
  {http://adsabs.harvard.edu/abs/1997ARA%26A..35....1D} {35, 1}

\bibitem[\protect\citeauthoryear{{Drew}}{{Drew}}{1989}]{Drew89}
{Drew} J.~E.,  1989, \mn@doi [\apjs] {10.1086/191374}, \href
  {http://adsabs.harvard.edu/abs/1989ApJS...71..267D} {71, 267}

\bibitem[\protect\citeauthoryear{{Falceta-Gon{\c c}alves} \&
  {Abraham}}{{Falceta-Gon{\c c}alves} \&
  {Abraham}}{2009}]{FalcetaGoncalvesAbraham09}
{Falceta-Gon{\c c}alves} D.,  {Abraham} Z.,  2009, \mn@doi [\mnras]
  {10.1111/j.1365-2966.2009.15364.x}, \href
  {http://adsabs.harvard.edu/abs/2009MNRAS.399.1441F} {399, 1441}

\bibitem[\protect\citeauthoryear{{Gayley}, {Owocki}  \& {Cranmer}}{{Gayley}
  et~al.}{1997}]{GayleyOwockiCranmer97}
{Gayley} K.~G.,  {Owocki} S.~P.,   {Cranmer} S.~R.,  1997, \mn@doi [\apj]
  {10.1086/303573}, \href {http://adsabs.harvard.edu/abs/1997ApJ...475..786G}
  {475, 786}

\bibitem[\protect\citeauthoryear{{Groh}, {Hillier}, {Madura}  \&
  {Weigelt}}{{Groh} et~al.}{2012}]{GrohP12}
{Groh} J.~H.,  {Hillier} D.~J.,  {Madura} T.~I.,   {Weigelt} G.,  2012, \mn@doi
  [\mnras] {10.1111/j.1365-2966.2012.20984.x}, \href
  {http://adsabs.harvard.edu/abs/2012MNRAS.423.1623G} {423, 1623}

\bibitem[\protect\citeauthoryear{{Hamaguchi} \& {Corcoran}}{{Hamaguchi} \&
  {Corcoran}}{2015}]{HamaguchiP15}
{Hamaguchi} K.,  {Corcoran} M.~F.,  2015, in EAS Publications Series. pp
  37--40, \mn@doi{10.1051/eas/1571006}

\bibitem[\protect\citeauthoryear{{Hamaguchi} et~al.,}{{Hamaguchi}
  et~al.}{2007}]{HamaguchiP07}
{Hamaguchi} K.,  et~al., 2007, \mn@doi [\apj] {10.1086/518101}, \href
  {http://adsabs.harvard.edu/abs/2007ApJ...663..522H} {663, 522}

\bibitem[\protect\citeauthoryear{{Hamaguchi} et~al.,}{{Hamaguchi}
  et~al.}{2014}]{HamaguchiP14}
{Hamaguchi} K.,  et~al., 2014, \mn@doi [\apj] {10.1088/0004-637X/784/2/125},
  \href {http://adsabs.harvard.edu/abs/2014ApJ...784..125H} {784, 125}

\bibitem[\protect\citeauthoryear{{Hillier}, {Davidson}, {Ishibashi}  \&
  {Gull}}{{Hillier} et~al.}{2001}]{HillierP01}
{Hillier} D.~J.,  {Davidson} K.,  {Ishibashi} K.,   {Gull} T.,  2001, \mn@doi
  [\apj] {10.1086/320948}, \href
  {http://adsabs.harvard.edu/abs/2001ApJ...553..837H} {553, 837}

\bibitem[\protect\citeauthoryear{{Hillier} et~al.,}{{Hillier}
  et~al.}{2006}]{HillierP06}
{Hillier} D.~J.,  et~al., 2006, \mn@doi [\apj] {10.1086/501225}, \href
  {http://adsabs.harvard.edu/abs/2006ApJ...642.1098H} {642, 1098}

\bibitem[\protect\citeauthoryear{{Kashi} \& {Soker}}{{Kashi} \&
  {Soker}}{2008}]{KashiSoker08}
{Kashi} A.,  {Soker} N.,  2008, \mn@doi [\mnras]
  {10.1111/j.1365-2966.2008.13883.x}, \href
  {http://adsabs.harvard.edu/abs/2008MNRAS.390.1751K} {390, 1751}

\bibitem[\protect\citeauthoryear{{Kashi} \& {Soker}}{{Kashi} \&
  {Soker}}{2009}]{KashiSoker09c}
{Kashi} A.,  {Soker} N.,  2009, \mn@doi [\mnras]
  {10.1111/j.1365-2966.2008.14331.x}, \href
  {http://adsabs.harvard.edu/abs/2009MNRAS.394..923K} {394, 923}

\bibitem[\protect\citeauthoryear{{Kashi}, {Soker}  \& {Akashi}}{{Kashi}
  et~al.}{2011}]{KashiSokerAkashi11}
{Kashi} A.,  {Soker} N.,   {Akashi} M.,  2011, \mn@doi [\mnras]
  {10.1111/j.1365-2966.2011.18340.x}, \href
  {http://adsabs.harvard.edu/abs/2011MNRAS.413.2658K} {413, 2658}

\bibitem[\protect\citeauthoryear{{Leutenegger}, {Cohen}, {Zsarg{\'o}},
  {Martell}, {MacArthur}, {Owocki}, {Gagn{\'e}}  \& {Hillier}}{{Leutenegger}
  et~al.}{2010}]{LeuteneggerP10}
{Leutenegger} M.~A.,  {Cohen} D.~H.,  {Zsarg{\'o}} J.,  {Martell} E.~M.,
  {MacArthur} J.~P.,  {Owocki} S.~P.,  {Gagn{\'e}} M.,   {Hillier} D.~J.,
  2010, \mn@doi [\apj] {10.1088/0004-637X/719/2/1767}, \href
  {http://adsabs.harvard.edu/abs/2010ApJ...719.1767L} {719, 1767}

\bibitem[\protect\citeauthoryear{{Madura}, {Gull}, {Owocki}, {Groh}, {Okazaki}
  \& {Russell}}{{Madura} et~al.}{2012}]{MaduraP12}
{Madura} T.~I.,  {Gull} T.~R.,  {Owocki} S.~P.,  {Groh} J.~H.,  {Okazaki}
  A.~T.,   {Russell} C.~M.~P.,  2012, \mn@doi [\mnras]
  {10.1111/j.1365-2966.2011.20165.x}, \href
  {http://adsabs.harvard.edu/abs/2012MNRAS.420.2064M} {420, 2064}

\bibitem[\protect\citeauthoryear{{Madura} et~al.,}{{Madura}
  et~al.}{2013}]{MaduraP13}
{Madura} T.~I.,  et~al., 2013, \mn@doi [\mnras] {10.1093/mnras/stt1871}, \href
  {http://adsabs.harvard.edu/abs/2013MNRAS.436.3820M} {436, 3820}

\bibitem[\protect\citeauthoryear{{Mehner}, {Davidson}, {Humphreys}, {Martin},
  {Ishibashi}, {Ferland}  \& {Walborn}}{{Mehner} et~al.}{2010}]{MehnerP10}
{Mehner} A.,  {Davidson} K.,  {Humphreys} R.~M.,  {Martin} J.~C.,  {Ishibashi}
  K.,  {Ferland} G.~J.,   {Walborn} N.~R.,  2010, \mn@doi [\apjl]
  {10.1088/2041-8205/717/1/L22}, \href
  {http://adsabs.harvard.edu/abs/2010ApJ...717L..22M} {717, L22}

\bibitem[\protect\citeauthoryear{{Moffat} \& {Corcoran}}{{Moffat} \&
  {Corcoran}}{2009}]{MoffatCorcoran09}
{Moffat} A.~F.~J.,  {Corcoran} M.~F.,  2009, \mn@doi [\apj]
  {10.1088/0004-637X/707/1/693}, \href
  {http://adsabs.harvard.edu/abs/2009ApJ...707..693M} {707, 693}

\bibitem[\protect\citeauthoryear{{Okazaki}, {Owocki}, {Russell}  \&
  {Corcoran}}{{Okazaki} et~al.}{2008}]{OkazakiP08}
{Okazaki} A.~T.,  {Owocki} S.~P.,  {Russell} C.~M.~P.,   {Corcoran} M.~F.,
  2008, \mn@doi [\mnras] {10.1111/j.1745-3933.2008.00496.x}, \href
  {http://adsabs.harvard.edu/abs/2008MNRAS.388L..39O} {388, L39}

\bibitem[\protect\citeauthoryear{{Owocki} \& {Gayley}}{{Owocki} \&
  {Gayley}}{1995}]{OwockiGayley95}
{Owocki} S.~P.,  {Gayley} K.~G.,  1995, \mn@doi [\apjl] {10.1086/309786}, \href
  {http://adsabs.harvard.edu/abs/1995ApJ...454L.145O} {454, L145}

\bibitem[\protect\citeauthoryear{{Parkin} \& {Sim}}{{Parkin} \&
  {Sim}}{2013}]{ParkinSim13}
{Parkin} E.~R.,  {Sim} S.~A.,  2013, \mn@doi [\apj]
  {10.1088/0004-637X/767/2/114}, \href
  {http://adsabs.harvard.edu/abs/2013ApJ...767..114P} {767, 114}

\bibitem[\protect\citeauthoryear{{Parkin}, {Pittard}, {Corcoran}, {Hamaguchi}
  \& {Stevens}}{{Parkin} et~al.}{2009}]{ParkinP09}
{Parkin} E.~R.,  {Pittard} J.~M.,  {Corcoran} M.~F.,  {Hamaguchi} K.,
  {Stevens} I.~R.,  2009, \mn@doi [\mnras] {10.1111/j.1365-2966.2009.14475.x},
  \href {http://adsabs.harvard.edu/abs/2009MNRAS.394.1758P} {394, 1758}

\bibitem[\protect\citeauthoryear{{Parkin}, {Pittard}, {Corcoran}  \&
  {Hamaguchi}}{{Parkin} et~al.}{2011}]{ParkinP11}
{Parkin} E.~R.,  {Pittard} J.~M.,  {Corcoran} M.~F.,   {Hamaguchi} K.,  2011,
  \mn@doi [\apj] {10.1088/0004-637X/726/2/105}, \href
  {http://adsabs.harvard.edu/abs/2011ApJ...726..105P} {726, 105}

\bibitem[\protect\citeauthoryear{{Pittard} \& {Corcoran}}{{Pittard} \&
  {Corcoran}}{2002}]{PittardCorcoran02}
{Pittard} J.~M.,  {Corcoran} M.~F.,  2002, \mn@doi [\aap]
  {10.1051/0004-6361:20020025}, \href
  {http://adsabs.harvard.edu/abs/2002A%26A...383..636P} {383, 636}

\bibitem[\protect\citeauthoryear{{Price}}{{Price}}{2007}]{Price07}
{Price} D.~J.,  2007, \mn@doi [\pasa] {10.1071/AS07022}, \href
  {http://adsabs.harvard.edu/abs/2007PASA...24..159P} {24, 159}

\bibitem[\protect\citeauthoryear{{Russell}}{{Russell}}{2013}]{Russell13}
{Russell} C.~M.~P.,  2013, PhD thesis, University of Delaware

\bibitem[\protect\citeauthoryear{{Russell}, {Corcoran}, {Okazaki}, {Madura}  \&
  {Owocki}}{{Russell} et~al.}{2011a}]{RussellP11a}
{Russell} C.~M.~P.,  {Corcoran} M.~F.,  {Okazaki} A.~T.,  {Madura} T.~I.,
  {Owocki} S.~P.,  2011a, Bulletin de la Societe Royale des Sciences de Liege,
  \href {http://adsabs.harvard.edu/abs/2011BSRSL..80..719R} {80, 719}

\bibitem[\protect\citeauthoryear{{Russell}, {Corcoran}, {Okazaki}, {Madura}  \&
  {Owocki}}{{Russell} et~al.}{2011b}]{RussellP11b}
{Russell} C.~M.~P.,  {Corcoran} M.~F.,  {Okazaki} A.~T.,  {Madura} T.~I.,
  {Owocki} S.~P.,  2011b, in {C.~Neiner, G.~Wade, G.~Meynet, \& G.~Peters} ed.,
   IAU Symposium Vol. 272, IAU Symposium. pp 630--631,
  \mn@doi{10.1017/S1743921311011641}

\bibitem[\protect\citeauthoryear{{Seward}, {Forman}, {Giacconi}, {Griffiths},
  {Harnden}, {Jones}  \& {Pye}}{{Seward} et~al.}{1979}]{SewardP79}
{Seward} F.~D.,  {Forman} W.~R.,  {Giacconi} R.,  {Griffiths} R.~E.,  {Harnden}
  Jr. F.~R.,  {Jones} C.,   {Pye} J.~P.,  1979, \mn@doi [\apjl]
  {10.1086/183108}, \href {http://adsabs.harvard.edu/abs/1979ApJ...234L..55S}
  {234, L55}

\bibitem[\protect\citeauthoryear{{Smith}, {Brickhouse}, {Liedahl}  \&
  {Raymond}}{{Smith} et~al.}{2001}]{SmithP01}
{Smith} R.~K.,  {Brickhouse} N.~S.,  {Liedahl} D.~A.,   {Raymond} J.~C.,  2001,
  \mn@doi [\apjl] {10.1086/322992}, \href
  {http://adsabs.harvard.edu/abs/2001ApJ...556L..91S} {556, L91}

\bibitem[\protect\citeauthoryear{{Soker} \& {Behar}}{{Soker} \&
  {Behar}}{2006}]{SokerBehar06}
{Soker} N.,  {Behar} E.,  2006, \mn@doi [\apj] {10.1086/508336}, \href
  {http://adsabs.harvard.edu/abs/2006ApJ...652.1563S} {652, 1563}

\bibitem[\protect\citeauthoryear{{Stevens} \& {Kallman}}{{Stevens} \&
  {Kallman}}{1990}]{StevensKallman90}
{Stevens} I.~R.,  {Kallman} T.~R.,  1990, \mn@doi [\apj] {10.1086/169486},
  \href {http://adsabs.harvard.edu/abs/1990ApJ...365..321S} {365, 321}

\bibitem[\protect\citeauthoryear{{Stevens} \& {Pollock}}{{Stevens} \&
  {Pollock}}{1994}]{StevensPollock94}
{Stevens} I.~R.,  {Pollock} A.~M.~T.,  1994, \mnras, \href
  {http://adsabs.harvard.edu/abs/1994MNRAS.269..226S} {269, 226}

\bibitem[\protect\citeauthoryear{{Townsend}}{{Townsend}}{2009}]{Townsend09}
{Townsend} R.~H.~D.,  2009, \mn@doi [\apjs] {10.1088/0067-0049/181/2/391},
  \href {http://adsabs.harvard.edu/abs/2009ApJS..181..391T} {181, 391}

\bibitem[\protect\citeauthoryear{{Wilms}, {Allen}  \& {McCray}}{{Wilms}
  et~al.}{2000}]{WilmsAllenMcCray00}
{Wilms} J.,  {Allen} A.,   {McCray} R.,  2000, \mn@doi [\apj] {10.1086/317016},
  \href {http://adsabs.harvard.edu/abs/2000ApJ...542..914W} {542, 914}

\makeatother
\end{thebibliography}




%
%


\bsp	
\label{lastpage}
\end{document}